\documentclass[journal]{IEEEtran}
\IEEEoverridecommandlockouts
\usepackage{lineno}
\usepackage{hyperref}
\usepackage{cite}
\usepackage{amsmath,amssymb,amsfonts,cases} 
\usepackage{amsmath}
\usepackage{amsthm}
\DeclareMathOperator*{\argmax}{argmax}
\DeclareMathOperator*{\argmin}{argmin}
\usepackage{graphicx}
\usepackage{textcomp}
\usepackage{xcolor}
\usepackage{graphicx}
\usepackage{float}
\usepackage{subfigure}
\usepackage{amsmath,mleftright}
\usepackage{amsfonts,amssymb}
\usepackage{mathrsfs}
\usepackage{mathtools}
\usepackage{algorithm}
\usepackage{algorithmicx}
\usepackage{algpseudocode}
\usepackage{bm}
\usepackage{multirow}
\usepackage{array}
\usepackage{amssymb}
\usepackage{amsmath}
\usepackage{cite}
\usepackage{url}
\usepackage{xcolor}
\usepackage{cite,graphicx,amsmath,amssymb}
\usepackage{subfigure}
\usepackage{fancyhdr}
\usepackage{mdwmath}
\usepackage{mdwtab}
\usepackage{caption}
\usepackage{amsthm}
\usepackage{setspace}
\usepackage{bm}
\usepackage{algorithm}
\usepackage{algpseudocode}
\usepackage{mathtools}
\usepackage{dsfont}
\usepackage{bbm}
\usepackage{framed}
\newtheorem{remark}{Remark}
\newtheorem{theorem}{Theorem}

\newtheorem{lemma}{Lemma}

\newtheorem{corollary}{Corollary}

\makeatletter
\newcommand{\biggg}{\bBigg@{3}}
\newcommand{\Biggg}{\bBigg@{3.5}}
\makeatother
\makeatletter
\renewcommand{\maketag@@@}[1]{\hbox{\m@th\normalsize\normalfont#1}}%
\makeatother
\def\BibTeX{{\rm B\kern-.05em{\sc i\kern-.025em b}\kern-.08em
    T\kern-.1667em\lower.7ex\hbox{E}\kern-.125emX}}
    \expandafter\def\expandafter\normalsize\expandafter{%
    \normalsize%
    \setlength\abovedisplayskip{4pt}%
    \setlength\belowdisplayskip{4pt}%
    \setlength\abovedisplayshortskip{2pt}%
    \setlength\belowdisplayshortskip{2pt}%
}
\begin{document}
\title{Linear Receive Beamforming for CAPA Systems}
\author{Chongjun Ouyang, Zhaolin Wang, Xingqi Zhang, and Yuanwei Liu\vspace{-10pt}
\thanks{C. Ouyang and Z. Wang are with the School of Electronic Engineering and Computer Science, Queen Mary University of London, London, E1 4NS, U.K. (email: \{c.ouyang, zhaolin.wang\}@qmul.ac.uk).}
\thanks{X. Zhang is with Department of Electrical and Computer Engineering, University of Alberta, Edmonton AB, T6G 2R3, Canada (email: xingqi.zhang@ualberta.ca).}
\thanks{Y. Liu is with the Department of Electrical and Electronic Engineering, The University of Hong Kong, Hong Kong (email: yuanwei@hku.hk).}}
\maketitle
\begin{abstract}
The performance of linear receive beamforming in continuous-aperture array (CAPA)-based uplink communications is analyzed. Three continuous beamforming techniques are proposed under the criteria of maximum-ratio combining (MRC), zero-forcing (ZF), and minimum mean-squared error (MMSE). \romannumeral1) For \emph{MRC beamforming}, a closed-form expression for the beamformer is derived to maximize per-user signal power. The achieved uplink rate and mean-squared error (MSE) in detecting received data symbols are analyzed. \romannumeral2) For \emph{ZF beamforming}, a closed-form beamformer is derived based on channel correlation to eliminate interference. As a further advance, its optimality in maximizing effective channel gain while ensuring zero inter-user interference is proven. \romannumeral3) \emph{MMSE beamforming} is established as the optimal linear receive approach for CAPAs in terms of maximizing per-user rate and minimizing MSE. Closed-form expressions are derived for the MMSE beamformer and the achievable sum-rate and sum-MSE. It is mathematically proven that all proposed beamformers lie within the signal subspace spanned by users' spatial responses. Numerical results demonstrate that CAPAs outperform conventional spatially-discrete arrays (SPDAs) by achieving higher sum-rates and lower sum-MSEs under the proposed linear beamforming techniques.
\end{abstract} 
\section{Introduction}
Since Jack Winters published his seminal paper in 1984 on multiple-input and multiple-output (MIMO) antenna systems for significant capacity gains in radio networks \cite{winters1984optimum}, multiple-antenna technology has greatly enhanced the reliability and spectral efficiency of modern cellular communication systems \cite{tse2005fundamentals,heath2018foundations}. In recent years, various multiple-antenna technologies and array architectures, such as massive MIMO \cite{marzetta2010noncooperative}, holographic MIMO \cite{pizzo2020spatially}, and gigantic MIMO \cite{bjornson2024enabling}, have emerged to improve network performance. Observing this 40-year evolution, multiple-antenna technology continues to progress toward \emph{larger} aperture sizes, \emph{denser} antenna configurations, \emph{higher} operational frequencies, and \emph{more flexible} aperture shapes. The ultimate goal of these architectures is to create a (approximately) continuous electromagnetic (EM) aperture, known as a \emph{continuous-aperture array (CAPA)} \cite{ouyang2024primer,liu2025near,liu2024capa}.

CAPA represents a significant shift from traditional spatially-discrete arrays (SPDAs) by acting as \emph{a single electrically large antenna with a continuous current distribution}, which is made of a virtually infinite number of radiating elements coupled with electronic circuits and a limited number of radio-frequency (RF) chains. CAPA serves as an ideal transmission aperture for RF signals, which provides continuous control over amplitude and phase for the current at each point where RF signals are modulated. Fully analog RF beamforming is supported across the entire communication bandwidth, enabling precise alignment of the radiation pattern to meet system requirements \cite{hu2018beyond,dardari2020communicating,gong2023holographic}. Compared to conventional SPDAs, CAPA can leverage spatial resources more effectively and flexibly to result in higher performance gains. Additionally, CAPA requires only as many RF ports as spatially multiplexed signals, making it an efficient and scalable solution \cite{bjornson2019massive}. While achieving a continuous aperture has long been a goal in antenna design, recent breakthroughs in materials science and array fabrication have made CAPA a practical reality. Notably, some CAPA prototypes have already reached the stage of commercialization \cite{liu2024capa}.
\subsection{Prior Works}\label{Prior_Work_Section}
Building on this background, research interest in the design and analysis of CAPA-based wireless communications has grown significantly. A major challenge in this field lies in the continuous integral linear operator-based signal model used by CAPAs, which fundamentally differs from the conventional matrix-based linear models applied to SPDAs \cite{di2024electromagnetic}. While some basic matrix operations, such as addition and multiplication, can be extended to the operator domain, several advanced mathematical operations that are frequently encountered in the theoretical study of MIMO, including inversion, square root, and eigen-decomposition (EVD), are not readily applicable. In the operator setting, their computation often becomes analytically intractable, and the associated theoretical analysis remains highly nontrivial. This fundamental difference motivates the research for CAPAs.

By leveraging the Fourier relationship between spatial and spatial-frequency (or wavenumber-domain) responses, the degrees of freedom (DoFs) offered by CAPAs were analyzed in \cite{poon2005degrees} through the lens of signal space. Beyond DoFs, studies have examined the signal-to-noise ratio (SNR) and power scaling laws achievable with CAPAs \cite{dardari2020communicating}. Research on the channel capacity of CAPA-based MIMO channels has also intensified. For example, \cite{gruber2008new} analyzed the Shannon information capacity of space-time wireless channels formed by paired CAPAs operating at a fixed frequency within a temporally bandlimited environment. This study was extended in \cite{wan2023mutual} to account for non-white EM interference, characterizing the capacity of CAPA-based MIMO channels using the Fredholm determinant. These analyses all assumed a free-space line-of-sight channel model. To incorporate multipath scattering, \cite{ouyang2024diversity} studied an isotropic Rayleigh fading model for CAPAs to explore the diversity-multiplexing tradeoff. These studies focus on single-user setups. Expanding on this, \cite{zhao2024continuous} derived the capacity region for both uplink and downlink channels where two single-antenna users are served by a CAPA.

In addition to performance analysis of CAPA-based communications, research has also progressed on beamforming design for CAPAs. The work in \cite{sanguinetti2022wavenumber} introduced the concept of wavenumber-division multiplexing by using Fourier series expansion to discretize the continuous spatial response into its wavenumber-domain components. Building on this approach, \cite{zhang2023pattern} designed transmit beamforming for CAPA-based downlink multiuser channels, proposing to use wavenumber-domain Fourier series to approximate the spatial response and recast the current optimization problem into a discrete form. This method has been applied to design transmit beamforming for CAPA-based uplink channels \cite{qian2024spectral}, simultaneous wireless information and power transfer (SWIPT) \cite{huang2024holographic}, and integrated sensing and communications (ISAC) \cite{liu2024holographic}.

In the aforementioned studies, the continuous current distribution was optimized by discretizing it through a Fourier series. This process converted the optimization variables from a \emph{continuous function} into a \emph{discrete matrix}. This approach enables approximate solutions via conventional optimization tools. However, due to the reliance on approximation, these methods cannot guarantee optimality and provide limited insights into system design. Additionally, achieving high precision with Fourier series typically requires a large number of expansion terms, which significantly increases computational complexity. To address these limitations, our previous work \cite{wang2024beamforming} employed the \emph{calculus of variations (CoV)} method to directly optimize the continuous current distribution for downlink CAPA systems. This approach resulted in both improved communication performance and a reduction in computational complexity by several orders of magnitude. We further extended this work to investigate the optimal linear downlink continuous beamforming design in \cite{wang2025optimal}.
\subsection{Motivation and Contributions}
The performance of linear transmit beamforming for CAPAs has been studied; see \cite{zhang2023pattern,qian2024spectral,huang2024holographic,liu2024holographic,wang2024beamforming,wang2025optimal}. In contrast, the effectiveness of linear receive beamforming in uplink CAPA-based multiuser channels has received limited attention. Our prior work \cite{zhao2024continuous} studied the uplink capacity region achievable with CAPAs using successive interference cancellation decoding, but this analysis was restricted to a two-user setup, limiting its generalizability. In general, exploring uplink beamforming design is particularly valuable, as it often allows for closed-form and analytically tractable solutions, which in turn facilitate discussions on system performance bounds. Furthermore, many theoretical results derived for the uplink can be directly leveraged to analyze downlink performance or guide downlink beamforming design via uplink-downlink duality \cite{heath2018foundations}. Additionally, uplink beamforming design enables an in-depth exploration of the interplay between the achievable data rate and the estimation error in recovering the received symbols. This relationship is instrumental in developing effective beamforming strategies for other CAPA systems \cite{wang2025beamforming}. 

Motivated by the above considerations, this paper exploits CAPAs in receiving signals from multiple users, analyzing the performance of linear receive beamforming with respect to uplink \emph{rate} and \emph{mean-squared error (MSE)} in decoding the received data symbols. We focus on beamforming methods based on the widely used criteria: maximum-ratio combining (MRC), zero-forcing (ZF), and minimum mean-squared error (MMSE). Our primary objective is to design continuous beamformers and prove their optimality under specific constraints by using the formalism of operators and integral transforms. The main contributions are listed as follows.
\begin{itemize}
  \item We derive a closed-form expression for the MRC beamformer in CAPAs to maximize per-user signal power and analyze the resulting sum-rate and sum-MSE. Besides, we propose a channel correlation-based ZF beamforming method for CAPAs to eliminate inter-user interference (IUI), and derive closed-form expressions for the beamformer. We also offer a function space interpretation of the ZF beamforming to demonstrate its optimality in maximizing received SNR while fully eliminating IUI.
  \item We prove that MMSE beamforming is the optimal uplink beamforming for CAPAs by demonstrating that maximizing the per-user uplink rate is equivalent to minimizing the per-user MSE. Using this result, we derive a closed-form expression for the optimal beamformer through an operator-based generalized Rayleigh quotient, and then analyze the achieved sum-rate and sum-MSE. Furthermore, we show that MMSE beamformer operates as a cascade of two linear filters: one for whitening the IUI and noise, and another for maximizing signal power.      
  \item We compare the performance of MRC, ZF, and MMSE beamforming for CAPAs and prove that MMSE/optimal beamforming converges to MRC beamforming in the low-SNR regime and to ZF beamforming in the high-SNR regime. We also compare the linear receive beamformers for SPDAs and CAPAs, showing that in both systems, the linear receive beamformers lie within the signal subspace spanned by users' spatial responses. 
  \item We present simulation results to validate our analytical findings and further explore the performance of the proposed CAPA-based linear receive beamforming techniques. The numerical results demonstrate that: \romannumeral1) CAPAs outperform SPDAs in sum-rate and sum-MSE under ZF and MMSE beamforming; \romannumeral2) in interference-dominated scenarios such as high-SNR regimes, SPDAs may outperform CAPAs with MRC beamforming. These findings highlight the importance of appropriate receive beamforming and interference management in CAPA-based uplink communications.
\end{itemize}

In contrast to our previous studies on linear transmit beamforming design \cite{wang2024beamforming,wang2025optimal}, this article derives closed-form optimal beamformer expressions. We also prove the optimality of the proposed ZF beamformer in maximizing effective channel gain while nullifying interference and establish equivalence between MSE-optimal and rate-optimal beamformers. These contributions were neither addressed in \cite{wang2024beamforming,wang2025optimal} nor trivially extendable from those works and existing SPDA-based results.
\subsection{Organization and Notations}
The remainder of this paper is organized as follows. Section \ref{Section: System Model} introduces the system model for uplink CAPA communications. In Sections \ref{Section: Maximum-Ratio Combining Beamforming}, \ref{Section: Zero-Forcing Beamforming}, and \ref{Section: Optimal Beamforming: Minimum Mean-Square Error Beamforming}, the sum-rate and sum-MSE performance of MRC, ZF, and MMSE beamforming are analyzed, respectively. Section \ref{Section: Performance Comparison and Further Discussion} offers a performance comparison between SPDAs and CAPAs and evaluates CAPA performance under various channel conditions. Section \ref{Section: Numerical Results} presents numerical results to confirm the accuracy of the derived findings. Finally, Section \ref{Section: Conclusion} concludes the paper.
\subsubsection*{Notations}
Throughout this paper, scalars, vectors, and matrices are denoted by non-bold, bold lower-case, and bold upper-case letters, respectively. For a matrix $\mathbf{A}$, $[\mathbf{A}]_{i,j}$, ${\mathbf{A}}^{\mathsf{T}}$, ${\mathbf{A}}^{*}$, and ${\mathbf{A}}^{\mathsf{H}}$ denote the $(i,j)$th entry, transpose, conjugate, and transpose conjugate of $\mathbf{A}$, respectively. For a square matrix $\mathbf{B}$, ${\mathbf{B}}^{-1}$ and ${\mathbf{B}}^{\frac{1}{2}}$ denotes the inverse and principal square root of $\mathbf{B}$, respectively. The notations $\lvert a\rvert$ and $\lVert \mathbf{a} \rVert$ represent the magnitude of scalar $a$ and the norm of vector $\mathbf{a}$, respectively. The notation $[{\mathbf{a}}]_i$ denotes the $i$th entry of vector $\mathbf{a}$, and ${\rm{diag}}({\mathbf{a}})$ returns a diagonal matrix whose diagonal elements are entries of $\mathbf{a}$. The identity matrix with dimensions $N\times N$ is represented by $\mathbf{I}_N$, and the zero matrix is denoted by $\mathbf{0}$. The matrix inequalities ${\mathbf{A}}\succ{\mathbf{B}}$ and ${\mathbf{A}}\succeq{\mathbf{B}}$ imply that $\mathbf{A}-\mathbf{B}$ is positive definite and positive semi-definite, respectively. The sets $\mathbbmss{C}$ and $\mathbbmss{R}$ stand for the complex and real spaces, respectively. Big-O notation is represented by $\mathcal{O}\left(\cdot\right)$, and the ceiling operator is shown by $\lceil\cdot\rceil$. The expectation operator is denoted by $\mathbbmss{E}\{\cdot\}$, and $\overset{\rm{d}}{=}$ and $\overset{\rm{d}}{\approx}$ denote equivalence and approximation in distribution, respectively. The Kronecker delta is denoted by $\delta_{i,j}$ with $\delta_{i,j}=\left\{\begin{smallmatrix}1&i=j\\0&i\ne j\end{smallmatrix}\right.$. The Dirac delta function on the space ${\mathbbmss{R}}^{N\times1}$ is denoted by $\delta(\cdot)$ with
\begin{align}\nonumber
\delta({\mathbf{x}})=\left\{\begin{matrix}0&{\mathbf{x}}\ne{\mathbf{0}}\\\infty&{\mathbf{x}}={\mathbf{0}}\end{matrix}\right.,\quad\int_{{\mathcal{V}}}\delta({\mathbf{x}}-{\mathbf{x}}_0){\rm{d}}{\mathbf{x}}=1,
\end{align}
where ${\mathbf{x}}\in{\mathbbmss{R}}^{N\times1}$ and ${\mathcal{V}}\subseteq{\mathbbmss{R}}^{N\times1}$ is any volume that contains the point ${\mathbf{x}}_0\in{\mathbbmss{R}}^{N\times1}$. Finally, ${\mathcal{CN}}({\bm\mu},\mathbf{X})$ is used to denote the circularly-symmetric complex Gaussian distribution with mean $\bm\mu$ and covariance matrix $\mathbf{X}$.
\section{System Model}\label{Section: System Model}
\begin{figure}[!t]
 \centering
\includegraphics[height=0.22\textwidth]{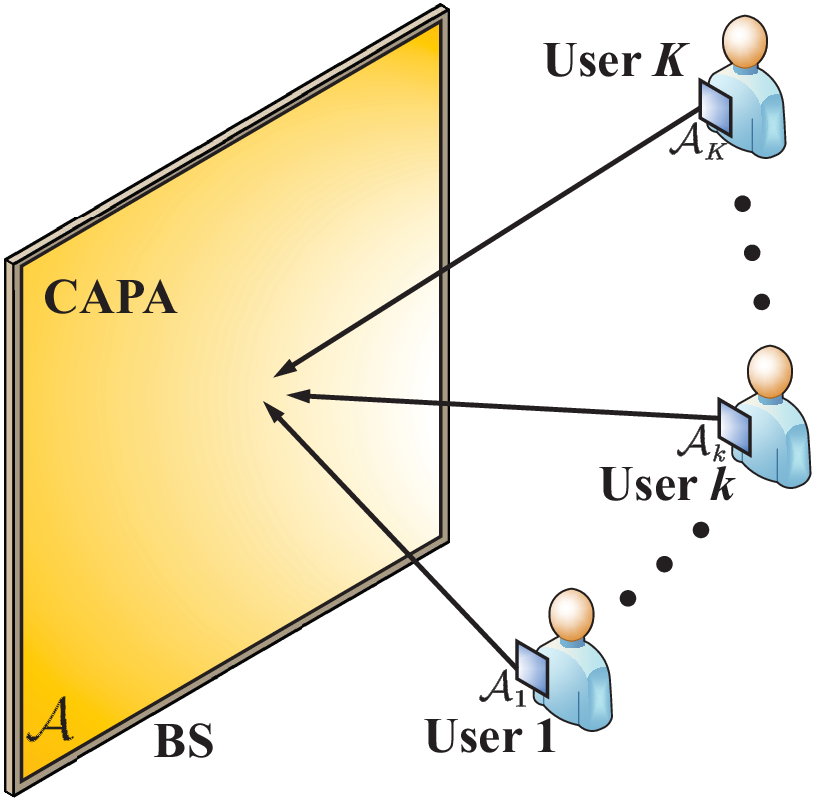}
\caption{Illustration of a CAPA-based uplink channel.}
\label{Figure: System_Model}
\end{figure}

Consider a single-cell uplink channel where $K$ users simultaneously transmit signals to a base station (BS) equipped with a CAPA, as depicted in {\figurename} {\ref{Figure: System_Model}}. Let $\mathcal{A}\subseteq{\mathbbmss{R}}^{3}$ and $\mathcal{A}_k\subseteq{\mathbbmss{R}}^{3}$ represent the antenna aperture at the BS and the aperture for user $k\in{\mathcal{K}}\triangleq\{1,\ldots,K\}$, respectively. It is assumed that the size of the antenna aperture at the BS is significantly larger than that of any user, i.e., $\lvert \mathcal{A}\rvert\gg \lvert \mathcal{A}_k\rvert$, $\forall k\in{\mathcal{K}}$. 

\subsection{Signal Model}
Let $x_k({\mathbf{s}})\in{\mathbbmss{C}}$ denote the continuous electrical signal generated by each user $k\in{\mathcal{K}}$ to convey their data, where ${\mathbf{s}}\in{\mathcal{A}_k}$. The resulting excited electric field $e_k(\mathbf{r})\in{\mathbbmss{C}}$ at point ${\mathbf{r}}\in{\mathcal{A}}$, due to $x_k({\mathbf{s}})$, is expressed as follows \cite{poon2005degrees}:
\begin{align}\nonumber
e_k(\mathbf{r})=\int_{{\mathcal{A}}_k} g_{k}({\mathbf{r}},{\mathbf{s}})x_k({\mathbf{s}}){\rm{d}}{\mathbf{s}}, 
\end{align}
where $g_{k}({\mathbf{r}},{\mathbf{s}})\in{\mathbbmss{C}}$ represents the spatial channel response between ${\mathbf{s}}\in{\mathcal{A}}_k$ and ${\mathbf{r}}\in{\mathcal{A}}$. The electrical signal $x_k({\mathbf{s}})\in{\mathbbmss{C}}$ is generated as follows:
\begin{align}\nonumber
x_k({\mathbf{s}})= s_k j_k({\mathbf{s}}),
\end{align}
where $s_k\sim{\mathcal{CN}}(0,1)$ is the normalized data symbol with ${\mathbbmss{E}}\{\lvert s_k\rvert^2\}=1$, and $j_k({\mathbf{s}})\in{\mathbbmss{C}}$ is the associated source current. Given that $\lvert \mathcal{A}\rvert\gg \lvert \mathcal{A}_k\rvert$ and $\lvert \mathcal{A}_k\rvert$ is typically negligible compared with the propagation distance, signal variations within $\mathcal{A}_k$ are also negligible, allowing us to simplify $e_k(\mathbf{r})$ as follows:
\begin{align}\nonumber
e_k(\mathbf{r})\approx g_{k}({\mathbf{r}},{\mathbf{s}}_k)s_k j_k({\mathbf{s}}_k)\lvert \mathcal{A}_k\rvert, 
\end{align}
where ${\mathbf{s}}_k\in{\mathcal{A}}_k$ represents the central point of ${\mathcal{A}}_k$. The transmit power is then given by
\begin{align}\nonumber
P_k=\int_{\mathcal{A}_k}\lvert j_k({\mathbf{s}}) \rvert^2 {\rm{d}}{\mathbf{s}}\approx \lvert j_k({\mathbf{s}}_k) \rvert^2 \lvert \mathcal{A}_k\rvert.
\end{align}
For brevity, we set $j_k({\mathbf{s}}_k)=\sqrt{\frac{P_k}{\lvert \mathcal{A}_k\rvert}}$ and define $h_{k}({\mathbf{r}})\triangleq g_{k}({\mathbf{r}},{\mathbf{s}}_k)\sqrt{\lvert \mathcal{A}_k\rvert}$ as the effective channel. It follows that
\begin{align}\nonumber
e_k(\mathbf{r})\approx \sqrt{P_k}h_{k}({\mathbf{r}})s_k.
\end{align}
The total observed electric field $y({\mathbf{r}})\in{\mathbbmss{C}}$ at point ${\mathbf{r}}\in{\mathcal{A}}$ is the sum of the information-carrying electric fields $\{e_k(\mathbf{r})\}_{k=1}^{K}$, along with a random noise field $n({\mathbf{r}})\in{\mathbbmss{C}}$, i.e.,
\begin{align}\label{Multiple_Access_Channel_Basic_Channel_Model_CAPA}
y({\mathbf{r}})=\sum_{k=1}^{K}e_k(\mathbf{r}) + n({\mathbf{r}})
 = \sum_{k=1}^{K}\sqrt{P_k}h_{k}({\mathbf{r}})s_k+n({\mathbf{r}}),
\end{align}
where $n({\mathbf{r}})$ represents the thermal noise modeled as a zero-mean complex Gaussian random process satisfying ${\mathbbmss{E}}\{n({\mathbf{r}}_1)n^{*}({\mathbf{r}}_2)\}=\sigma^2 \delta({\mathbf{r}}_1-{\mathbf{r}}_2)$ with $\sigma^2$ reflecting the noise intensity. The noise field $n(\mathbf{r})$ and the data symbols $\{s_k\}_{k=1}^{K}$ are assumed to be uncorrelated. 

After observing $y(\mathbf{r})$, the BS employs $K$ parallel linear beamformers $\{{w}_k({\mathbf{r}})\in{\mathbbmss{C}}\}_{k=1}^{K}$ along with maximum-likelihood (ML) decoding to recover the transmitted data symbols $\{s_k\}_{k=1}^{K}$. The output after processing $y(\mathbf{r})$ through beamformer ${w}_k({\mathbf{r}})$ is given by
\begin{equation}\label{Linear_Receive_Beamforming}
\begin{split}
\hat{y}_k&=\int_{{\mathcal{A}}}{w}_k^{*}(\mathbf{r})y(\mathbf{r}){\rm{d}}{\mathbf{r}}=\sqrt{P_k}s_k\int_{{\mathcal{A}}}{w}_k^{*}(\mathbf{r})h_k(\mathbf{r}){\rm{d}}{\mathbf{r}}\\
&+{\underbrace{\sum\nolimits_{k'\in{{\mathcal{K}}_k}}\sqrt{P_{k'}}s_{k'}\int_{{\mathcal{A}}}{w}_k^{*}(\mathbf{r})h_{k'}(\mathbf{r}){\rm{d}}{\mathbf{r}}}_{\mathsf{inter-user~ interference}}}+n_k,
\end{split}
\end{equation}
where ${\mathcal{K}}_k\triangleq{\mathcal{K}}\setminus\{k\}$ and $n_k=\int_{\mathcal{A}}{w}_k^{*}(\mathbf{r})n(\mathbf{r}){\rm{d}}\mathbf{r}$. Since $n(\mathbf{r})$ is a zero-mean complex Gaussian process with ${\mathbbmss{E}}\{n({\mathbf{r}}_1)n^{*}({\mathbf{r}}_2)\}=\sigma^2 \delta({\mathbf{r}}_1-{\mathbf{r}}_2)$, $n_k$ is a complex Gaussian random variable, which satisfies $n_k\sim{\mathcal{CN}}(0,\sigma^2\int_{{\mathcal{A}}}\lvert w_k(\mathbf{r})\rvert^2{\rm{d}}{\mathbf{r}})$ \cite[\textbf{Lemma 1}]{zhao2024continuous}. 

For analytical tractability, we define the channel vector and beamformer vector at position ${\mathbf{r}}\in{\mathcal{A}}$ as
\begin{align}
&{\mathbf{h}}({\mathbf{r}})\triangleq[h_1({\mathbf{r}}),\ldots,h_K({\mathbf{r}})]\in{\mathbbmss{C}}^{1\times K},\nonumber\\
&{\mathbf{w}}({\mathbf{r}})\triangleq[w_1({\mathbf{r}}),\ldots,w_K({\mathbf{r}})]\in{\mathbbmss{C}}^{1\times K},\nonumber
\end{align}
respectively. We consider the BS knows full channel information to establish theoretical performance bounds. The impact of channel uncertainty will be addressed in future work. Furthermore, we assume the following throughout this article.
\subsubsection*{\textbf{Assumption-1}} 
The spatial channel responses of the $K$ users are mutually non-parallel, i.e., the channels' squared-correlation coefficient satisfies
\begin{align}\label{Definition_Channel_Squared_Correlation_Coefficient}
\rho_{k,k'}\triangleq\frac{\left\lvert\int_{{\mathcal{A}}}{h}_k^{*}(\mathbf{r})h_{k'}(\mathbf{r}){\rm{d}}{\mathbf{r}}\right\rvert^2}{a_ka_{k'}}\left\{
\begin{array}{ll}
=1       & {k      =      k'},\\
\in[0,1)     & {k \ne k'},
\end{array} \right.
\end{align}
where $a_k\triangleq\int_{{\mathcal{A}}}\lvert h_k(\mathbf{r})\rvert^2{\rm{d}}{\mathbf{r}}$ denotes the channel gain of user $k\in{\mathcal{K}}$. This assumption is justified by practical systems where, due to scattering and random user locations, the likelihood of perfectly parallel spatial channel responses between different users is near zero. Under this assumption, the channel correlation matrix ${\mathbf{R}}\in{\mathbbmss{C}}^{K\times K}$ defined below will be full-rank:
\begin{equation}\label{Channel_Correlation_Matrix_No_Power}
\begin{split}
{\mathbf{R}}&\triangleq\left[ \begin{smallmatrix}
	\int_{{\mathcal{A}}}\lvert h_1(\mathbf{r})\rvert^2{\rm{d}}{\mathbf{r}}&		\cdots&		\int_{{\mathcal{A}}}h_1^{*}(\mathbf{r})h_K(\mathbf{r}){\rm{d}}{\mathbf{r}}\\
	\vdots&			\ddots&		\vdots\\
	\int_{{\mathcal{A}}}h_K^{*}(\mathbf{r})h_1(\mathbf{r}){\rm{d}}{\mathbf{r}}&		\cdots&		\int_{{\mathcal{A}}}\lvert h_K(\mathbf{r})\rvert^2{\rm{d}}{\mathbf{r}}\\
\end{smallmatrix} \right]\\
&=\left[r_{k_1,k_2}\right]_{k_1,k_2\in{\mathcal{K}}}= \int_{\mathcal{A}}{\mathbf{h}}^{\mathsf{H}}({\mathbf{r}}){\mathbf{h}}({\mathbf{r}}){\rm{d}}{\mathbf{r}}\succ{\mathbf{0}}.
\end{split}
\end{equation}
Here, we use $r_{k_1,k_2}=\int_{{\mathcal{A}}}h_{k_1}^{*}(\mathbf{r})h_{k_2}(\mathbf{r}){\rm{d}}{\mathbf{r}}=[{\mathbf{R}}]_{k_1,k_2}\in{\mathbbmss{C}}$ to denote the $(k_1,k_2)$th element of $\mathbf{R}$ for $k_1,k_2\in{\mathcal{K}}$. Note that the $k$th diagonal element of $\mathbf{R}$ represents the channel gain for user $k\in{\mathcal{K}}$, i.e., $r_{k,k}=[{\mathbf{R}}]_{k,k}=\int_{{\mathcal{A}}}\lvert h_k(\mathbf{r})\rvert^2{\rm{d}}{\mathbf{r}}=a_k$. 
\subsection{Performance Metric}
Referring to \eqref{Linear_Receive_Beamforming}, the signal-to-interference-plus-noise ratio (SINR) for decoding $s_k$ is given by
\begin{align}\label{SINR_Linear_Beamforming}
\gamma_k=\frac{\frac{P_k}{\sigma^2}\left\lvert\int_{{\mathcal{A}}}{w}_k^{*}(\mathbf{r})h_k(\mathbf{r}){\rm{d}}{\mathbf{r}}\right\rvert^2}
{\sum\limits_{k'\in{{\mathcal{K}}_k}}\frac{P_{k'}}{\sigma^2}\left\lvert\int_{{\mathcal{A}}}{w}_k^{*}(\mathbf{r})h_{k'}(\mathbf{r}){\rm{d}}{\mathbf{r}}\right\rvert^2+\int_{{\mathcal{A}}}\lvert w_k(\mathbf{r})\rvert^2{\rm{d}}{\mathbf{r}}}.
\end{align}
We now discuss the performance metrics of our interest.
\subsubsection{Uplink Rate}
According to \eqref{SINR_Linear_Beamforming}, the uplink rate for each user $k\in{\mathcal{K}}$ is given by 
\begin{align}\label{Per_User_Rate}
{\mathcal{R}}_k=\log_2(1+\gamma_k).
\end{align}
\subsubsection{Mean-Squared Error}
The estimation error for $s_k$ based on the observation $\hat{y}_k$ is measured by the MSE as follows:
\begin{align}\nonumber
{\mathsf{MSE}}_k=\min_{\beta_k\in{\mathbbmss{C}}}{\mathbbmss{E}}\{\lvert \beta_k \hat{y}_k - s_k\rvert^2\},
\end{align}
where a scalar filter $\beta_k$ is introduced to mitigate the noise enhancement effect of $w_k(\mathbf{r})$. The optimal solution for $\beta_k$ and the corresponding MSE are given below.
\vspace{-5pt}
\begin{lemma}\label{Lemma_MSE_Fundamental}
Given $w_k(\mathbf{r})$, the optimal solution for $\beta_k$ is
\begin{align}
\beta_k^{\star}&=\argmin_{\beta_k\in{\mathbbmss{C}}}{\mathbbmss{E}}\{\lvert \beta_k \hat{y}_k - s_k\rvert^2\}\nonumber\\
&=\frac{\frac{\sqrt{P_k}}{\sigma^2}\int_{{\mathcal{A}}}{w}_k(\mathbf{r})h_k^{*}(\mathbf{r}){\rm{d}}{\mathbf{r}}}
{\sum_{k'=1}^{K}\frac{P_{k'}}{\sigma^2}\lvert\int_{{\mathcal{A}}}{w}_k^{*}(\mathbf{r})h_{k'}(\mathbf{r}){\rm{d}}{\mathbf{r}}\rvert^2+\int_{{\mathcal{A}}}\lvert w_k(\mathbf{r})\rvert^2{\rm{d}}{\mathbf{r}}},\label{MSE_Fundamental}
\end{align}
and the resulting MSE is
\begin{align}\label{Per_User_MSE}
{\mathsf{MSE}}_k={\mathbbmss{E}}\{\lvert \beta_k^{\star} \hat{y}_k - s_k\rvert^2\}=\frac{1}{1+\gamma_k}.
\end{align}
\end{lemma}
\vspace{-5pt}
\begin{IEEEproof}
Please refer to Appendix \ref{Proof_Lemma_MSE_Fundamental} for more details.
\end{IEEEproof}
The preceding analysis shows that for any arbitrary ${w}_k(\mathbf{r})$, the effective beamformer minimizing the MSE is given by
\begin{align}\nonumber
{w}_k^{\star}(\mathbf{r})=\beta_k^{\star}{w}_k(\mathbf{r}).
\end{align}
Additionally, substituting ${w}_k(\mathbf{r})={w}_k^{\star}(\mathbf{r})$ into \eqref{Per_User_Rate} leaves the SINR and sum-rate unchanged.
\vspace{-5pt}
\begin{remark}\label{Remark_MMSE_Rate}
By comparing \eqref{Per_User_Rate} with \eqref{Per_User_MSE}, we conclude the beamformer that maximizes the rate also minimizes the MSE. In other words, under the considered model, maximizing the uplink rate is equivalent to minimizing the MSE.
\end{remark}
\vspace{-5pt}
Subsequently, we analyze the uplink rate and MSE achieved by three linear beamforming schemes for CAPAs: MRC beamforming, ZF beamforming, and MMSE beamforming. 
\section{Maximum-Ratio Combining Beamforming}\label{Section: Maximum-Ratio Combining Beamforming}
\subsection{Analysis of the SINR}
For MRC beamforming, the beamformer $w_k(\mathbf{r})$ is aligned with the spatial channel response $h_k(\mathbf{r})$. The main results are summarized below.
\vspace{-5pt}
\begin{theorem}[MRC Beamforming for CAPAs]\label{Theorem_MRC_General}
Given the channel responses $\{h_k({\mathbf{r}})\}_{k=1}^{K}$, the MRC beamformers for CAPAs are expressed as follows:
\begin{align}\label{CAPA_MRC_Closed_Form}
\boxed{
w_k(\mathbf{r})=h_k(\mathbf{r})\triangleq w_{{\mathsf{MRC}},k}(\mathbf{r}),~\forall k\in{\mathcal{K}}.}
\end{align}
This can also be written in the following form:
\begin{align}\label{CAPA_MRC_Closed_Form_Matrix}
\boxed{
{\mathbf{w}}(\mathbf{r})={\mathbf{h}}(\mathbf{r}){\mathbf{I}}_K\triangleq {\mathbf{w}}_{{\mathsf{MRC}}}(\mathbf{r}).}
\end{align}
\end{theorem}
\vspace{-5pt} 
Inserting \eqref{CAPA_MRC_Closed_Form} into \eqref{SINR_Linear_Beamforming} yields the per-user SINR achieved by MRC beamforming:
\begin{align}
\gamma_k&=\frac{P_k\lvert\int_{{\mathcal{A}}}\lvert h_k(\mathbf{r})\rvert^2{\rm{d}}{\mathbf{r}}\rvert^2}
{\sum_{k'\in{\mathcal{K}}_k}P_{k'}\lvert\int_{{\mathcal{A}}}{h}_k^{*}(\mathbf{r})h_{k'}(\mathbf{r}){\rm{d}}{\mathbf{r}}\rvert^2+\sigma^2\int_{{\mathcal{A}}}\lvert h_k(\mathbf{r})\rvert^2{\rm{d}}{\mathbf{r}}}\nonumber\\
&=\frac{P_k}{\sigma^2}\frac{a_k}
{\frac{1}{a_k}{\mathbf{r}}_{k,k}^{\mathsf{H}}{\mathbf{P}}_k{\mathbf{r}}_{k,k}+1}\triangleq \gamma_{{\mathsf{MRC}},k},\label{SINR_MRC}
\end{align}
where ${{\mathbf{r}}}_{k,k}=[r_{1,k},\ldots,r_{k-1,k},r_{k+1,k},\ldots,r_{K,k}]^{\mathsf{T}}\in{\mathbbmss{C}}^{K_1\times1}$, ${{\mathbf{P}}_k}={\rm{diag}}([\frac{P_1}{\sigma^2},\ldots,\frac{P_{k-1}}{\sigma^2},\frac{P_{k+1}}{\sigma^2},\ldots,\frac{P_K}{\sigma^2}])\in{\mathbbmss{C}}^{K_1\times K_1}$, and $K_1\triangleq K-1$.
\subsection{Sum-Rate Performance}
Based on \eqref{SINR_MRC}, the SINR achieved by MRC beamforming for user $k$ can be rewritten as follows:
\begin{align}\nonumber
\gamma_{{\mathsf{MRC}},k}=\frac{P_k}{\sigma^2}a_k\left(1-\alpha_{{\mathsf{MRC}},k}\right),
\end{align}
where $\alpha_{{\mathsf{MRC}},k}\triangleq \frac{{\mathbf{r}}_{k,k}^{\mathsf{H}}{\mathbf{P}}_k{\mathbf{r}}_{k,k}}{a_k+{\mathbf{r}}_{k,k}^{\mathsf{H}}{\mathbf{P}}_k{\mathbf{r}}_{k,k}}\in[0,1)$. Consequently, the sum-rate is given by
\begin{align}\nonumber
{\mathcal{R}}=\sum_{k=1}^{K}{\mathcal{R}}_k=\sum_{k=1}^{K}\log_2\left(1+\frac{P_k}{\sigma^2}a_k(1-\alpha_{{\mathsf{MRC}},k})\right),
\end{align}
where $\frac{P_k}{\sigma^2}a_k$ is the \emph{single-user SNR} as if there were no IUI, and $\alpha_{{\mathsf{MRC}},k}$ can be interpreted as the \emph{SNR loss factor} due to the IUI. These results indicate that the sum-rate depends on both the channel gain and the channel correlation factor, which aligns with the findings in \cite{zhao2024channel,zhao2024continuous,ouyang2024primer}. 
\subsection{Mean-Squared Error Performance}
Next, we analyze the MSE performance of MRC beamforming in estimating $s_k$. Referring to \eqref{MSE_Fundamental}, the optimal scalar filter under MRC beamforming is given by
\begin{align}\nonumber
\beta_k^{\star}=\frac{\sqrt{P_k}}
{\sum_{k'=1}^{K}P_{k}\frac{\lvert r_{k,k'}\rvert^2}{a_k}+\sigma^2}\triangleq \beta_{{\mathsf{MRC}},k}.
\end{align}
Consequently, the effective MRC beamformer that minimizes the MSE for estimating $s_k$ is given by
\begin{align}
w_{{\mathsf{MRC}},k}^{\star}({\mathbf{r}})=\beta_{{\mathsf{MRC}},k}w_{{\mathsf{MRC}},k}(\mathbf{r})=\frac{\sqrt{P_k}h_k(\mathbf{r})}
{\sum_{k'=1}^{K}P_{k}\frac{\lvert r_{k,k'}\rvert^2}{a_k}+\sigma^2},\nonumber
\end{align}
and the associated MSE is given by
\begin{align}\nonumber
{\mathsf{MSE}}_k=\frac{1}{1+\gamma_{{\mathsf{MRC}},k}}=\frac{1}{1+\frac{P_k}{\sigma^2}a_k(1-\alpha_{{\mathsf{MRC}},k})}.
\end{align}
\section{Zero-Forcing Beamforming}\label{Section: Zero-Forcing Beamforming}
\subsection{Analysis of the SINR}
\subsubsection{General Discussion}
In ZF beamforming, the beamformers are designed to satisfy the following zero-interference conditions:
\begin{align}\label{ZFBF_Condition}
\int_{{\mathcal{A}}}{w}_k^{*}(\mathbf{r})h_{k'}(\mathbf{r}){\rm{d}}{\mathbf{r}}=\delta_{k,k'},~\forall k,k'\in{\mathcal{K}},
\end{align}
i.e., $\int_{\mathcal{A}}{\mathbf{w}}^{{\mathsf{H}}}({\mathbf{r}}){\mathbf{h}}({\mathbf{r}}){\rm{d}}{\mathbf{r}}={\mathbf{I}}_K$. For SPDAs, a common method to achieve zero interference is through the pseudoinverse of the channel matrix \cite{heath2018foundations}. However, for CAPAs, where the channel response is modeled as a continuous operator with infinite dimensionality, the conventional pseudoinverse-based ZF beamforming amethod is not feasible. To address this, we propose a \emph{channel correlation-based ZF beamforming} method to nullify IUI. The main results are summarized as follows.
\vspace{-5pt}
\begin{theorem}[ZF Beamforming for CAPAs]\label{Theorem_ZFBF_General}
Given the channel responses $\{h_k({\mathbf{r}})\}_{k=1}^{K}$, the following beamformers satisfy the zero-interference conditions in \eqref{ZFBF_Condition}:
\begin{align}\label{ZFBF_General}
\boxed{w_k({\mathbf{r}})=\sum_{k'=1}^{K}[{\mathbf{R}}^{-1}]_{k',k}h_{k'}({\mathbf{r}})\triangleq w_{{\mathsf{ZF}},k}(\mathbf{r}),~\forall k\in{\mathcal{K}},}
\end{align}
where ${\mathbf{R}}$ is the channel correlation matrix defined in \eqref{Channel_Correlation_Matrix_No_Power}. These beamformers can also be represented as follows:
\begin{align}\label{ZFBF_General_Matrix}
\boxed{{\mathbf{w}}({\mathbf{r}})={\mathbf{h}}({\mathbf{r}}){\mathbf{R}}^{-1}\triangleq{{\mathbf{w}}}_{{\mathsf{ZF}}}(\mathbf{r}).}
\end{align}
\end{theorem}
\vspace{-5pt} 
\begin{IEEEproof}
Substituting \eqref{ZFBF_General_Matrix} into $\int_{\mathcal{A}}{\mathbf{w}}^{{\mathsf{H}}}({\mathbf{r}}){\mathbf{h}}({\mathbf{r}}){\rm{d}}{\mathbf{r}}$ gives
\begin{align}
\int_{\mathcal{A}}{\mathbf{w}}^{{\mathsf{H}}}({\mathbf{r}}){\mathbf{h}}({\mathbf{r}}){\rm{d}}{\mathbf{r}}
=({\mathbf{R}}^{-1})^{\mathsf{H}}\int_{{\mathcal{A}}}{{\mathbf{h}}}^{\mathsf{H}}(\mathbf{r}){\mathbf{h}}({\mathbf{r}}){\rm{d}}{\mathbf{r}}=({\mathbf{R}}^{-1})^{\mathsf{H}}{\mathbf{R}},\nonumber
\end{align}
which, together with the fact that $({\mathbf{R}}^{-1})^{\mathsf{H}}={\mathbf{R}}^{-1}$, yields $\int_{\mathcal{A}}{\mathbf{w}}^{{\mathsf{H}}}({\mathbf{r}}){\mathbf{h}}({\mathbf{r}}){\rm{d}}{\mathbf{r}}={\mathbf{I}}_K$. This completes the proof.
\end{IEEEproof}
Based on \textbf{Theorem \ref{Theorem_ZFBF_General}}, the SINR or SNR achieved by ZF beamforming in \eqref{ZFBF_General} is given by
\begin{align}
\gamma_k=\frac{P_k}{\sigma^2\int_{{\mathcal{A}}}\lvert w_{{\mathsf{ZF}},k}(\mathbf{r})\rvert^2{\rm{d}}{\mathbf{r}}}\triangleq \gamma_{{\mathsf{ZF}},k}.\label{SINR_ZF_Beamforming_Step1}
\end{align}
From \eqref{ZFBF_General} and \eqref{ZFBF_General_Matrix}, we express the norm of $w_{{\mathsf{ZF}},k}(\mathbf{r})$ as $\int_{{\mathcal{A}}}\lvert w_{{\mathsf{ZF}},k}(\mathbf{r})\rvert^2{\rm{d}}{\mathbf{r}}=\left[\int_{{\mathcal{A}}}{{\mathbf{w}}}_{{\mathsf{ZF}}}^{\mathsf{H}}(\mathbf{r})
{{\mathbf{w}}}_{{\mathsf{ZF}}}(\mathbf{r}){\rm{d}}{\mathbf{r}}\right]_{k,k}$, where
\begin{align}
\int_{{\mathcal{A}}}{{\mathbf{w}}}_{{\mathsf{ZF}}}^{\mathsf{H}}(\mathbf{r})
{{\mathbf{w}}}_{{\mathsf{ZF}}}(\mathbf{r}){\rm{d}}{\mathbf{r}}&=({\mathbf{R}}^{-1})^{\mathsf{H}}\int_{{\mathcal{A}}}{{\mathbf{h}}}^{\mathsf{H}}(\mathbf{r})
{{\mathbf{h}}}(\mathbf{r}){\rm{d}}{\mathbf{r}}{\mathbf{R}}^{-1}\nonumber\\
&=({\mathbf{R}}^{-1})^{\mathsf{H}}{\mathbf{R}}{\mathbf{R}}^{-1}=({\mathbf{R}}^{-1})^{\mathsf{H}},\nonumber
\end{align}
and ${\mathbf{R}}=\int_{{\mathcal{A}}}{{\mathbf{h}}}^{\mathsf{H}}(\mathbf{r})
{{\mathbf{h}}}(\mathbf{r}){\rm{d}}{\mathbf{r}}$. Since ${\mathbf{R}}={\mathbf{R}}^{\mathsf{H}}$, it holds that
\begin{align}\nonumber
\int_{{\mathcal{A}}}\lvert w_{{\mathsf{ZF}},k}(\mathbf{r})\rvert^2{\rm{d}}{\mathbf{r}}=\left[({\mathbf{R}}^{-1})^{\mathsf{H}}\right]_{k,k}=[{\mathbf{R}}^{-1}]_{k,k},
\end{align}
which yields
\begin{align}\label{ZF_SINR_Type1}
\gamma_{{\mathsf{ZF}},k}=\frac{P_k}{\sigma^2[{\mathbf{R}}^{-1}]_{k,k}},~\forall k\in{\mathcal{K}}.
\end{align}
\subsubsection{A Function Space Perspective}\label{Subsubsection: A Function Space Perspective}
It is important to note that \textbf{Theorem \ref{Theorem_ZFBF_General}} provides just one feasible solution to the equations formulated in \eqref{ZFBF_Condition}. However, since the continuous spatial response of CAPAs can be considered as a vector with infinitely large dimensions, there may be infinite solutions to \eqref{ZFBF_Condition}. This raises an intriguing question: \emph{does the ZF solution presented in \textbf{Theorem \ref{Theorem_ZFBF_General}} simultaneously nullify the IUI while maximizing the received SNR for each user?} Specifically, by referring to \eqref{SINR_ZF_Beamforming_Step1}, we seek to determine whether the solution in \textbf{Theorem \ref{Theorem_ZFBF_General}} satisfies the following condition:
\begin{align}\label{ZF_Optimal_Question}
\argmin_{\int_{{\mathcal{A}}}{w}_k^{*}(\mathbf{r})h_{k'}(\mathbf{r}){\rm{d}}{\mathbf{r}}=\delta_{k,k'}}
{\int_{{\mathcal{A}}}\lvert w_{k}(\mathbf{r})\rvert^2{\rm{d}}{\mathbf{r}}}\overset{?}{=}w_{{\mathsf{ZF}},k}(\mathbf{r}).
\end{align}To address this question, we provide a more detailed analysis of the ZF beamformer presented in \textbf{Theorem \ref{Theorem_ZFBF_General}} from a \emph{function space} perspective, and demonstrate that the ZF beamformer given in \eqref{ZFBF_General} is indeed the solution to the problem given in the left-hand side of \eqref{ZF_Optimal_Question}. 

For ease of explanation, we define ${\mathbf{h}}_k({\mathbf{r}})\triangleq[h_{k'}({\mathbf{r}})]_{k'\in{\mathcal{K}}_k}\in{\mathbbmss{C}}^{1\times K_1}$, ${{\mathbf{r}}}_{k,k'}\triangleq\int_{\mathcal{A}}{\mathbf{h}}_k^{\mathsf{H}}({\mathbf{r}})h_{k'}({\mathbf{r}}){\rm{d}}{\mathbf{r}}\in{\mathbbmss{C}}^{K_1\times1}$, ${{\mathbf{R}}}_k\triangleq \int_{\mathcal{A}}{\mathbf{h}}_k^{\mathsf{H}}({\mathbf{r}}){\mathbf{h}}_k({\mathbf{r}}){\rm{d}}{\mathbf{r}}=[{{\mathbf{r}}}_{k,k'}]_{k'\in{\mathcal{K}}_k}\in{\mathbbmss{C}}^{K_1\times K_1}$, and ${\overline{\mathbf{R}}}_k\triangleq\left[\begin{smallmatrix}
r_{k,k}&{{\mathbf{r}}}_{k,k}^{\mathsf{H}}\\
{{\mathbf{r}}}_{k,k}&{\mathbf{R}}_k
\end{smallmatrix}
\right]\in{\mathbbmss{C}}^{K \times K}$. Note that ${\overline{\mathbf{R}}}_k$ is an elementary transformation of $\mathbf{R}$, where the $k$th column is removed and placed as the first column, followed by moving the $k$th row to the first row. Therefore, it holds that ${\overline{\mathbf{R}}}_k={\overline{\mathbf{R}}}_k^{\mathsf{H}}$, and the first column of the matrix ${\overline{\mathbf{R}}}_k^{-1}$ takes the form as follows:
\begin{align}\label{Matrix_Inversion_Result_1}
{\overline{\mathbf{r}}}_{k,1}
=[[{\mathbf{R}}^{-1}]_{k,k},[[{\mathbf{R}}^{-1}]_{k',k}]_{k'\in{\mathcal{K}}_k}]^{\mathsf{T}}\in{\mathbbmss{C}}^{K\times1}.
\end{align}
By applying the blockwise inversion formula \cite{bernstein2009matrix}, ${\overline{\mathbf{r}}}_{k,1}$ can be calculated as follows:
\begin{align}\label{Matrix_Inversion_Result_2}
{\overline{\mathbf{r}}}_{k,1}=\frac{1}{r_{k,k}-{{\mathbf{r}}}_{k,k}^{\mathsf{H}}{\mathbf{R}}_k^{-1}{{\mathbf{r}}}_{k,k}}
\left[\begin{smallmatrix}1\\-{\mathbf{R}}_k^{-1}{{\mathbf{r}}}_{k,k}\end{smallmatrix}\right].
\end{align}
A full derivation of \eqref{Matrix_Inversion_Result_2} is provided in Appendix \ref{Proof_Lemma_Matrix_Blockwise_Inversion}. 

For clarity, we denote
\begin{align}\label{Inverse_Channel_No_User_k_Correlation_Matrix_No_Power}
{\mathbf{R}}_k^{-1}= {{\left[ \begin{smallmatrix}
	{\overline{r}}_{1,1}&		\cdots&		{\overline{r}}_{1,k-1}&	{\overline{r}}_{1,k+1}&	\cdots&		{\overline{r}}_{1,K}\\
	\vdots&		\ddots&		\vdots&	\vdots&	\ddots&		\vdots\\
	{\overline{r}}_{k-1,1}&		\cdots&		{\overline{r}}_{k-1,k-1}&	{\overline{r}}_{k-1,k+1}&	\cdots&		{\overline{r}}_{k-1,K}\\
    {\overline{r}}_{k+1,1}&		\cdots&		{\overline{r}}_{k+1,k-1}&	{\overline{r}}_{k+1,k+1}&	\cdots&		{\overline{r}}_{k+1,K}\\
	\vdots&		\ddots&		\vdots&	\vdots&	\ddots&		\vdots\\
	{\overline{r}}_{K,1}&		\cdots&		{\overline{r}}_{K,k-1}&	{\overline{r}}_{K,k+1}&	\cdots&		{\overline{r}}_{K,K}\\
\end{smallmatrix} \right]}}.
\end{align}
Substituting \eqref{Matrix_Inversion_Result_1}, \eqref{Matrix_Inversion_Result_2}, and \eqref{Inverse_Channel_No_User_k_Correlation_Matrix_No_Power} into \eqref{ZFBF_General} gives
\begin{align}
w_{{\mathsf{ZF}},k}(\mathbf{r})&=\frac{[{{h}}_k({\mathbf{r}}),{\mathbf{h}}_k({\mathbf{r}})]}{r_{k,k}-{{\mathbf{r}}}_{k,k}^{\mathsf{H}}{\mathbf{R}}_k^{-1}{{\mathbf{r}}}_{k,k}}
\left[\begin{smallmatrix}1\\-{\mathbf{R}}_k^{-1}{{\mathbf{r}}}_{k,k}\end{smallmatrix}\right]\nonumber\\
&=\frac{h_k({\mathbf{r}})-\sum\limits_{k_1\in{\mathcal{K}}_k}\sum\limits_{k_2\in{\mathcal{K}}_k}{\overline{r}}_{k_1,k_2}r_{k_2,k}h_{k_1}({\mathbf{r}})}{r_{k,k}-{{\mathbf{r}}}_{k,k}^{\mathsf{H}}{\mathbf{R}}_k^{-1}{{\mathbf{r}}}_{k,k}}
\label{ZFBF_Signal_Space_1}.
\end{align}
Inserting $r_{k_2,k}=\int_{{\mathcal{A}}}h_{k_2}^{*}(\mathbf{r}')h_k(\mathbf{r}'){\rm{d}}{\mathbf{r}}'$ into \eqref{ZFBF_Signal_Space_1} gives 
\begin{align}\label{ZFBF_Signal_Space_2}
w_{{\mathsf{ZF}},k}(\mathbf{r})=\frac{h_k({\mathbf{r}})-
\int_{{\mathcal{A}}}P_{k}({\mathbf{r}},{\mathbf{r}}')h_k({\mathbf{r}}'){\rm{d}}{\mathbf{r}}'}{r_{k,k}-{{\mathbf{r}}}_{k,k}^{\mathsf{H}}{\mathbf{R}}_k^{-1}{{\mathbf{r}}}_{k,k}}
,
\end{align}
where 
\begin{align}
P_{k}({\mathbf{r}},{\mathbf{r}}')&\triangleq\sum\nolimits_{k_1\in{\mathcal{K}}_k}\sum\nolimits_{k_2\in{\mathcal{K}}_k}{\overline{r}}_{k_1,k_2}h_{k_1}({\mathbf{r}})h_{k_2}^{*}(\mathbf{r}')\nonumber\\
&={\mathbf{h}}_k({\mathbf{r}}){\mathbf{R}}_k^{-1}{\mathbf{h}}_k^{\mathsf{H}}({\mathbf{r}}'). \label{ZF_Projection_Kernel_Most_Original}
\end{align}
Since the term ${r_{k,k}-{{\mathbf{r}}}_{k,k}^{\mathsf{H}}{\mathbf{R}}_k^{-1}{{\mathbf{r}}}_{k,k}}$ in \eqref{ZFBF_Signal_Space_2} does not influence the SINR achieved by the ZF beamformer, we can omit it and simplify the ZF beamformer as follows:
\begin{align}
w_{{\mathsf{ZF}},k}(\mathbf{r})=h_k({\mathbf{r}})-
\int_{{\mathcal{A}}}P_{k}({\mathbf{r}},{\mathbf{r}}')h_k({\mathbf{r}}'){\rm{d}}{\mathbf{r}}'\label{ZFBF_Signal_Space_Used_To_Analyze}.
\end{align}

Building on \eqref{ZFBF_Signal_Space_Used_To_Analyze}, we derive a function space interpretation of the ZF beamformer $w_{{\mathsf{ZF}},k}(\mathbf{r})$. Let ${\mathcal{W}}_k$ denote the interference subspace spanned by $\{h_{k'}({\mathbf{r}})\}_{k'\in{\mathcal{K}}_k}$, and let $\{\psi_{k'}({\mathbf{r}})\}_{k'\in{\mathcal{K}}_k}$ represent an orthonormal basis for ${\mathcal{W}}_k$. A straightforward method to construct this basis is to apply the \emph{Gram-Schmidt process} to $\{h_{k'}({\mathbf{r}})\}_{k'\in{\mathcal{K}}_k}$, whose details are given in Appendix \ref{Proof_Gram-Schmidt}. By definition, $\{\psi_{k'}({\mathbf{r}})\}_{k'\in{\mathcal{K}}_k}$ are mutually orthogonal, i.e.,
\begin{align}
&\int_{{\mathcal{A}}}\psi_{k_1}^{*}({\mathbf{r}})\psi_{k_2}({\mathbf{r}}){\rm{d}}{\mathbf{r}}=\delta_{k_1,k_2}\nonumber\\
&\Leftrightarrow\int_{{\mathcal{A}}}{\bm\psi}_k^{\mathsf{H}}({\mathbf{r}}){\bm\psi}_k({\mathbf{r}}){\rm{d}}{\mathbf{r}}={\mathbf{I}}_{K_1},\label{Orthonormal_Basis}
\end{align}
where ${\bm\psi}_k({\mathbf{r}})\triangleq[\psi_{k'}({\mathbf{r}})]_{k'\in{\mathcal{K}}_k}\in{\mathbbmss{C}}^{1\times K_1}$. Any function $f_{{\mathcal{W}}_k}({\mathbf{r}})\in{\mathcal{W}}_k$ can be \emph{uniquely} expressed as a linear combination of the basis functions $\{\psi_{k'}({\mathbf{r}})\}_{k'\in{\mathcal{K}}_k}$. For each $h_{k'}({\mathbf{r}})$ ($k'\ne k$), this yields
\begin{align}\label{Function_Space_Expansion}
h_{k'}({\mathbf{r}})=\sum\nolimits_{k''\in{\mathcal{K}}_k}\phi_{k',k''}\psi_{k''}({\mathbf{r}})={\bm\psi}_k({\mathbf{r}}){\bm\phi}_{k,k'},
\end{align}
where ${\bm\phi}_{k,k'}\triangleq[\phi_{k',k''}]_{k''\in{\mathcal{K}}_k}\in{\mathbbmss{C}}^{{K_1}\times1}$ with $\phi_{k',k''}=\int_{\mathcal{A}}h_{k'}({\mathbf{r}})\psi_{k''}^{*}({\mathbf{r}}){\rm{d}}{\mathbf{r}}$ being the projection of $h_{k'}({\mathbf{r}})$ onto $\psi_{k''}({\mathbf{r}})$. Combining \eqref{Orthonormal_Basis} and \eqref{Function_Space_Expansion}, we rewrite \eqref{ZF_Projection_Kernel_Most_Original} as follows:
\begin{align}\label{ZF_Projection_Kernel}
P_{k}({\mathbf{r}},{\mathbf{r}}')={\bm\psi}_k({\mathbf{r}}){\bm\psi}_k^{\mathsf{H}}({\mathbf{r}}')=
\sum\nolimits_{k'\in{\mathcal{K}}_k}\psi_{k'}({\mathbf{r}})\psi_{k'}^{*}(\mathbf{r}').
\end{align}
A complete derivation of \eqref{ZF_Projection_Kernel} is included in Appendix \ref{Proof_Lemma_ZF_Projection_Kernel}.

\begin{figure}[!t]
 \centering
\includegraphics[height=0.18\textwidth]{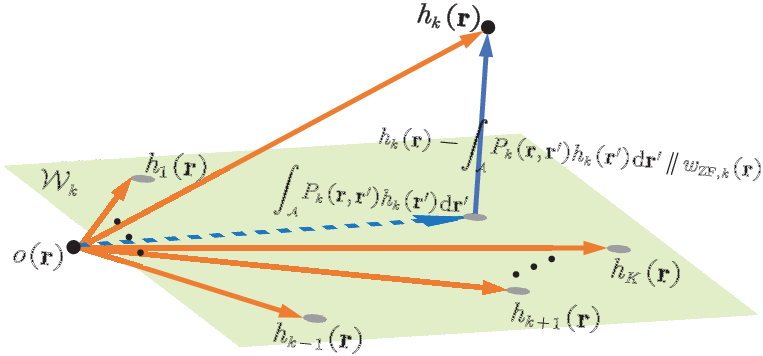}
\caption{Illustration of ZF beamforming for CAPAs, where $o(\mathbf{r})$ represents the original point of the entire function space defined on $\mathbf{r}\in{\mathcal{A}}$, and the green hyperplane represents the interference subspace $\mathcal{W}_k$.}
\label{Subspace}
\end{figure}

It follows from \eqref{ZF_Projection_Kernel} that for an arbitrary function $f_{\mathcal{A}}(\mathbf{r})$ defined on $\mathcal{A}$, it holds that
\begin{align}
\int_{{\mathcal{A}}}P_{k}({\mathbf{r}},{\mathbf{r}}')f_{\mathcal{A}}(\mathbf{r}'){\rm{d}}{\mathbf{r}}'
=\sum_{k_2\in{\mathcal{K}}_k}\psi_{k_2}({\mathbf{r}})\int_{{\mathcal{A}}}\psi_{k_2}^{*}(\mathbf{r}')f_{\mathcal{A}}(\mathbf{r}'){\rm{d}}{\mathbf{r}}',\nonumber
\end{align}
where $\int_{{\mathcal{A}}}\psi_{k_2}^{*}(\mathbf{r}')f_{\mathcal{A}}(\mathbf{r}'){\rm{d}}{\mathbf{r}}'$ represents the projection of $f_{\mathcal{A}}(\mathbf{r})$ along the direction of $\psi_{k_2}(\mathbf{r})$.
\vspace{-5pt}
\begin{remark}\label{remark_subspace_zf}
The above arguments imply that $P_{k}({\mathbf{r}},{\mathbf{r}}')$ acts as a projection operator that maps $f_{\mathcal{A}}({\mathbf{r}})$ onto the interference subspace ${\mathcal{W}}_k$. Additionally, $\int_{{\mathcal{A}}}P_{k}({\mathbf{r}},{\mathbf{r}}')f_{\mathcal{A}}(\mathbf{r}'){\rm{d}}{\mathbf{r}}'$ can be interpreted as the projection of $f_{\mathcal{A}}(\mathbf{r}')$ onto ${\mathcal{W}}_k$, which is orthogonal to $f_{\mathcal{A}}(\mathbf{r})-\int_{{\mathcal{A}}}P_{k}({\mathbf{r}},{\mathbf{r}}')f_{\mathcal{A}}(\mathbf{r}'){\rm{d}}{\mathbf{r}}'$.
\end{remark}
\vspace{-5pt}
According to \textbf{Remark \ref{remark_subspace_zf}}, $\int_{{\mathcal{A}}}P_{k}({\mathbf{r}},{\mathbf{r}}')h_k({\mathbf{r}}'){\rm{d}}{\mathbf{r}}'$ represents the projection of $h_k({\mathbf{r}})$ onto the interference subspace ${\mathcal{W}}_k$. Consequently, the proposed ZF beamformer $w_{{\mathsf{ZF}},k}(\mathbf{r})=h_k({\mathbf{r}})-
\int_{{\mathcal{A}}}P_{k}({\mathbf{r}},{\mathbf{r}}')h_k({\mathbf{r}}'){\rm{d}}{\mathbf{r}}'$ can be interpreted as a function orthogonal to ${\mathcal{W}}_k$ while \emph{remaining maximally aligned} with user $k$'s spatial channel response, as shown in {\figurename} {\ref{Subspace}}. 
\vspace{-5pt}
\begin{remark}
The results in {\figurename} {\ref{Subspace}} demonstrate that the ZF solution derived in \textbf{Theorem \ref{Theorem_ZFBF_General}} maximizes the effective channel gain or received SNR for each user while ensuring zero IUI, providing the solution to the problem defined in \eqref{ZF_Optimal_Question}. 
\end{remark}
\vspace{-5pt}

\subsection{Sum-Rate Performance}
Using \eqref{ZF_SINR_Type1}, \eqref{Matrix_Inversion_Result_1}, and \eqref{Matrix_Inversion_Result_2}, the SINR achieved by ZF beamforming becomes
\begin{align}\nonumber
\gamma_{{\mathsf{ZF}},k}
=\frac{P_k}{\sigma^2}\frac{1}{[{\overline{\mathbf{r}}}_{k,1}]_{1}}
=\frac{P_k(a_k-{{\mathbf{r}}}_{k,k}^{\mathsf{H}}{\mathbf{R}}_k^{-1}{{\mathbf{r}}}_{k,k})}{\sigma^2},
\end{align}
where $a_k=r_{k,k}$ represents the channel gain. Consequently, the sum-rate achieved by ZF beamforming is given by
\begin{align}\nonumber
{\mathcal{R}}=\sum_{k=1}^{K}{\mathcal{R}}_k=\sum_{k=1}^{K}\log_2\left(1+\frac{P_k}{\sigma^2}a_{k}\left(1-\alpha_{{\mathsf{ZF}},k}\right)\right),
\end{align}
where $\alpha_{{\mathsf{ZF}},k}\triangleq \frac{1}{a_{k}}{{{\mathbf{r}}}_{k,k}^{\mathsf{H}}{\mathbf{R}}_k^{-1}{{\mathbf{r}}}_{k,k}}$. This term quantifies the SNR loss due to IUI cancellation via ZF beamforming. We prove in Appendix \ref{Proof_Lemma_ZF_SNR_Gain_Loss} that $\frac{{{\mathbf{r}}}_{k,k}^{\mathsf{H}}{\mathbf{R}}_k^{-1}{{\mathbf{r}}}_{k,k}}{a_{k}}=\frac{{{\mathbf{r}}}_{k,k}^{\mathsf{H}}{\mathbf{R}}_k^{-1}{{\mathbf{r}}}_{k,k}}{r_{k,k}}\in[0,1)$.
\subsection{Mean-Squared Error Performance}
Next, we analyze the MSE performance of ZF beamforming in estimating $s_k$. By substituting \eqref{ZFBF_Condition}, \eqref{SINR_ZF_Beamforming_Step1} and \eqref{ZF_SINR_Type1} into \eqref{MSE_Fundamental}, the optimal scalar filter under ZF beamforming is given by
\begin{align}\nonumber
\beta_k^{\star}=\frac{\sqrt{P_k}}
{P_{k}+\sigma^2[{\mathbf{R}}^{-1}]_{k,k}}\triangleq \beta_{{\mathsf{ZF}},k}.
\end{align}
Consequently, the effective ZF beamformer that minimizes the MSE for estimating $s_k$ is given by
\begin{align}
w_{{\mathsf{ZF}},k}^{\star}({\mathbf{r}})=\beta_{{\mathsf{ZF}},k}w_{{\mathsf{ZF}},k}({\mathbf{r}})=\frac{\sqrt{P_k}\sum_{k'=1}^{K}[{\mathbf{R}}^{-1}]_{k',k}h_{k'}({\mathbf{r}})}
{P_{k}+\sigma^2[{\mathbf{R}}^{-1}]_{k,k}}.\nonumber
\end{align}
The associated MSE is then
\begin{align}\nonumber
{\mathsf{MSE}}_k=\frac{1}{1+\gamma_{{\mathsf{ZF}},k}}=\frac{1}{1+\frac{P_k}{\sigma^2}a_k(1-\alpha_{{\mathsf{ZF}},k})}.
\end{align}
\section{Optimal Beamforming: Minimum Mean-Squared Error Beamforming}\label{Section: Optimal Beamforming: Minimum Mean-Square Error Beamforming}
As discussed in \textbf{Remark \ref{Remark_MMSE_Rate}}, maximizing the uplink per-user rate is equivalent to minimizing the MSE. In this context, we now turn our focus to the \emph{rate-optimal beamforming} design, or equivalently, the \emph{MSE-optimal beamforming} design. To maintain consistency with the terminology used in SPDAs, we refer to this as \emph{MMSE beamforming}.
\subsection{Analysis of the SINR}
According to \eqref{Per_User_Rate}, the optimal beamformer that maximizes the per-user uplink rate can be expressed as follows:
\begin{align}\label{SINR_Rayleigh_Objective}
w_{{\mathsf{MMSE}},k}({\mathbf{r}})=\argmax_{{w}_k(\mathbf{r})}\gamma_k
=\argmax_{{w}_k(\mathbf{r})}\frac{\frac{P_{k}}{\sigma^2}f_{\mathsf{nm}}({w}_k(\mathbf{r}))}{f_{\mathsf{dm}}({w}_k(\mathbf{r}))},
\end{align}
where 
\begin{align}
f_{\mathsf{nm}}({w}_k(\mathbf{r}))&\triangleq\iint_{{\mathcal{A}}^2}w_k^{*}(\mathbf{r})h_{k}(\mathbf{r})h_{k}^{*}(\mathbf{r}')w_k({\mathbf{r}}'){\rm{d}}{\mathbf{r}}'{\rm{d}}{\mathbf{r}},
\label{Channel_No_User_k_Correlation_Matrix_With_Power_Trans_Result3_Before}\\
f_{\mathsf{dm}}({w}_k(\mathbf{r}))&\triangleq\iint_{{\mathcal{A}}^2}w_k^{*}(\mathbf{r})C_k({\mathbf{r}},{\mathbf{r}}')w_k({\mathbf{r}}'){\rm{d}}{\mathbf{r}}'{\rm{d}}{\mathbf{r}},
\label{Channel_No_User_k_Correlation_Matrix_With_Power_Trans_Result2_Before}
\end{align}
and $C_k({\mathbf{r}},{\mathbf{r}}')\triangleq \delta({\mathbf{r}}-{\mathbf{r}}')+\sum_{k'\in{\mathcal{K}}_k}\frac{P_{k'}}{\sigma^2}h_{k'}(\mathbf{r})h_{k'}^{*}(\mathbf{r}')$. The optimization problem in \eqref{SINR_Rayleigh_Objective} is equivalent to maximizing the following generalized Rayleigh quotient:
\begin{align}\label{Rayleigh_Quotient_Operator}
\gamma_k=
\frac{\frac{P_{k}}{\sigma^2}\iint_{{\mathcal{A}}^2}w_k^{*}(\mathbf{r})h_{k}(\mathbf{r})h_{k}^{*}(\mathbf{r}')w_k({\mathbf{r}}'){\rm{d}}{\mathbf{r}}'{\rm{d}}{\mathbf{r}}}
{\iint_{{\mathcal{A}}^2}w_k^{*}(\mathbf{r})C_k({\mathbf{r}},{\mathbf{r}}')w_k({\mathbf{r}}'){\rm{d}}{\mathbf{r}}'{\rm{d}}{\mathbf{r}}}.
\end{align}
By referring to \eqref{Rayleigh_Quotient_Operator}, the Rayleigh quotient for CAPAs is based on a continuous operator, which makes it significantly more challenging to address than its SPDA counterpart that is represented in matrix form.
\subsubsection{Optimal Linear Receive Beamforming}
To address the problem formulated in \eqref{SINR_Rayleigh_Objective}, we first define the correlation matrix ${\mathbf{C}}_k\in{\mathbbmss{C}}^{{K_1}\times{K_1}}$ as follows:
\begin{align}\label{Channel_No_User_k_Correlation_Matrix_With_Power}
{\mathbf{C}}_k\triangleq{\mathbf{P}}_k^{\frac{1}{2}}\int_{\mathcal{A}}{\mathbf{h}}_k^{\mathsf{H}}({\mathbf{r}}){\mathbf{h}}_k({\mathbf{r}}){\rm{d}}{\mathbf{r}}{\mathbf{P}}_k^{\frac{1}{2}}
={\mathbf{P}}_k^{\frac{1}{2}}{\mathbf{R}}_k{\mathbf{P}}_k^{\frac{1}{2}}\succ{\mathbf{0}},
\end{align}
where ${{\mathbf{P}}_k}={\rm{diag}}([\frac{P_{k'}}{\sigma^2}]_{k'\in{\mathcal{K}}_k})$ is the power allocation matrix. The EVD of ${\mathbf{C}}_k$ is ${\mathbf{U}}_{{\mathbf{C}}_k}{\bm\Lambda}_{{\mathbf{C}}_k}{\mathbf{U}}_{{\mathbf{C}}_k}^{\mathsf{H}}$, where ${\mathbf{U}}_{{\mathbf{C}}_k}{\mathbf{U}}_{{\mathbf{C}}_k}^{\mathsf{H}}={\mathbf{U}}_{{\mathbf{C}}_k}^{\mathsf{H}}{\mathbf{U}}_{{\mathbf{C}}_k}={\mathbf{I}}_{K_1}$ and ${\bm\Lambda}_{{\mathbf{C}}_k}={\rm{diag}}([\lambda_{{\mathbf{C}}_k,k'}]_{k'\in{\mathcal{K}}_k})\in{\mathbbmss{C}}^{{K_1}\times{K_1}}$ contains positive eigenvalues $\lambda_{{\mathbf{C}}_k,k'}>0$ for $k'\in{\mathcal{K}}_k$. For convenience, we define the following auxiliary operators:
\begin{align}
{{B}}_{k}({\mathbf{r}},{\mathbf{r}}')&\triangleq\delta({\mathbf{r}}-{\mathbf{r}}')-{\mathbf{h}}_k({\mathbf{r}}){\mathbf{P}}_k^{\frac{1}{2}}{\mathbf{B}}_k{\mathbf{P}}_k^{\frac{1}{2}}{\mathbf{h}}_k^{\mathsf{H}}({\mathbf{r}}'),
\label{MMSE_Proof_Pilot1}\\
{\overline{B}}_{k}({\mathbf{r}},{\mathbf{r}}')&\triangleq\delta({\mathbf{r}}-{\mathbf{r}}')
-{\mathbf{h}}_k({\mathbf{r}}){\mathbf{P}}_k^{\frac{1}{2}}{\overline{{\mathbf{B}}}_{k}}{\mathbf{P}}_k^{\frac{1}{2}}{\mathbf{h}}_k^{\mathsf{H}}({\mathbf{r}}'),\label{MMSE_Proof_Pilot2}
\end{align}
where ${\mathbf{B}}_k={\mathbf{U}}_{{{{\mathbf{C}}}_{k}}}{{\bm\Lambda}}_{{{\mathbf{B}}}_{k}}{\mathbf{U}}_{{{\mathbf{C}}}_{k}}^{\mathsf{H}}\in{\mathbbmss{C}}^{{K_1}\times{K_1}}$, ${\overline{{\mathbf{B}}}_{k}}={\mathbf{U}}_{{{{\mathbf{C}}}_{k}}}{{\bm\Lambda}}_{\overline{{\mathbf{B}}}_{k}}{\mathbf{U}}_{{{\mathbf{C}}}_{k}}^{\mathsf{H}}\in{\mathbbmss{C}}^{{K_1}\times{K_1}}$, ${{\bm\Lambda}}_{{{\mathbf{B}}}_{k}}={\rm{diag}}([\lambda_{{\mathbf{B}}_k,k'}]_{k'\in{\mathcal{K}}_k})\in{\mathbbmss{C}}^{{K_1}\times{K_1}}$, and ${{\bm\Lambda}}_{{\overline{\mathbf{B}}}_{k}}={\rm{diag}}([\lambda_{\overline{\mathbf{B}}_k,k'}]_{k'\in{\mathcal{K}}_k})\in{\mathbbmss{C}}^{{K_1}\times{K_1}}$. Furthermore, we set ${{\lambda}}_{{\mathbf{B}_{k}},k'}=\frac{1+\sqrt{1+\lambda_{{\mathbf{C}_{k}},k'}}}{\lambda_{{\mathbf{C}_{k}},k'}}>0$ and ${{\lambda}}_{{{\overline{\mathbf{B}}}_{k}},k'}=\frac{1+\sqrt{1+\lambda_{{\mathbf{C}_{k}},k'}}}{\lambda_{{\mathbf{C}_{k}},k'}\sqrt{1+\lambda_{{\mathbf{C}_{k}},k'}}}>0$ for $k'\in{\mathcal{K}}_k$. Here, ${\mathbf{U}}_{{{{\mathbf{C}}}_{k}}}{{\bm\Lambda}}_{{{\mathbf{B}}}_{k}}{\mathbf{U}}_{{{\mathbf{C}}}_{k}}^{\mathsf{H}}$ and ${\mathbf{U}}_{{{{\mathbf{C}}}_{k}}}{{\bm\Lambda}}_{\overline{{\mathbf{B}}}_{k}}{\mathbf{U}}_{{{\mathbf{C}}}_{k}}^{\mathsf{H}}$ denote the EVDs of ${\mathbf{B}}_k$ and ${\overline{{\mathbf{B}}}_{k}}$, respectively. Using these definitions, we derive
\begin{align}
&\iint_{{\mathcal{A}}^2}\!{\overline{B}}_{k}({\mathbf{r}}_1,{\mathbf{r}})C_{k}({\mathbf{r}},{\mathbf{r}}'){\overline{B}}_{k}({\mathbf{r}}',{\mathbf{r}}_2){\rm{d}}{\mathbf{r}}'{\rm{d}}{\mathbf{r}}
=\delta({\mathbf{r}}_1-{\mathbf{r}}_2),\label{Channel_No_User_k_Correlation_Matrix_With_Power_Trans1}\\
&\int_{{\mathcal{A}}}{\overline{B}}_{k}({\mathbf{r}}_1,{\mathbf{r}})B_{k}({\mathbf{r}},{\mathbf{r}}_2){\rm{d}}{\mathbf{r}}
=\int_{{\mathcal{A}}}B_{k}({\mathbf{r}}_2,{\mathbf{r}}){\overline{B}}_{k}({\mathbf{r}},{\mathbf{r}}_1){\rm{d}}{\mathbf{r}}\nonumber\\
&\qquad\qquad\qquad\qquad\qquad~=\delta({\mathbf{r}}_1-{\mathbf{r}}_2).\label{Channel_No_User_k_Correlation_Matrix_With_Power_Trans2}
\end{align}
A full derivation of \eqref{Channel_No_User_k_Correlation_Matrix_With_Power_Trans1} and \eqref{Channel_No_User_k_Correlation_Matrix_With_Power_Trans2} is given in Appendix \ref{Section: Analysis of the SINR: Preliminaries1}.

From \eqref{Channel_No_User_k_Correlation_Matrix_With_Power_Trans1}, ${\overline{B}}_{k}({\mathbf{r}},{\mathbf{r}}')$ can be viewed as the \emph{inverse square root} of $C_{k}({\mathbf{r}},{\mathbf{r}}')$. From \eqref{Channel_No_User_k_Correlation_Matrix_With_Power_Trans2}, ${\overline{B}}_{k}({\mathbf{r}},{\mathbf{r}}')$ and $B_{k}({\mathbf{r}},{\mathbf{r}}')$ are \emph{mutually invertible}. To leverage these facts, we define the auxiliary function $u_k({\mathbf{r}})\triangleq\int_{{\mathcal{A}}} B_{k}({\mathbf{r}},{\mathbf{r}}'){w}_k({\mathbf{r}}'){\rm{d}}{\mathbf{r}}'$. For any ${u}_k(\mathbf{r})$, the corresponding $w_k(\mathbf{r})$ is recovered as follows:
\begin{align}\label{Equivalent_Trans_Problem_2}
w_k({\mathbf{r}})=\int_{{\mathcal{A}}} {\overline{B}}_{k}({\mathbf{r}},{\mathbf{r}}'){u}_k({\mathbf{r}}'){\rm{d}}{\mathbf{r}}'.
\end{align}
From \eqref{Equivalent_Trans_Problem_2}, we rewrite $f_{\mathsf{dm}}({w}_k(\mathbf{r}))$ and $f_{\mathsf{nm}}({w}_k(\mathbf{r}))$ as follows:
\begin{align}
f_{\mathsf{dm}}({w}_k(\mathbf{r}))&=\int_{{\mathcal{A}}}\lvert{u}_k({\mathbf{r}})\rvert^2{\rm{d}}{\mathbf{r}},\label{Channel_No_User_k_Correlation_Matrix_With_Power_Trans_Result2}\\
f_{\mathsf{nm}}({w}_k(\mathbf{r}))&=\left\lvert\int_{{\mathcal{A}}}{u}_k^{*}({\mathbf{r}}){v}_k({\mathbf{r}}){\rm{d}}{\mathbf{r}}\right\rvert^2,
\label{Channel_No_User_k_Correlation_Matrix_With_Power_Trans_Result3}
\end{align}
where ${v}_k({\mathbf{r}})\triangleq \int_{{\mathcal{A}}}{\overline{B}}_{k}({\mathbf{r}},{\mathbf{r}}')h_{k}({\mathbf{r}}'){\rm{d}}{\mathbf{r}}'$. Appendix \ref{Proof_Lemma_Channel_No_User_k_Correlation_Matrix_With_Power_Trans_Result23} provides a detailed derivation of \eqref{Channel_No_User_k_Correlation_Matrix_With_Power_Trans_Result2} and \eqref{Channel_No_User_k_Correlation_Matrix_With_Power_Trans_Result3}.

Combining \eqref{Channel_No_User_k_Correlation_Matrix_With_Power_Trans_Result2} with \eqref{Channel_No_User_k_Correlation_Matrix_With_Power_Trans_Result3}, we reformulate \eqref{SINR_Rayleigh_Objective} as follows:
\begin{align}\label{Rayleigh_Quotient_Operator_Transformation}
u_k^{\star}({\mathbf{r}})=\argmax_{{u}_k(\mathbf{r})}
\frac{\frac{P_{k}}{\sigma^2}\lvert\int_{{\mathcal{A}}}{u}_k^{*}({\mathbf{r}}){v}_k({\mathbf{r}}){\rm{d}}{\mathbf{r}}\rvert^2}
{\int_{{\mathcal{A}}}\lvert{u}_k({\mathbf{r}})\rvert^2{\rm{d}}{\mathbf{r}}},
\end{align}
where the variable to be optimized shifts from $w_k({\mathbf{r}})=\int_{{\mathcal{A}}} {\overline{B}}_{k}({\mathbf{r}},{\mathbf{r}}'){u}_k({\mathbf{r}}'){\rm{d}}{\mathbf{r}}'$ to $u_k({\mathbf{r}})=\int_{{\mathcal{A}}} B_{k}({\mathbf{r}},{\mathbf{r}}'){w}_k({\mathbf{r}}'){\rm{d}}{\mathbf{r}}'$. The mutual convertibility of ${u}_k(\mathbf{r})$ and $w_k(\mathbf{r})$ ensures \emph{equivalence} between problems \eqref{Rayleigh_Quotient_Operator} and \eqref{Rayleigh_Quotient_Operator_Transformation}. Examining \eqref{Rayleigh_Quotient_Operator_Transformation} reveals that its optimal solution is independent of $\int_{{\mathcal{A}}}\lvert{u}_k({\mathbf{r}})\rvert^2{\rm{d}}{\mathbf{r}}$. Thus, the optimal $u_k^{\star}({\mathbf{r}})$ can be selected as follows:
\begin{align}\nonumber
u_k^{\star}({\mathbf{r}})={v}_k({\mathbf{r}}).
\end{align}
Using the transformation in \eqref{Equivalent_Trans_Problem_2} and setting $u_k({\mathbf{r}})=u_k^{\star}({\mathbf{r}})$, the optimal solution to \eqref{Rayleigh_Quotient_Operator} can be written as follows:
\begin{align}\nonumber
w_{{\mathsf{MMSE}},k}({\mathbf{r}})
=\int_{{\mathcal{A}}} {\overline{B}}_{k}({\mathbf{r}},{\mathbf{r}}'){v}_k({\mathbf{r}}'){\rm{d}}{\mathbf{r}}',
\end{align}
which, together with ${v}_k({\mathbf{r}})= \int_{{\mathcal{A}}}{\overline{B}}_{k}({\mathbf{r}},{\mathbf{r}}')h_{k}({\mathbf{r}}'){\rm{d}}{\mathbf{r}}'$, gives
\begin{align}\label{Rayleigh_Quotient_Optimal_Solution_Step2_Eq1}
w_{{\mathsf{MMSE}},k}({\mathbf{r}})
=\int_{{\mathcal{A}}} {\overline{B}}_{k}({\mathbf{r}},{\mathbf{r}}')
{\overline{B}}_{k}({\mathbf{r}}',{\mathbf{r}}'')h_{k}({\mathbf{r}}''){\rm{d}}{\mathbf{r}}''{\rm{d}}{\mathbf{r}}'.
\end{align}
Calculating the integral in \eqref{Rayleigh_Quotient_Optimal_Solution_Step2_Eq1}, we derive the \emph{optimal uplink beamformer} that maximizes the SINR in \eqref{SINR_Rayleigh_Objective} as follows. 
\vspace{-5pt}
\begin{theorem}[Rate-Optimal MMSE Beamforming for CAPAs]\label{Theorem_MMSE_Rate_Optimal}
Given the channel responses $\{h_k({\mathbf{r}})\}_{k=1}^{K}$, the following beamformers maximize the per-user rate given in \eqref{Per_User_Rate}:
\begin{align}\label{Rate_Optimal_Beamformer_General_Matrix}
\boxed{w_{{\mathsf{MMSE}},k}({\mathbf{r}})=h_k({\mathbf{r}})-{\mathbf{h}}_k({\mathbf{r}})({\mathbf{P}}_k^{-1}+{\mathbf{R}}_k)^{-1}{\mathbf{r}}_{k,k},~\forall k\in{\mathcal{K}}.}
\end{align}
\end{theorem}
\vspace{-5pt} 
\begin{IEEEproof}
Please refer to Appendix \ref{Proof_Theorem_MMSE_Rate_Optimal} for more details.
\end{IEEEproof}
\subsubsection{Further Discussion}
We now provide a more in-depth interpretation by analyzing \eqref{Rayleigh_Quotient_Optimal_Solution_Step2_Eq1}. The optimal beamformer $w_{{\mathsf{MMSE}},k}({\mathbf{r}})$ comprises two main components: ${\overline{B}}_{k}({\mathbf{r}},{\mathbf{r}}')$ and ${\overline{B}}_{k}({\mathbf{r}}',{\mathbf{r}}'')h_{k}({\mathbf{r}}'')$. Referring to \eqref{Multiple_Access_Channel_Basic_Channel_Model_CAPA}, we denote $z_k({\mathbf{r}})=\sum_{k'\in{\mathcal{K}}_k}\sqrt{P_{k'}}h_{k'}({\mathbf{r}})s_{k'}+n({\mathbf{r}})$ as the interference-plus-noise term when decoding $s_{k}$. Given that $\{s_k\sim{\mathcal{CN}}(0,1)\}_{k=1}^{K}$ are uncorrelated with $n({\mathbf{r}})$, the correlation function of the complex Gaussian random process $\check{n}_k({\mathbf{r}}')\triangleq\int_{\mathcal{A}}{\overline{B}}_{k}^{*}({\mathbf{r}},{\mathbf{r}}')z_k({\mathbf{r}}){\rm{d}}{\mathbf{r}}
=\int_{\mathcal{A}}{\overline{B}}_{k}({\mathbf{r}}',{\mathbf{r}})z_k({\mathbf{r}}){\rm{d}}{\mathbf{r}}$ is given by
\begin{equation}\nonumber
\begin{split}
&{\mathbbmss{E}}\{\check{n}_k({\mathbf{r}}_1')\check{n}_k^{*}({\mathbf{r}}_2')\}\\
&=\iint_{\mathcal{A}^2}{\overline{B}}_{k}({\mathbf{r}}_1',{\mathbf{r}})\sigma^2 C_k({\mathbf{r}},{\mathbf{r}}'){\overline{B}}_{k}({\mathbf{r}}',{\mathbf{r}}_2'){\rm{d}}{\mathbf{r}}{\rm{d}}{\mathbf{r}}',
\end{split}
\end{equation}
where $C_k({\mathbf{r}},{\mathbf{r}}')= \delta({\mathbf{r}}-{\mathbf{r}}')+\sum_{k'\in{\mathcal{K}}_k}\frac{P_{k'}}{\sigma^2}h_{k'}(\mathbf{r})h_{k'}^{*}(\mathbf{r}')$. It follows from \eqref{Channel_No_User_k_Correlation_Matrix_With_Power_Trans1} that
\begin{align}\nonumber
{\mathbbmss{E}}\{\check{n}_k({\mathbf{r}}_1')\check{n}_k^{*}({\mathbf{r}}_2')\}=\sigma^2\delta({\mathbf{r}}_1'-{\mathbf{r}}_2').
\end{align}
This implies that $\check{n}_k({\mathbf{r}}')=\int_{\mathcal{A}}{\overline{B}}_{k}({\mathbf{r}}',{\mathbf{r}})z_k({\mathbf{r}}){\rm{d}}{\mathbf{r}}$ is a white Gaussian noise process.
\vspace{-5pt}
\begin{remark}\label{Remark_Optimal_Beamforming_Whitening}
The above results suggest that the invertible operator ${\overline{B}}_{k}({\mathbf{r}},{\mathbf{r}}')$ in the optimal beamformer $w_{{\mathsf{MMSE}},k}({\mathbf{r}})$ acts as a filter to whiten the interference-plus-noise term.
\end{remark}
\vspace{-5pt}
After passing the received signal $y(\mathbf{r})$ through the whitening filter ${\overline{B}}_{k}^{*}({\mathbf{r}},{\mathbf{r}}')={\overline{B}}_{k}({\mathbf{r}}',{\mathbf{r}})$, the output signal is
\begin{equation}\label{Output_Signal_Whitening_Filtering}
\begin{split}
y_k(\mathbf{r})&\triangleq\int_{{\mathcal{A}}}{\overline{B}}_{k}({\mathbf{r}}',{\mathbf{r}})y(\mathbf{r}){\rm{d}}{\mathbf{r}}\\
&=\sqrt{P_k}s_k\int_{{\mathcal{A}}}{\overline{B}}_{k}({\mathbf{r}}',{\mathbf{r}}'')h_{k}({\mathbf{r}}''){\rm{d}}{\mathbf{r}}''
+\check{n}_k({\mathbf{r}}').
\end{split}
\end{equation}
From \eqref{Output_Signal_Whitening_Filtering}, we observe that the output of the whitening filter is equivalent to the received signal of a single-user channel, with $s_k$ being the data symbol and $\int_{{\mathcal{A}}}{\overline{B}}_{k}({\mathbf{r}}',{\mathbf{r}}'')h_{k}({\mathbf{r}}''){\rm{d}}{\mathbf{r}}''$ representing the effective channel. 
\vspace{-5pt}
\begin{remark}\label{Remark_Optimal_Beamforming_MRC_Beamforming}
By observing \eqref{Output_Signal_Whitening_Filtering}, we deduce that the subsequent rate-optimal beamformer to decode $s_k$ from $y_k(\mathbf{r})$ should be the MRC beamformer that is aligned with the effective channel ${\overline{B}}_{k}({\mathbf{r}}',{\mathbf{r}}'')h_{k}({\mathbf{r}}'')$, which is the remaining part of the optimal beamformer $w_{{\mathsf{MMSE}},k}({\mathbf{r}})$ as per \eqref{Rayleigh_Quotient_Optimal_Solution_Step2_Eq1}.
\end{remark}
\vspace{-5pt}
In summary, the structure of the optimal uplink beamformer consists of two filters: the first one whitens the interference-plus-noise term, while the second applies MRC (matched filtering). This structure mirrors that of SPDAs \cite[Section 8.3.3]{tse2005fundamentals} and is also known as the \emph{receive Wiener filter} \cite{verdu1998multiuser}. 
\subsection{Sum-Rate Performance}
Upon utilizing the equivalence between \eqref{SINR_Rayleigh_Objective} and \eqref{Rayleigh_Quotient_Operator_Transformation}, the SINR achieved by the optimal uplink beamformer $w_{{\mathsf{MMSE}},k}({\mathbf{r}})$ can be expressed as follows:
\begin{subequations}\label{MMSE_SINR_Maximization_Final}
\begin{align}
\max_{{w}_k(\mathbf{r})}\gamma_k&=\max_{{u}_k(\mathbf{r})}
\frac{P_{k}}{\sigma^2}\frac{\lvert\int_{{\mathcal{A}}}{u}_k^{*}({\mathbf{r}}){v}_k({\mathbf{r}}){\rm{d}}{\mathbf{r}}\rvert^2}
{\int_{{\mathcal{A}}}\lvert{u}_k({\mathbf{r}})\rvert^2{\rm{d}}{\mathbf{r}}}\\&=\frac{P_{k}}{\sigma^2}\int_{{\mathcal{A}}}\lvert{v}_k({\mathbf{r}})\rvert^2{\rm{d}}{\mathbf{r}}
\triangleq \gamma_{{\mathsf{MMSE}},k}.
\end{align}
\end{subequations}
Using ${v}_k({\mathbf{r}})= \int_{{\mathcal{A}}}{\overline{B}}_{k}({\mathbf{r}},{\mathbf{r}}')h_{k}({\mathbf{r}}'){\rm{d}}{\mathbf{r}}'$ and ${\overline{B}}_{k}^{*}({\mathbf{r}},{\mathbf{r}}')={\overline{B}}_{k}({\mathbf{r}}',{\mathbf{r}})$, we derive a closed-form $\gamma_{{\mathsf{MMSE}},k}$ as follows.
\vspace{-5pt}
\begin{corollary}\label{Corollary_SINR_Per_User_Optimal_Beamforming}
Given the channel responses $\{h_k({\mathbf{r}})\}_{k=1}^{K}$, the SINR of each user $k\in{\mathcal{K}}$ achieved by the optimal linear receive beamforming can be written as follows:
\begin{align}\label{SINR_Per_User_Optimal_Beamforming_Final_Result}
\gamma_{{\mathsf{MMSE}},k}=\frac{P_{k}}{\sigma^2}a_k-\frac{P_{k}}{\sigma^2}{\mathbf{r}}_{k,k}^{\mathsf{H}}({\mathbf{P}}_k^{-1}+{\mathbf{R}}_k)^{-1}{\mathbf{r}}_{k,k}.
\end{align}
\end{corollary}
\vspace{-5pt}
\begin{IEEEproof}
Please refer to Appendix \ref{Proof_Theorem_MMSE_Rate_Optimal} for more details.
\end{IEEEproof}
As a result, the sum-rate achieved by the optimal uplink beamforming can be written as follows:
\begin{align}\nonumber
{\mathcal{R}}=\sum_{k=1}^{K}{\mathcal{R}}_k=\sum_{k=1}^{K}\log_2\left(1+\frac{P_k}{\sigma^2}a_{k}\left(1-\alpha_{{\mathsf{MMSE}},k}\right)\right),
\end{align}
where $\alpha_{{\mathsf{MMSE}},k}\triangleq \frac{1}{a_{k}}{\mathbf{r}}_{k,k}^{\mathsf{H}}({\mathbf{P}}_k^{-1}+{\mathbf{R}}_k)^{-1}{\mathbf{r}}_{k,k}$ accounts for the SNR loss factor caused by IUI. We prove in Appendix \ref{Proof_MMSE_Power_Loss_Factor} that $\alpha_{{\mathsf{MMSE}},k}\in[0,1)$.
\subsection{Mean-Squared Error Performance}
Next, we analyze the MSE performance of the optimal beamforming in estimating $s_k$. As discussed earlier, the rate-optimal uplink beamforming also minimizes the MSE in estimating $s_k$, which is referred to as \emph{MMSE beamforming}. By substituting \eqref{Rate_Optimal_Beamformer_General_Matrix} into \eqref{MSE_Fundamental}, we derive the optimal scalar filter $\beta_k^{\star}$ and the MMSE beamformer that minimizes the MSE for estimating $s_k$, which is presented in the following lemma.
\vspace{-5pt}
\begin{lemma}\label{Lemma_Optimal_MMSE_Indirectky}
Under MMSE beamforming, the optimal scalar filter $\beta_k^{\star}$ and the effective beamformer that minimizes ${\mathsf{MSE}}_k$ are given by
\begin{align}
&\beta_k^{\star}=\frac{\sqrt{P_k}}
{P_{k}(a_k-{\mathbf{r}}_{k,k}^{\mathsf{H}}({\mathbf{P}}_k^{-1}+{\mathbf{R}}_k)^{-1}{\mathbf{r}}_{k,k})+\sigma^2}\triangleq \beta_{{\mathsf{MMSE}},k},\nonumber\\
&w_{{\mathsf{MMSE}},k}^{\star}({\mathbf{r}})=\frac{\sqrt{P_k}(h_k({\mathbf{r}})-{\mathbf{h}}_k({\mathbf{r}})({\mathbf{P}}_k^{-1}+{\mathbf{R}}_k)^{-1}{\mathbf{r}}_{k,k})}
{P_{k}(a_k-{\mathbf{r}}_{k,k}^{\mathsf{H}}({\mathbf{P}}_k^{-1}+{\mathbf{R}}_k)^{-1}{\mathbf{r}}_{k,k})+\sigma^2},\nonumber
\end{align}
respectively, where $k\in{\mathcal{K}}$. 
\end{lemma}
\vspace{-5pt}
\begin{IEEEproof}
Please refer to Appendix \ref{Proof_Lemma_Optimal_MMSE_Indirectky} for more details.
\end{IEEEproof}
Based on \eqref{Per_User_MSE} and \eqref{SINR_Per_User_Optimal_Beamforming_Final_Result}, the MMSE for estimating $s_k$ can be written as follows:
\begin{align}\nonumber
{\mathsf{MSE}}_k=\frac{1}{1+\gamma_{{\mathsf{MMSE}},k}}=\frac{1}{1+\frac{P_k}{\sigma^2}a_{k}\left(1-\alpha_{{\mathsf{MMSE}},k}\right)}.
\end{align}
The results in \textbf{Lemma \ref{Lemma_Optimal_MMSE_Indirectky}} are partly based on those derived from rate maximization, such as \textbf{Theorem \ref{Theorem_MMSE_Rate_Optimal}}. However, from the perspective of minimizing the MSE, the MSE-optimal beamformer $w_{{\mathsf{MMSE}},k}^{\star}({\mathbf{r}})$ can also be directly calculated by solving the following MSE minimization problem:
\begin{align}\label{MMSE_Formulation}
\argmin\nolimits_{w_k({\mathbf{r}})}{\overline{\mathsf{MSE}}}_k\triangleq{\mathbbmss{E}}\left\{\left\lvert \hat{y}_k - s_k\right\rvert^2\right\},
\end{align}
where $\hat{y}_k=\int_{{\mathcal{A}}}{w}_k^{*}(\mathbf{r})y(\mathbf{r}){\rm{d}}{\mathbf{r}}$ is given in \eqref{Linear_Receive_Beamforming}.
\vspace{-5pt}
\begin{theorem}[MSE-Optimal MMSE Beamforming for CAPAs]\label{Theorem_MMSE_Uplink_Direct_Solution}
Given the channel responses $\{h_k({\mathbf{r}})\}_{k=1}^{K}$, the solution to the MSE minimization problem formulated in \eqref{MMSE_Formulation} is given by
\begin{align}\label{MMSE_Uplink_Direct_Solution_Eq2}
w_{{\mathsf{MMSE}},k}^{\diamond}({\mathbf{r}})=\frac{\sqrt{P_k}}{\sigma^2}(h_k({\mathbf{r}})-{\mathbf{h}}({\mathbf{r}}){\mathbf{P}}^{\frac{1}{2}}{\tilde{\mathbf{C}}}{\mathbf{P}}^{\frac{1}{2}}{\mathbf{r}}_k),
\end{align}
where $k\in{\mathcal{K}}$, ${{\mathbf{P}}}={\rm{diag}}([\frac{P_k}{\sigma^2}]_{k\in{\mathcal{K}}})\in{\mathbbmss{C}}^{K\times K}$, ${\tilde{\mathbf{C}}}=({\mathbf{I}}_{K}+{\mathbf{P}}^{\frac{1}{2}}{\mathbf{R}}{\mathbf{P}}^{\frac{1}{2}})^{-1}\in{\mathbbmss{C}}^{K\times K}$, and ${\mathbf{R}}$ is defined in \eqref{Channel_Correlation_Matrix_No_Power} with ${\mathbf{r}}_k=\int_{\mathcal{A}}{\mathbf{h}}^{\mathsf{H}}({\mathbf{r}}){{h}}_k({\mathbf{r}}){\rm{d}}{\mathbf{r}}\in{\mathbbmss{C}}^{K\times 1}$ being its $k$th column.
\end{theorem}
\vspace{-5pt}
\begin{IEEEproof}
Please refer to Appendix \ref{Proof_Theorem_MMSE_Uplink_Direct_Solution} for more details.
\end{IEEEproof}
Upon comparing $w_{{\mathsf{MMSE}},k}^{\star}({\mathbf{r}})$ with $w_{{\mathsf{MMSE}},k}^{\diamond}({\mathbf{r}})$, the following corollary can be found.
\vspace{-5pt}
\begin{corollary}\label{Corollary_MMSE_Uplink_Direct_Indirect_Comparison}
Given the channel responses $\{h_k({\mathbf{r}})\}_{k=1}^{K}$, it holds that $w_{{\mathsf{MMSE}},k}^{\diamond}({\mathbf{r}})=w_{{\mathsf{MMSE}},k}^{\star}({\mathbf{r}})$ for $k\in{\mathcal{K}}$. 
\end{corollary}
\vspace{-5pt}
\begin{IEEEproof}
Please refer to Appendix \ref{Proof_Corollary_MMSE_Uplink_Direct_Indirect_Comparison} for more details.
\end{IEEEproof}
\vspace{-5pt}
\begin{remark}
The results in \textbf{Corollary \ref{Corollary_MMSE_Uplink_Direct_Indirect_Comparison}} suggest that the MMSE beamformer obtained by solving the MSE minimization problem \eqref{MMSE_Formulation} is consistent with that derived from the rate maximization problem \eqref{SINR_Rayleigh_Objective}. This finding further reinforces the conclusion in \textbf{Remark \ref{Remark_MMSE_Rate}} that maximizing the uplink per-user rate is equivalent to minimizing the MSE.
\end{remark}
\vspace{-5pt}
Since $\frac{\sqrt{P_k}}{\sigma^2}$ in $w_{{\mathsf{MMSE}},k}^{\diamond}({\mathbf{r}})$ does not affect the SINR, the MMSE beamformer can also be expressed as follows.
\vspace{-5pt}
\begin{corollary}\label{Corollary_MMSE_Uplink_Direct_Indirect_Comparison_Beamforming_Rewritten}
Given the channel responses $\{h_k({\mathbf{r}})\}_{k=1}^{K}$, the beamformers that maximize the per-user rate given in \eqref{Per_User_Rate} can be also written as follows:
\begin{align}\nonumber
\boxed{w_{{\mathsf{MMSE}},k}(\mathbf{r})
=\sum_{k'=1}^{K}[({\mathbf{I}}_K+{\mathbf{P}}{\mathbf{R}})^{-1}]_{k',k}h_{k'}({\mathbf{r}}),~\forall k\in{\mathcal{K}}.}
\end{align}
These beamformers can also be represented as follows:
\begin{align}\label{MMSE_General_Another_Matrix}
\boxed{{{\mathbf{w}}}(\mathbf{r})={\mathbf{h}}({\mathbf{r}})({\mathbf{I}}_K+{\mathbf{P}}{\mathbf{R}})^{-1}\triangleq {{\mathbf{w}}}_{{\mathsf{MMSE}}}(\mathbf{r}).}
\end{align}
\end{corollary}
\vspace{-5pt}
\begin{IEEEproof}
Please refer to Appendix \ref{Proof_Corollary_MMSE_Uplink_Direct_Indirect_Comparison_Beamforming_Rewritten} for more details.
\end{IEEEproof}
\section{Performance Comparison and Evaluation}\label{Section: Performance Comparison and Further Discussion}
\subsection{Comparison Among Beamforming for CAPAs}\label{Subsection: Performance Comparison Among Beamforming for CAPAs}
We now proceed to compare the performance of the three CAPA beamforming methods discussed.
\subsubsection{General Discussion}
Since MMSE beamforming is designed to maximize per-user SINR, it follows that
\begin{align}\nonumber
\gamma_{{\mathsf{MMSE}},k}\geq \gamma_{{\mathsf{MRC}},k},~\gamma_{{\mathsf{MMSE}},k}\geq \gamma_{{\mathsf{ZF}},k},~\forall k\in{\mathcal{K}}. 
\end{align}
To explore this further, we focus on the SINR for user $1$ in a two-user scenario where $K=2$. Under this setup, the following parameters hold
\begin{align}\nonumber
{\mathbf{R}}_1=r_{2,2}=a_2,~{\mathbf{r}}_{1,1}=r_{1,2}.
\end{align}
This yields $\gamma_{{\mathsf{MRC}},1}=\frac{P_1 a_1}{\sigma^2}\left(1-\frac{{P_2 a_2}\rho}{{\sigma^2}+{P_2 a_2}\rho}\right)$, $\gamma_{{\mathsf{ZF}},1}=\frac{P_1a_1}{\sigma^2}\left(1-\rho\right)$, and $\gamma_{{\mathsf{MMSE}},1}=\frac{P_1a_1}{\sigma^2}
\left(1-\frac{{P_2a_2}\rho}{{\sigma^2}+{P_2a_2}}\right)$, where $\rho=\rho_{1,2}$, as defined in \eqref{Definition_Channel_Squared_Correlation_Coefficient},  denotes the squared channel correlation between the two users. It is clear from $\rho\in[0,1)$ that $\gamma_{{\mathsf{MMSE}},1}\geq\gamma_{{\mathsf{MRC}},1}$ and $\gamma_{{\mathsf{MMSE}},1}\geq\gamma_{{\mathsf{ZF}},1}$, which confirms the superiority of MMSE beamforming. Additionally, if $\rho>1-\frac{\sigma^2}{P_2a_2}$, then $\gamma_{{\mathsf{MRC}},1}>\gamma_{{\mathsf{ZF}},1}$, which indicates that when users’ spatial responses are highly correlated, MRC beamforming may outperform ZF beamforming due to the lower SNR loss from preserving IUI.

We then compare the computational complexity of the three beamforming methods. Each CAPA beamformer (MRC, ZF, and MMSE) can be represented as a linear combination of the users’ spatial responses, as shown in \eqref{CAPA_MRC_Closed_Form_Matrix}, \eqref{ZFBF_General_Matrix}, and \eqref{MMSE_General_Another_Matrix}. However, the weighting coefficients for ZF and MMSE beamforming involve matrix inversions of ${\mathbf{R}}^{-1}\in{\mathbbmss{C}}^{K\times K}$ and $({\mathbf{P}}^{-1}+{\mathbf{R}})^{-1}\in{\mathbbmss{C}}^{K\times K}$, respectively. Therefore, the computational complexities for MRC, ZF, and MMSE beamforming are ${\mathcal{O}}(1)$, ${\mathcal{O}}(K^3)$, and ${\mathcal{O}}(K^3)$, respectively. 
\subsubsection{Asymptotic Discussion}\label{Subsubsection: Asymptotic Discussion}
At very low SNR (i.e., when $P_1,\ldots,P_K$ are significantly smaller than $\sigma^2$), it has $\frac{P_{k}}{\sigma^2}\rightarrow0$, $\forall k\in{\mathcal{K}}$, so $({\mathbf{I}}_K+{\mathbf{P}}{\mathbf{R}})^{-1}\rightarrow({\mathbf{I}}_{K})^{-1}={\mathbf{I}}_K$. Consequently, the optimal uplink beamforming in \eqref{MMSE_General_Another_Matrix} simplifies to MRC beamforming as shown in \eqref{CAPA_MRC_Closed_Form_Matrix}:
\begin{align}\nonumber
\lim\nolimits_{\frac{P_{1}}{\sigma^2},\ldots,\frac{P_{K}}{\sigma^2}\rightarrow0}{{\mathbf{w}}}_{{\mathsf{MMSE}}}(\mathbf{r})={\mathbf{h}}({\mathbf{r}}){\mathbf{I}}_K
={{\mathbf{w}}}_{{\mathsf{MRC}}}(\mathbf{r}).
\end{align}
Additionally, at high SNR (i.e., $\frac{P_{k}}{\sigma^2}\rightarrow\infty$, $\forall k\in{\mathcal{K}}$), we find
\begin{align}\nonumber
\lim\nolimits_{\frac{P_{1}}{\sigma^2},\ldots,\frac{P_{K}}{\sigma^2}\rightarrow\infty}({\mathbf{I}}_K+{\mathbf{P}}{\mathbf{R}})^{-1}\simeq
({\mathbf{P}}{\mathbf{R}})^{-1},
\end{align}
so the optimal uplink beamforming in \eqref{MMSE_General_Another_Matrix} aligns with ZF beamforming as shown in \eqref{ZFBF_General_Matrix}:
\begin{equation}\nonumber
\begin{split}
\lim\nolimits_{\frac{P_{1}}{\sigma^2},\ldots,\frac{P_{K}}{\sigma^2}\rightarrow\infty}{{\mathbf{w}}}_{{\mathsf{MMSE}}}(\mathbf{r})&\simeq
{\mathbf{h}}({\mathbf{r}}){\mathbf{R}}^{-1}{\mathbf{P}}^{-1}\\
&\parallel{\mathbf{h}}({\mathbf{r}}){\mathbf{R}}^{-1}={{\mathbf{w}}}_{{\mathsf{ZF}}}(\mathbf{r}),
\end{split}
\end{equation}
where $\parallel$ denotes element-wise parallel.

\begin{figure}[!t]
 \centering
\includegraphics[height=0.18\textwidth]{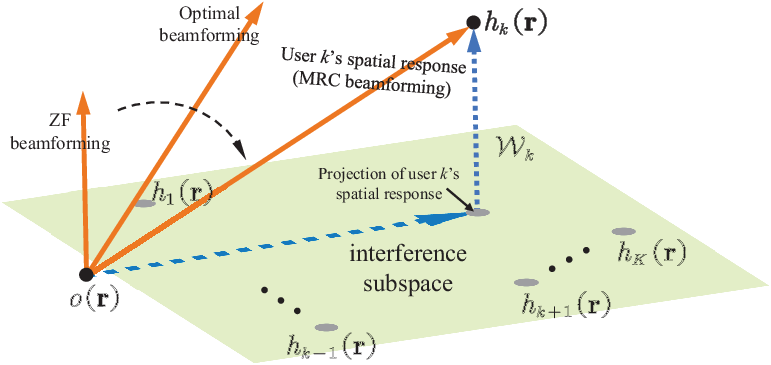}
\caption{The optimal beamforming goes from being the ZF beamforming at high SNR to the MRC beamforming at low SNR.}
\label{Relationship_Beamforming}
\end{figure}

These findings reveal that MRC and ZF beamforming represent the two extremes of optimal beamforming strategies. At high SNR, where IUI dominates over Gaussian noise, ZF beamforming excels. Conversely, at low SNR, where IUI is minimal, MRC beamforming becomes the preferred method to enhance each user's signal power. Thus, optimal (or MMSE) beamforming resembles ZF beamforming under high IUI conditions (i.e., high SNR) and resembles MRC beamforming when interference is low (i.e., low SNR), as shown in {\figurename} {\ref{Relationship_Beamforming}}.
\subsection{Comparison with Beamforming for SPDAs}
To further validate the derived results, we next simplify them to reflect conventional multiuser channels employing linear receive beamforming with SPDAs. For this purpose, we assume that the continuous surface $\mathcal{A}$ is discretized into $M$ spatially discrete antenna elements. Let ${\mathcal{S}}_m\subseteq {\mathcal{A}}$ and ${\mathbf{r}}_{m}^{\mathsf{s}}\in{\mathcal{S}}_m$ denote the aperture and center location of the $m$th antenna element for $m=1,\ldots,M$. Each SPDA element has an aperture size $\lvert{\mathcal{S}}_m\rvert=\lvert{\mathcal{S}}\rvert$ that is small relative to the propagation distance, with signal variations within each ${\mathcal{S}}_m$ considered negligible. Under these assumptions, the processed uplink signal at the BS after receive beamforming, as described in \eqref{Linear_Receive_Beamforming}, can be expressed in vector form as follows:
\begin{equation}\label{Linear_Receive_Beamforming_SPDA}
\begin{split}
\hat{y}_k&=\int_{\cup_{m=1}^{M}{\mathcal{S}}_m}{w}_k^{*}(\mathbf{r})y(\mathbf{r}){\rm{d}}{\mathbf{r}}\approx\sqrt{P_k}s_k\hat{\mathbf{w}}_k^{\mathsf{H}}\hat{\mathbf{h}}_k\\
&+\sum\nolimits_{k'\in {\mathcal{K}}_k}\sqrt{P_{k'}}s_{k'}\hat{\mathbf{w}}_k^{\mathsf{H}}\hat{\mathbf{h}}_{k'}+\hat{n}_k,
\end{split}
\end{equation}
where $\hat{\mathbf{h}}_k=[\sqrt{\lvert{\mathcal{S}}\rvert} h_k({\mathbf{r}}_1^{\mathsf{s}}),\ldots,\sqrt{\lvert{\mathcal{S}}\rvert} h_k({\mathbf{r}}_M^{\mathsf{s}})]^{\mathsf{T}}\in{\mathbbmss{C}}^{M\times1}$ is the channel vector for user $k$, $\hat{\mathbf{w}}_k=[w_k({\mathbf{r}}_1^{\mathsf{s}}),\ldots,w_k({\mathbf{r}}_M^{\mathsf{s}})]^{\mathsf{T}}\in{\mathbbmss{C}}^{M\times1}$ is the beamformer to recover the data symbol from user $k$, and $\hat{n}_k\sim{\mathcal{CN}}(0,\sigma^2\lVert \hat{\mathbf{w}}_k\rVert^2)$ is the noise component. 

\begin{table*}[!t]
\centering
\caption{Comparison of the Linear Receive Beamforming Methods for CAPAs and SPDAs.}
\label{table_compare}
\resizebox{0.7\textwidth}{!}{
\begin{tabular}{|c|cc|cc|}
\hline
\multirow{2}{*}{Type} & \multicolumn{2}{c|}{CAPA}                    & \multicolumn{2}{c|}{SPDA}                    \\ \cline{2-5} 
                      & \multicolumn{1}{c|}{Beamformer} & Complexity & \multicolumn{1}{c|}{Beamformer} & Complexity \\ \hline
MRC                   & \multicolumn{1}{c|}{${\mathbf{w}}_{{\mathsf{MRC}}}(\mathbf{r})={\mathbf{h}}(\mathbf{r}){\mathbf{I}}_K$}          & ${\mathcal{O}}(1)$          & \multicolumn{1}{c|}{${\mathbf{W}}_{{\mathsf{MRC}}}=\hat{\mathbf{H}}$}          & ${\mathcal{O}}(1)$          \\ \hline
ZF                    & \multicolumn{1}{c|}{${{\mathbf{w}}}_{{\mathsf{ZF}}}(\mathbf{r})={\mathbf{h}}({\mathbf{r}}){\mathbf{R}}^{-1}$}          & ${\mathcal{O}}(K^3)$          & \multicolumn{1}{c|}{${\mathbf{W}}_{{\mathsf{ZF}}}=\hat{\mathbf{H}}\hat{\mathbf{R}}^{-1}$}          & ${\mathcal{O}}(K^3)$           \\ \hline
MMSE                  & \multicolumn{1}{c|}{${{\mathbf{w}}}_{{\mathsf{MMSE}}}(\mathbf{r})={\mathbf{h}}({\mathbf{r}})({\mathbf{I}}_K+{\mathbf{P}}{\mathbf{R}})^{-1}$}          & ${\mathcal{O}}(K^3)$           & \multicolumn{1}{c|}{${\mathbf{W}}_{{\mathsf{MMSE}}}=\hat{\mathbf{H}}({\mathbf{I}}_K+\hat{\mathbf{P}}\hat{\mathbf{R}})^{-1}$}          & ${\mathcal{O}}(K^3)$           \\ \hline
\end{tabular}}
\vspace{-10pt}
\end{table*}

Referring to \eqref{Linear_Receive_Beamforming_SPDA}, we apply MRC, ZF, and optimal/MMSE beamforming techniques to recover $s_k$ from $\hat{y}_k$. Based on the derivations of CAPA beamformers in \eqref{CAPA_MRC_Closed_Form_Matrix}, \eqref{ZFBF_General_Matrix}, and \eqref{MMSE_General_Another_Matrix}, the beamformers corresponding to SPDAs are given by
\begin{align}
{\mathbf{W}}_{{\mathsf{MRC}}}=\hat{\mathbf{H}},~
{\mathbf{W}}_{{\mathsf{ZF}}}=\hat{\mathbf{H}}\hat{\mathbf{R}}^{-1},~
{\mathbf{W}}_{{\mathsf{MMSE}}}=\hat{\mathbf{H}}({\mathbf{I}}_K+\hat{\mathbf{P}}\hat{\mathbf{R}})^{-1},\nonumber
\end{align}
where $\hat{\mathbf{H}}=[\hat{\mathbf{h}}_{1},\ldots,\hat{\mathbf{h}}_{K}]\in{\mathbbmss{C}}^{M\times K}$, $\hat{{\mathbf{P}}}={\rm{diag}}([\frac{P_1}{\sigma^2},\ldots,\frac{P_K}{\sigma^2}])\in{\mathbbmss{C}}^{K\times K}$, and $\hat{\mathbf{R}}=\hat{\mathbf{H}}^{\mathsf{H}}\hat{\mathbf{H}}\in{\mathbbmss{C}}^{K\times K}$. Note that the $k$th column of ${\mathbf{W}}_{{\mathsf{MRC}}}$, ${\mathbf{W}}_{{\mathsf{ZF}}}$, and ${\mathbf{W}}_{{\mathsf{MMSE}}}$ corresponds to the beamformer for user $k\in{\mathcal{K}}$. For a clearer comparison, Table \ref{table_compare} (on the top of next page) compares the linear receive beamforming for CAPAs and SPDAs.
\vspace{-5pt}
\begin{remark}
From Table \ref{table_compare}, we observe that the beamformers for CAPAs share a similar form with those for SPDAs\footnote{The results in Table \ref{table_compare} may prompt the question of whether they can be directly extended from conventional SPDA-based designs. As discussed in the mathematical derivations and Section \ref{Prior_Work_Section}, such an extension is far from straightforward. This is because the analysis requires tackling several challenging operator-based optimization problems and rigorously proving the optimality of the proposed solutions. In summary, while the expressions in Table \ref{table_compare} might appear intuitive at first glance, they are in fact the outcome of nontrivial and rigorous derivations developed throughout this work.}. Both can be represented as a weighted sum or linear combination of the spatial responses of all users. The weighted coefficients depend on the channel correlation matrix ${\mathbf{R}}=[\int_{{\mathcal{A}}}h_{k_1}^{*}(\mathbf{r})h_{k_2}(\mathbf{r}){\rm{d}}{\mathbf{r}}]_{k_1,k_2\in{\mathcal{K}}}$ (for CAPAs) or $\hat{\mathbf{R}}=[\hat{\mathbf{h}}_{k_1}^{\mathsf{H}}\hat{\mathbf{h}}_{k_2}]_{k_1,k_2\in{\mathcal{K}}}$ (for SPDAs). By substituting the continuous channel responses $\{h_k(\mathbf{\mathbf{r}})\}_{k=1}^{K}$ in CAPAs with their SPDA counterparts $\{\hat{\mathbf{h}}_k\}_{k=1}^{K}$ and updating the channel correlation matrix accordingly, a similar structure emerges for SPDAs.
\end{remark}
\vspace{-5pt}
\vspace{-5pt}
\begin{remark}
The above results indicate that, for both CAPAs and SPDAs, the considered linear beamformers lie in the signal subspace spanned by the spatial responses of all users.
\end{remark}
\vspace{-5pt}
Next, we compare the SINR achieved by CAPAs and SPDAs. By following our previous derivation steps for CAPAs, the SINRs achieved by SPDAs can be expressed as follows:
\begin{align}\label{SPDA_General_SINR}
{\hat{\gamma}}_{\bullet,k}=\frac{P_k}{\sigma^2}{\hat{a}}_k\left(1-\hat{\alpha}_{\bullet,k}\right),~\bullet\in\{{\mathsf{MRC}},{\mathsf{ZF}},{\mathsf{MMSE}}\},
\end{align}
where $\hat{a}_k=\lVert\hat{\mathbf{h}}_k\rVert^2$, $\hat{\alpha}_{{\mathsf{MRC}},k}=\frac{\hat{\mathbf{r}}_{k,k}^{\mathsf{H}}\hat{\mathbf{P}}_k\hat{\mathbf{r}}_{k,k}}{\hat{a}_k+\hat{\mathbf{r}}_{k,k}^{\mathsf{H}}\hat{\mathbf{P}}_k\hat{\mathbf{r}}_{k,k}}$, $\hat{\alpha}_{{\mathsf{ZF}},k}=\frac{\hat{\mathbf{r}}_{k,k}^{\mathsf{H}}\hat{\mathbf{R}}_k^{-1}\hat{\mathbf{r}}_{k,k}}{\hat{a}_k}$, and $\hat{\alpha}_{{\mathsf{MMSE}},k}=\frac{\hat{\mathbf{r}}_{k,k}^{\mathsf{H}}(\hat{\mathbf{P}}_k^{-1}+\hat{\mathbf{R}}_k)^{-1}\hat{\mathbf{r}}_{k,k}}{\hat{a}_k}$. Here, $\hat{\mathbf{r}}_{k,k}=\hat{\mathbf{H}}_k^{\mathsf{H}}\hat{\mathbf{h}}_k$ with $\hat{\mathbf{H}}_k=[\hat{\mathbf{h}}_1,\ldots,\hat{\mathbf{h}}_{k-1},\hat{\mathbf{h}}_{k+1},\ldots,\hat{\mathbf{h}}_{K}]\in{\mathbbmss{C}}^{M\times {K_1}}$, $\hat{\mathbf{P}}_k={\rm{diag}}([\frac{P_{k'}}{\sigma^2}]_{k'\in{\mathcal{K}}_k})\in{\mathbbmss{C}}^{K_1\times {K_1}}$, and $\hat{\mathbf{R}}_k=\hat{\mathbf{H}}_k^{\mathsf{H}}\hat{\mathbf{H}}_k\in{\mathbbmss{C}}^{K_1\times {K_1}}$. Notably, the SINRs achieved by SPDAs are characterized by the channel gain $\hat{a}_k=\lVert\hat{\mathbf{h}}_k\rVert^2$ and the SNR loss factor $\hat{\alpha}_{\bullet,k}$. A more detailed performance comparison between CAPAs and SPDAs will be presented in the simulation section.
\subsection{Performance Evaluation}
After the above comparison, this part evaluates the performance of the three designed beamformers by focusing on per-user SINR. The receive CAPA is modeled as a planar array in the $x$-$y$ plane centered at the origin, with
\begin{align}\label{CAPA-Aperture_Setup}
{\mathcal{A}}=\left\{[r_x,r_y,0]^{\mathsf{T}}\left|\lvert r_x\rvert\leq {L_x}/{2},\lvert r_y\rvert\leq {L_y}/{2}\right.\right\}.
\end{align}
\subsubsection{Deterministic Channels}
For deterministic spatial channel responses $\{h_k({\mathbf{r}})\}_{k\in{\mathcal{K}}}$, the per-user SINRs for MRC, ZF, and MMSE beamformers depend on the correlation matrix ${\mathbf{R}}$ in \eqref{Channel_Correlation_Matrix_No_Power}. Calculating ${\mathbf{R}}$ involves $\frac{K(K+1)}{2}$ integrals due to ${\mathbf{R}}={\mathbf{R}}^{\mathsf{H}}$. These integrals are approximated using the Gauss-Legendre quadrature rule as follows \cite{wang2024beamforming}:
\begin{align}
[{\mathbf{R}}]_{k_1,k_2}&=\int_{{\mathcal{A}}}h_{k_1}^{*}(\mathbf{r})h_{k_2}(\mathbf{r}){\rm{d}}{\mathbf{r}}\nonumber\\
&=\int_{-\frac{L_x}{2}}^{\frac{L_x}{2}}\int_{-\frac{L_y}{2}}^{\frac{L_y}{2}}h_{k_1}^{*}(r_x,r_y)
h_{k_2}(r_x,r_y){\rm{d}}{{r}}_x{\rm{d}}r_y\nonumber\\
&\approx \frac{L_xL_y}{4}\sum_{i_x=1}^{I}\sum_{i_y=1}^{I}\omega_{i_x}\omega_{i_y}h_{k_1}^{*}\left(\frac{\theta_{i_x}L_x}{2},\frac{\theta_{i_y}L_y}{2}\right)\nonumber\\
&\times h_{k_2}\left(\frac{\theta_{i_x}L_x}{2},\frac{\theta_{i_y}L_y}{2}\right),~\forall k_1,k_2\in{\mathcal{K}},\label{Gauss_Quadrature_Rule}
\end{align}
where $\{\omega_i\}$ and $\{\theta_i\}$ are Gauss-Legendre weights and abscissas. This approximation becomes exact as $I\rightarrow\infty$. Our prior work \cite{wang2024beamforming} demonstrates that $I=10$ suffices for accurate channel correlation computations. 
\subsubsection{Statistical Channels}
For statistical channels, we model $\{h_k({\mathbf{r}})\}_{k\in{\mathcal{K}}}$ as random processes over ${\mathbf{r}}\in{\mathcal{A}}$ \cite{pizzo2020spatially}. For brevity, we focus on the statistics of the per-user SINR under \emph{isotropic Rayleigh fading} by assuming independent and identically distributed (i.i.d.) small-scale fading across users. As shown in \cite{ouyang2024diversity}, the channel response $h_k({\mathbf{r}})$ satisfies:
\begin{align}\label{Small_Scale_Fading_Basic}
h_k({\mathbf{r}})\overset{\rm{d}}{=}\eta_k^{\frac{1}{2}}\sum\nolimits_{\ell=1}^{\infty}\sigma_{\ell}^{1/2}\Psi_{k,\ell}\varphi_{\ell}({\mathbf{r}}),
\end{align} 
where $\eta_k$ models the geometric attenuation and shadow fading, $\{\Psi_{k,\ell}\sim{\mathcal{CN}}(0,1)\}$ are i.i.d. complex Gaussian variables, $\{\varphi_{\ell}({\mathbf{r}})\}$ form an orthonormal basis on ${\mathbf{r}}\in{\mathcal{A}}$, and $\{\sigma_{1}\geq\cdots\geq\sigma_{\infty}\geq0\}$ are eigenvalues of the normalized autocorrelation function $\frac{1}{\eta_k}{\mathbbmss{E}}\{h_k({\mathbf{r}})h_k^{*}({\mathbf{r}}')\}$. 

For electrically large arrays ($\min\{L_x,L_y\}\gg \lambda$), the eigenvalues exhibit \cite{ouyang2024diversity}
\begin{align}\nonumber
\sigma_{1}\approx\sigma_{2}\approx\cdots\approx\sigma_{{\mathsf{DoF}}}\approx1\gg\sigma_{{\mathsf{DoF}}+1}\geq\cdots\sigma_{\infty}\geq0,
\end{align}
where ${\mathsf{DoF}}$ (referred to as the effective DoF) scales proportionally with $L_xL_y$ and inversely with $\lambda$. This justifies approximating \eqref{Small_Scale_Fading_Basic} as follows \cite{ouyang2024diversity}:
\begin{align}\nonumber
h_k({\mathbf{r}})\overset{\rm{d}}{=}\eta_k^{\frac{1}{2}}\sum\nolimits_{\ell=1}^{\infty}\sigma_{\ell}^{1/2}\Psi_{k,\ell}\varphi_{\ell}({\mathbf{r}})
\approx \eta_k^{\frac{1}{2}}{\bm\varphi}({\mathbf{r}}){\mathbf{g}}_k,
\end{align}
where ${\mathbf{g}}_k=[\Psi_{k,\ell},\ldots,\Psi_{k,{\mathsf{DoF}}}]^{\mathsf{T}}\sim{\mathcal{CN}}({\mathbf{0}},{\mathbf{I}}_{{\mathsf{DoF}}})$ and ${\bm\varphi}({\mathbf{r}})=[\varphi_{1}({\mathbf{r}}),\ldots,\varphi_{{\mathsf{DoF}}}({\mathbf{r}})]\in{\mathbbmss{C}}^{1\times {\mathsf{DoF}}}$ with $\int_{{\mathcal{A}}}{\bm\varphi}^{\mathsf{H}}({\mathbf{r}}){\bm\varphi}({\mathbf{r}}){\rm{d}}{\mathbf{r}}={\mathbf{I}}_{{\mathsf{DoF}}}$. Accordingly, the channel correlation matrix becomes
\begin{align}\nonumber
{\mathbf{R}}=\int_{\mathcal{A}}{\mathbf{h}}^{\mathsf{H}}({\mathbf{r}}){\mathbf{h}}({\mathbf{r}}){\rm{d}}{\mathbf{r}}\overset{\rm{d}}{\approx}{{\bm{\Lambda}}}^{\frac{1}{2}}
{\mathbf{G}}^{\mathsf{H}}{\mathbf{G}}
{{\bm{\Lambda}}}^{\frac{1}{2}},
\end{align}
where ${\bm{\Lambda}}={\rm{diag}}([\eta_1,\ldots,\eta_K])\in{\mathbbmss{C}}^{K\times K}$ and ${\mathbf{G}}=[{\mathbf{g}}_1,\ldots,{\mathbf{g}}_K]\in{\mathbbmss{C}}^{{\mathsf{DoF}}\times K}$ contains $K{\mathsf{DoF}}$ i.i.d. ${\mathcal{CN}}(0,1)$ entries. Under this model, the per-user SINR achieved by the CAPA under MRC, ZF, and MMSE beamforming satisfies:
\begin{align}\label{Small_Scale_Fading_Basic4}
{{\gamma}}_{\bullet,k}\overset{\rm{d}}{\approx}\frac{P_k}{\sigma^2}{\check{a}}_k\left(1-\check{\alpha}_{\bullet,k}\right),~\bullet\in\{{\mathsf{MRC}},{\mathsf{ZF}},{\mathsf{MMSE}}\},
\end{align}
where $\check{a}_k=\eta_k\lVert{\mathbf{g}}_k\rVert^2$, $\check{\alpha}_{{\mathsf{MRC}},k}=\frac{\check{\mathbf{r}}_{k,k}^{\mathsf{H}}{\mathbf{P}}_k\check{\mathbf{r}}_{k,k}}{\check{a}_k+\check{\mathbf{r}}_{k,k}^{\mathsf{H}}{\mathbf{P}}_k\check{\mathbf{r}}_{k,k}}$, $\check{\alpha}_{{\mathsf{ZF}},k}=\frac{\check{\mathbf{r}}_{k,k}^{\mathsf{H}}\check{\mathbf{R}}_k^{-1}\check{\mathbf{r}}_{k,k}}{\check{a}_k}$, and $\check{\alpha}_{{\mathsf{MMSE}},k}=\frac{\check{\mathbf{r}}_{k,k}^{\mathsf{H}}({\mathbf{P}}_k^{-1}+\check{\mathbf{R}}_k)^{-1}\check{\mathbf{r}}_{k,k}}{\check{a}_k}$. Here, $\check{\mathbf{r}}_{k,k}=\eta_k^{{1}/{2}}\check{\mathbf{H}}_k^{\mathsf{H}}{\mathbf{g}}_k\in{\mathbbmss{C}}^{K_1\times 1}$ with $\check{\mathbf{H}}_k=[\eta_1^{{1}/{2}}{\mathbf{g}}_1,\ldots,\eta_{k-1}^{{1}/{2}}{\mathbf{g}}_{k-1},\eta_{k+1}^{{1}/{2}}{\mathbf{g}}_{k+1},\ldots,\eta_K^{{1}/{2}}{\mathbf{g}}_K]\in{\mathbbmss{C}}^{{\mathsf{DoF}}\times K_1}$, and $\check{\mathbf{R}}_k=\check{\mathbf{H}}_k^{\mathsf{H}}\check{\mathbf{H}}_k\in{\mathbbmss{C}}^{{K_1}\times {K_1}}$. 

By comparing \eqref{Small_Scale_Fading_Basic4} with \eqref{SPDA_General_SINR}, we conclude that for electrically large arrays (noting that CAPAs are typically electrically large \cite{liu2024capa}), the per-user SINR distribution for CAPAs closely approximates that of a $K$-user SPDA-based uplink system with \emph{uncorrelated Rayleigh fading} and ${\mathsf{DoF}}$ receiver antennas. Existing results for SPDAs \cite{heath2018foundations,ngo2013energy} thus apply to characterize these statistics. Due to the page limitations, further discussions and extensions to other fading types are skipped here and left as a potential direction for future work.
\section{Numerical Results}\label{Section: Numerical Results}
Numerical results are presented to validate the derived analytical findings and demonstrate the advantages of CAPAs over traditional SPDAs in terms of both \emph{sum-rate} and \emph{sum-MSE}. The simulations adopt the following parameter setup unless otherwise specified \cite{wang2024beamforming}. The receive CAPA is configured as a planar array within the $x$-$y$ plane centered at the coordinate origin. The aperture ${\mathcal{A}}$ follows \eqref{CAPA-Aperture_Setup} with $L_x=L_y=\sqrt{\lvert{\mathcal{A}}\rvert}$ and $\lvert{\mathcal{A}}\rvert=0.25~{{\text{m}}^{\text{2}}}$. A total of $K=8$ users are randomly distributed within the following region:
\begin{align}\nonumber
{\mathcal{U}}=\left\{ [r_x,r_y,0]^{\mathsf{T}}\left|\begin{array}{c}
	\lvert r_x\rvert\leq U_x,\lvert r_y\rvert\leq U_y,\\
	U_y,U_{z,\min}\leq r_z\leq U_{z,\max}\\
\end{array}\right. \right\},
\end{align}
where $U_x=U_y=5$ m, $U_{z,\min}=15$ m, and $U_{z,\max}=30$ m. The aperture size for each user $k\in{\mathcal{K}}$ is set as $\lvert{\mathcal{A}}_k\rvert=\frac{\lambda^2}{4\pi}$ with $\lambda$ denoting the wavelength, which equals that of an isotropic antenna. The transmit power is set to $P_k=40~{\text{mA}}^{\text{2}}$ for $k\in{\mathcal{K}}$, and the noise power is $\sigma^2=5.6\times {10^{-3}}~{\text{V}}^{\text{2}}/{\text{m}}^{\text{2}}$. 

A free-space line-of-sight model is used to characterize the spatial response for each user $k\in{\mathcal{K}}$, which yields \cite{ouyang2024primer}
\begin{align}\nonumber
g_{k}({\mathbf{r}},{\mathbf{s}}_k)=\frac{-{\rm{j}} k_0 \eta}{4\pi\lVert {\mathbf{r}}-{\mathbf{s}}_k\rVert}{\rm{e}}^{-{\rm{j}}k_0\lVert {\mathbf{r}}-{\mathbf{s}}_k\rVert},
\end{align}
where ${\mathbf{s}}_k\in{\mathcal{U}}$ represents the central point of user $k$, $k_0=\frac{2\pi}{\lambda}$ is the wavenumber, and $\eta=120\pi~{\Omega}$ is the free-space impedance. The signal frequency is set at $f=2.4$ GHz. While the line-of-sight channel serves as an example, the proposed beamforming methods in Table \ref{table_compare} generalize to other channel models.

For comparison, the performance of CAPA is evaluated alongside a conventional uplink multiuser SPDA system. In this setup, the continuous surface $\mathcal{A}$ is uniformly discretized into $M=\lceil\frac{L_x}{d}\rceil\times\lceil\frac{L_y}{d}\rceil$ antenna elements with effective aperture area $\lvert{\mathcal{S}}\rvert=\frac{\lambda^2}{4\pi}$ and spacing $d=\frac{\lambda}{2}$. All results are averaged over $500$ random channel realizations. The channel correlation matrix $\mathbf{R}$ in \eqref{Channel_Correlation_Matrix_No_Power} is computed using the Gauss-Legendre quadrature rule \eqref{Gauss_Quadrature_Rule} with $I=20$.

\begin{figure}[!t]
    \centering
    \subfigure[Sum-rate.]
    {
        \includegraphics[width=0.4\textwidth]{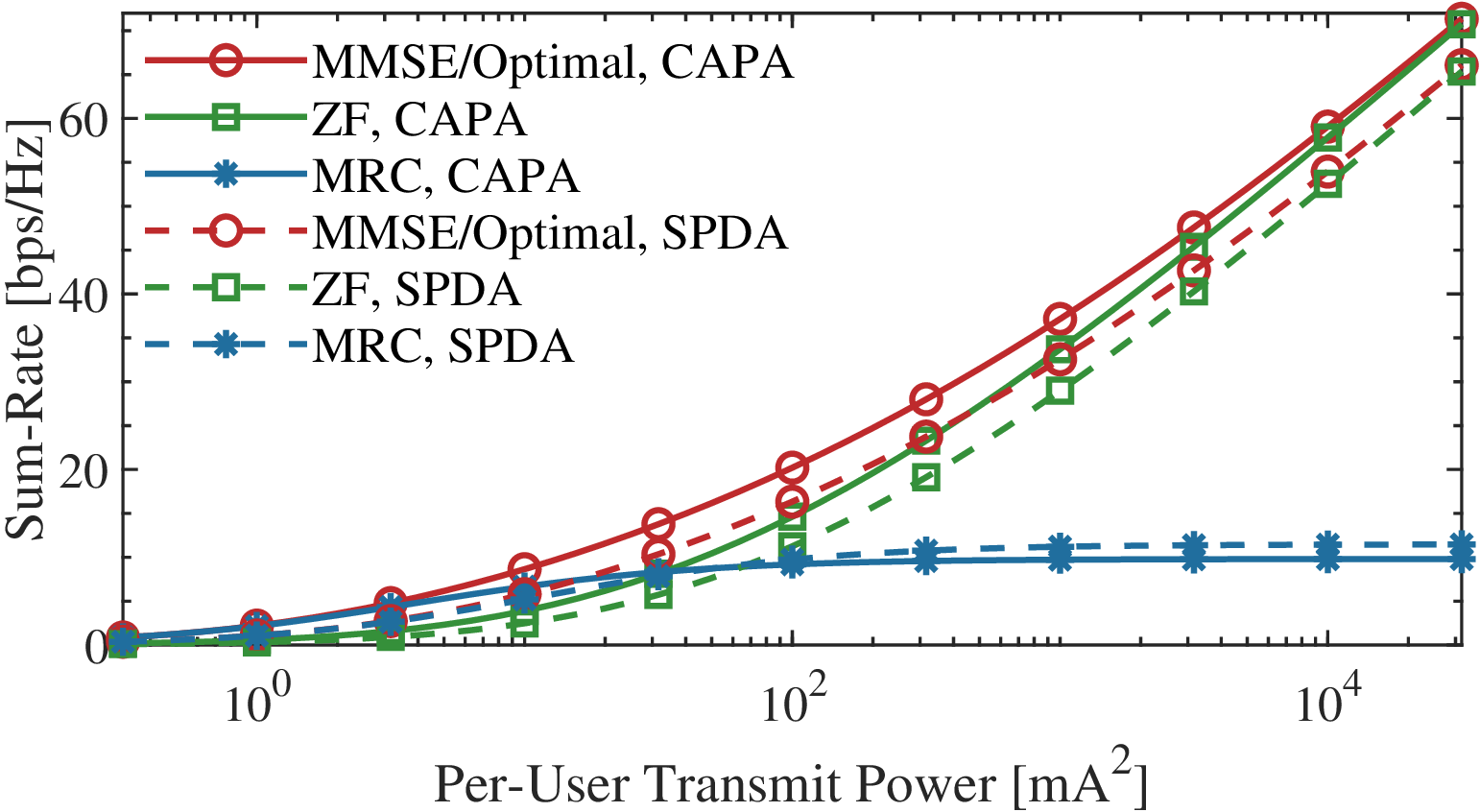}
	   \label{fig1a}	
    }\\
    \subfigure[Effective channel gain.]
    {
        \includegraphics[width=0.4\textwidth]{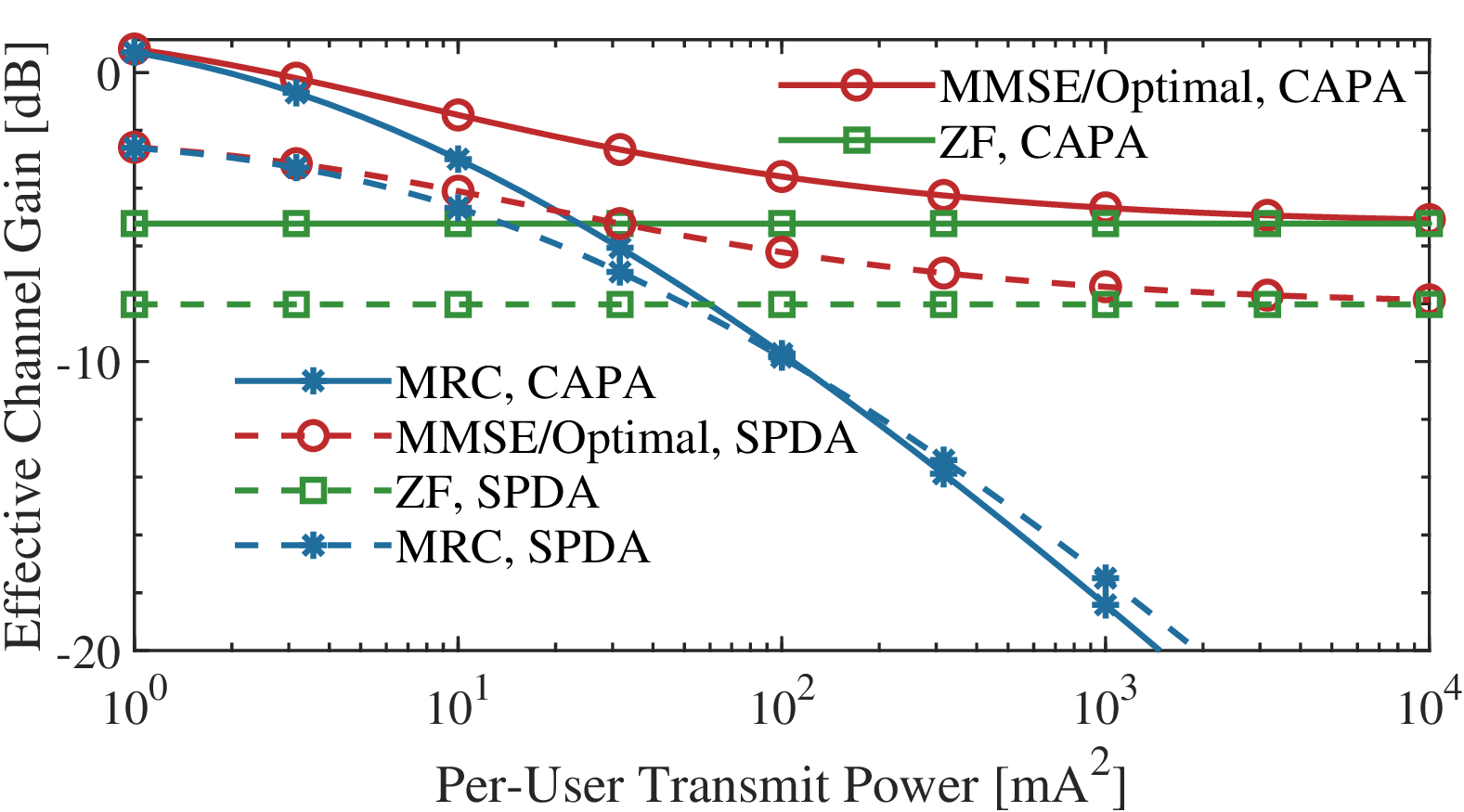}
	   \label{fig2a}	
    }\\
   \subfigure[Sum-MSE.]
    {
        \includegraphics[width=0.4\textwidth]{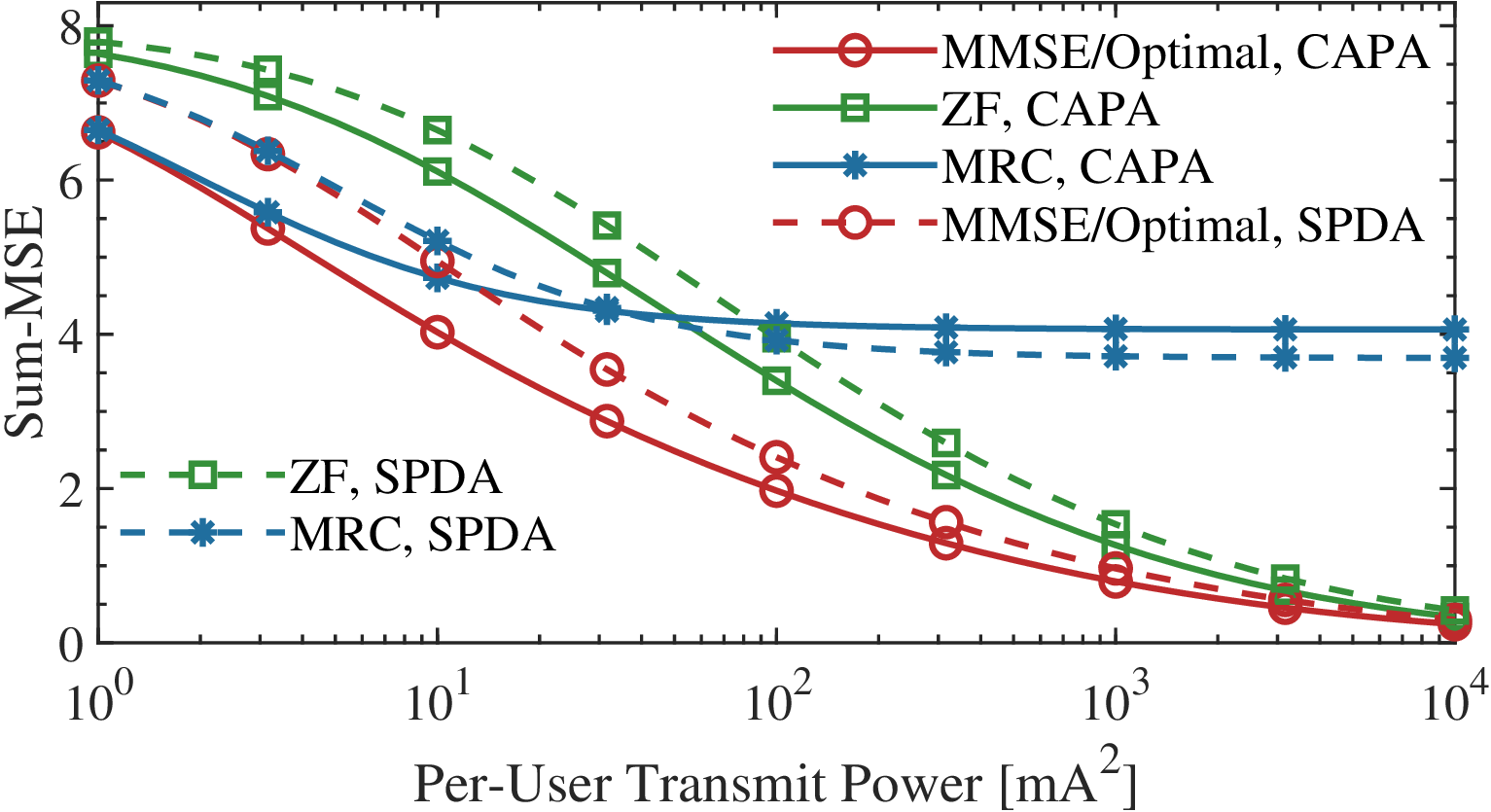}
	   \label{fig2b}	
    }
\caption{Sum-rate, average effective channel gain, and sum-MSE versus per-user transmit power $P_1=\ldots=P_K=P$.}
    \label{Figure2}
\end{figure}

\subsection{Uplink Performance Versus Transmit Power}
In {\figurename} {\ref{fig1a}}, the sum-rates achieved by different linear receive beamforming methods are shown as a function of per-user transmit power. For comparison, results are provided for both CAPAs and traditional SPDAs by assuming all users have the same transmit power $P_k=P$ for $k\in{\mathcal{K}}$. As illustrated, the sum-rates achieved by all linear beamformers increase with transmit power. In the low-SNR regime, the sum-rate achieved by MRC beamforming is nearly the same as that achieved by \emph{MMSE or optimal beamforming}, both of which outperform ZF beamforming. This is because Gaussian noise dominates IUI at low SNRs, making MRC beamforming---which amplifies each user's signal power---more effective for improving SINR than ZF beamforming, which focuses on nullifying IUI. However, the high-SNR sum-rate achieved by ZF beamforming closely approaches that of MMSE beamforming. This is explained by the fact that at high SNRs, IUI surpasses Gaussian noise as the dominant factor, making IUI cancellation crucial for improving sum-rate. These observations are consistent with the discussions in Section \ref{Subsubsection: Asymptotic Discussion}.

The sum-rates achieved by CAPAs and SPDAs are next compared. As observed in {\figurename} {\ref{fig1a}}, CAPAs outperform SPDAs in terms of the sum-rates achieved by both ZF and MMSE beamforming. However, for MRC beamforming, CAPA outperforms SPDA in the low-SNR regime, while SPDA outperforms CAPA in the high-SNR regime. This can be explained as follows: with a larger \emph{effective aperture area}, CAPAs fully utilize available spatial resources, enhancing the strength of both useful and interference signals. As MRC beamforming maximizes the useful signal's power without mitigating interference, it performs worse in interference-dominated high-SNR scenarios when using CAPAs, where interference is also amplified by the larger aperture area. In contrast, ZF and MMSE beamforming effectively manage IUI, enabling CAPAs to outperform SPDAs in these methods, as observed in {\figurename} {\ref{fig1a}}. 

To further examine this, {\figurename} {\ref{fig2a}} plots the average effective channel gain $\frac{1}{K}\sum_{k=1}^{K}a_k(1-\alpha_{{\mathsf{loss}},k})$ versus per-user power, where $\alpha_k$ and $\alpha_{{\mathsf{loss}},k}$ represent the channel gain and SNR loss factor, respectively. Since ZF beamforming inherently nullifies IUI, the transmit power has no impact on its effective channel gain. Conversely, the effective channel gain for MRC and MMSE beamforming decreases with transmit power due to rising interference power. As shown in {\figurename} {\ref{fig2a}}, the high-SNR effective channel gain achieved by MRC in SPDAs is higher than in CAPAs due to the reduced interference power resulting from SPDAs' smaller aperture area, which leads to reduced IUI. This observation explains why MRC beamforming in SPDAs outperforms that in CAPAs in the high-SNR regime.

{\figurename} {\ref{fig2b}} illustrates the sum-MSE as a function of transmit power. As expected, the MSE decreases for all considered schemes as per-user power increases. Additionally, in the low-SNR regime, CAPAs achieve a lower sum-MSE than SPDAs across all three linear beamforming methods. In the high-SNR regime, however, CAPAs outperform SPDAs only with ZF and MMSE beamforming. This outcome can be explained similarly to the results shown in {\figurename} {\ref{fig1a}}.

\begin{figure}[!t]
    \centering
    \subfigure[Sum-rate.]
    {
        \includegraphics[width=0.4\textwidth]{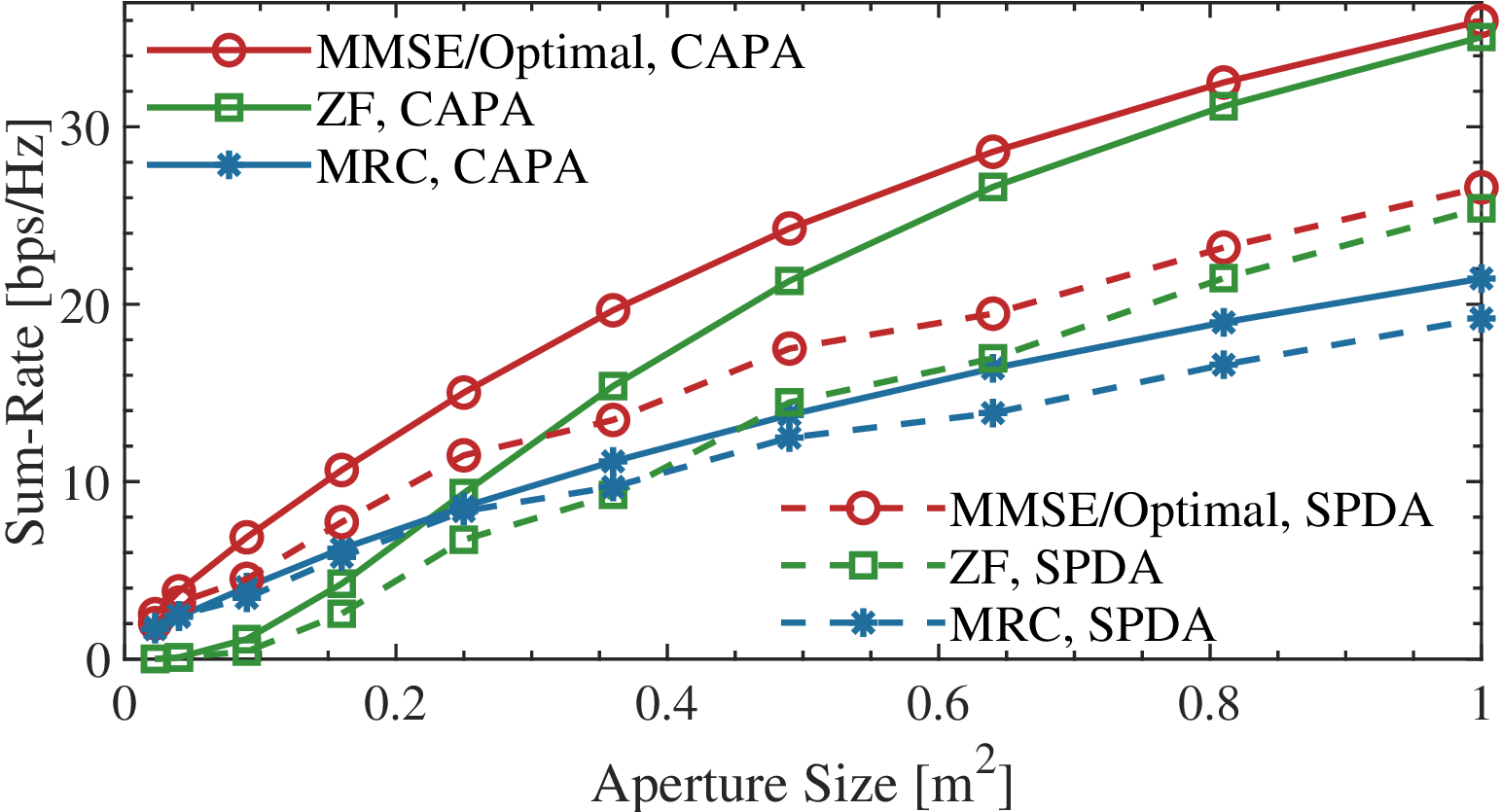}
	   \label{fig5a}	
    }\\
   \subfigure[Sum-MSE.]
    {
        \includegraphics[width=0.4\textwidth]{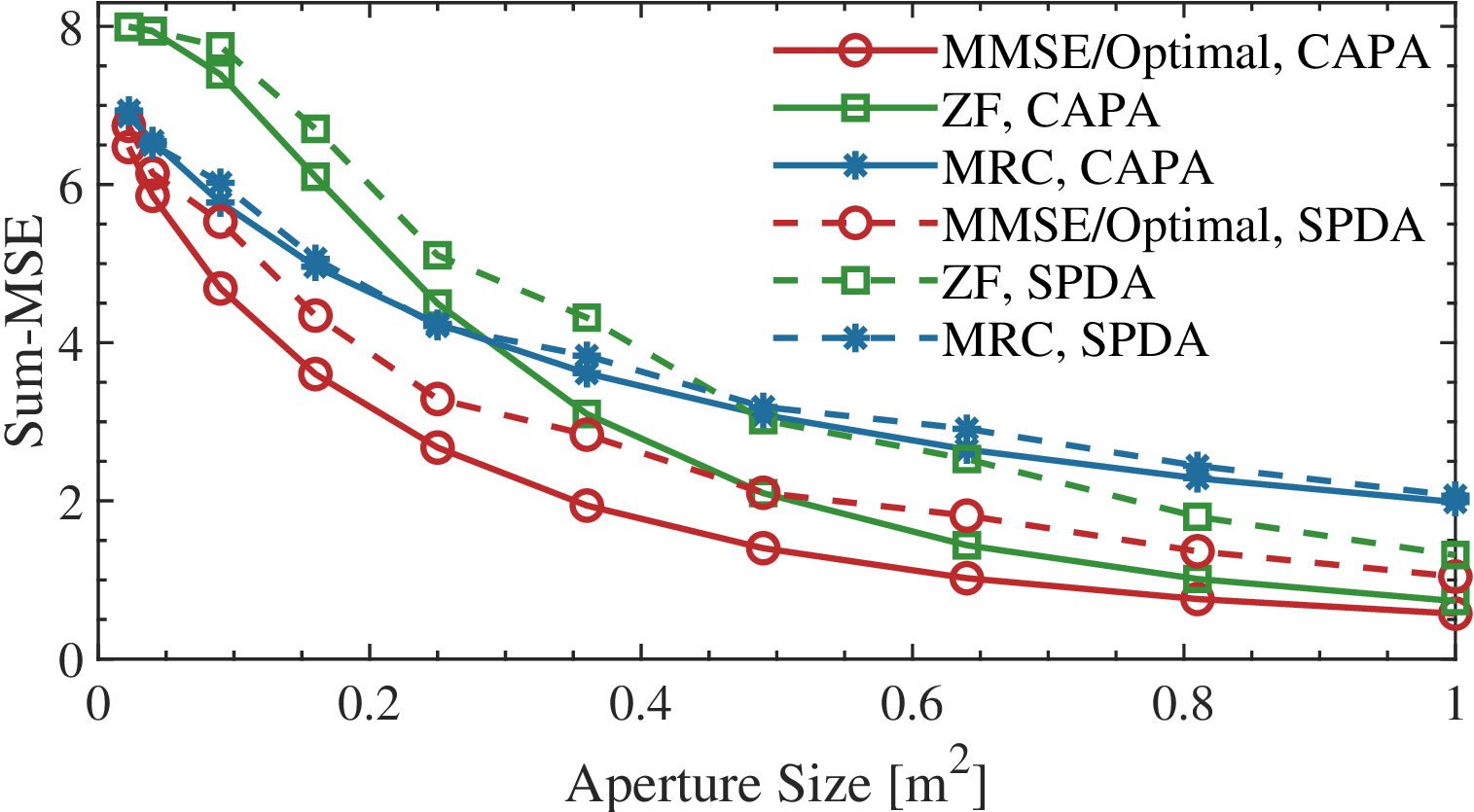}
	   \label{fig5b}	
    }
\caption{Sum-rate and sum-MSE versus aperture size $\lvert{\mathcal{A}}\rvert$.}
    \label{Figure5}
\end{figure}

\subsection{Uplink Performance Versus Aperture Size}
In {\figurename} {\ref{Figure5}}, we compare the sum-rate and sum-MSE performances of both CAPA and SPDA based on the aperture size $\lvert{\mathcal{A}}\rvert$. As shown in {\figurename} {\ref{fig5a}}, increasing the aperture size leads to higher sum-rates for both CAPA and SPDA due to the associated increase in channel gain. Additionally, {\figurename} {\ref{fig5a}} demonstrates that for all considered aperture sizes and beamforming methods, CAPA achieves a higher sum-rate than SPDA, which can be attributed to CAPA’s more effective utilization of spatial resources. Interestingly, when the aperture size is small, ZF beamforming performs worse than MRC beamforming; however, this trend reverses as the aperture size grows. This is because under our setup, system performance with a smaller aperture is more influenced by Gaussian noise, making MRC beamforming preferable, as it maximizes signal power. Conversely, with a larger aperture size, IUI becomes the dominant performance constraint, making ZF beamforming, with its superior IUI cancellation, more effective than MRC beamforming. Similar trends are observed in the sum-MSE performance shown in {\figurename} {\ref{fig5b}}.

\subsection{Uplink Performance Versus Number of Users}
\begin{figure}[!t]
    \centering
    \subfigure[Sum-rate.]
    {
        \includegraphics[width=0.4\textwidth]{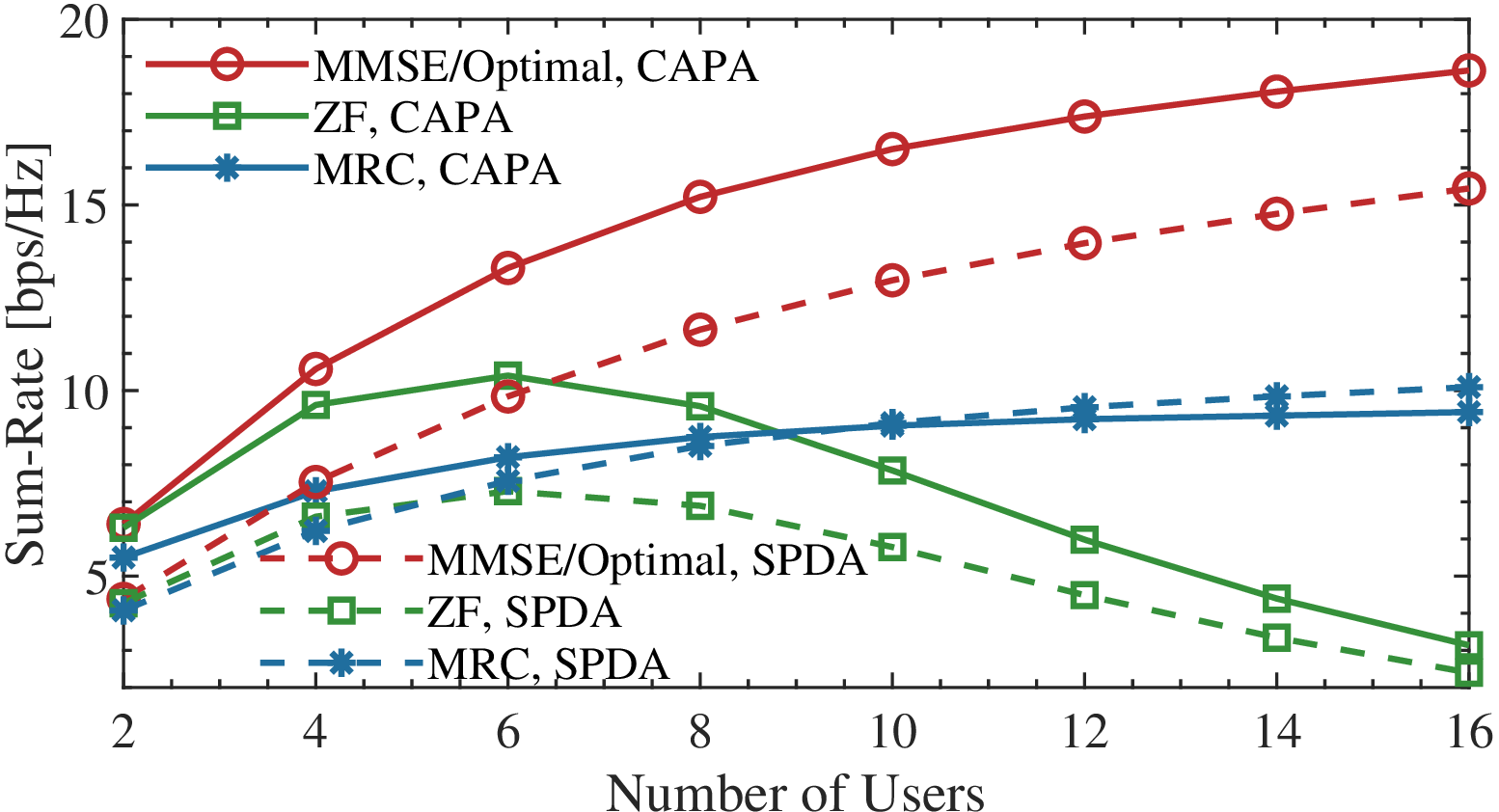}
	   \label{fig3a}	
    }\\
   \subfigure[Average rate.]
    {
        \includegraphics[width=0.4\textwidth]{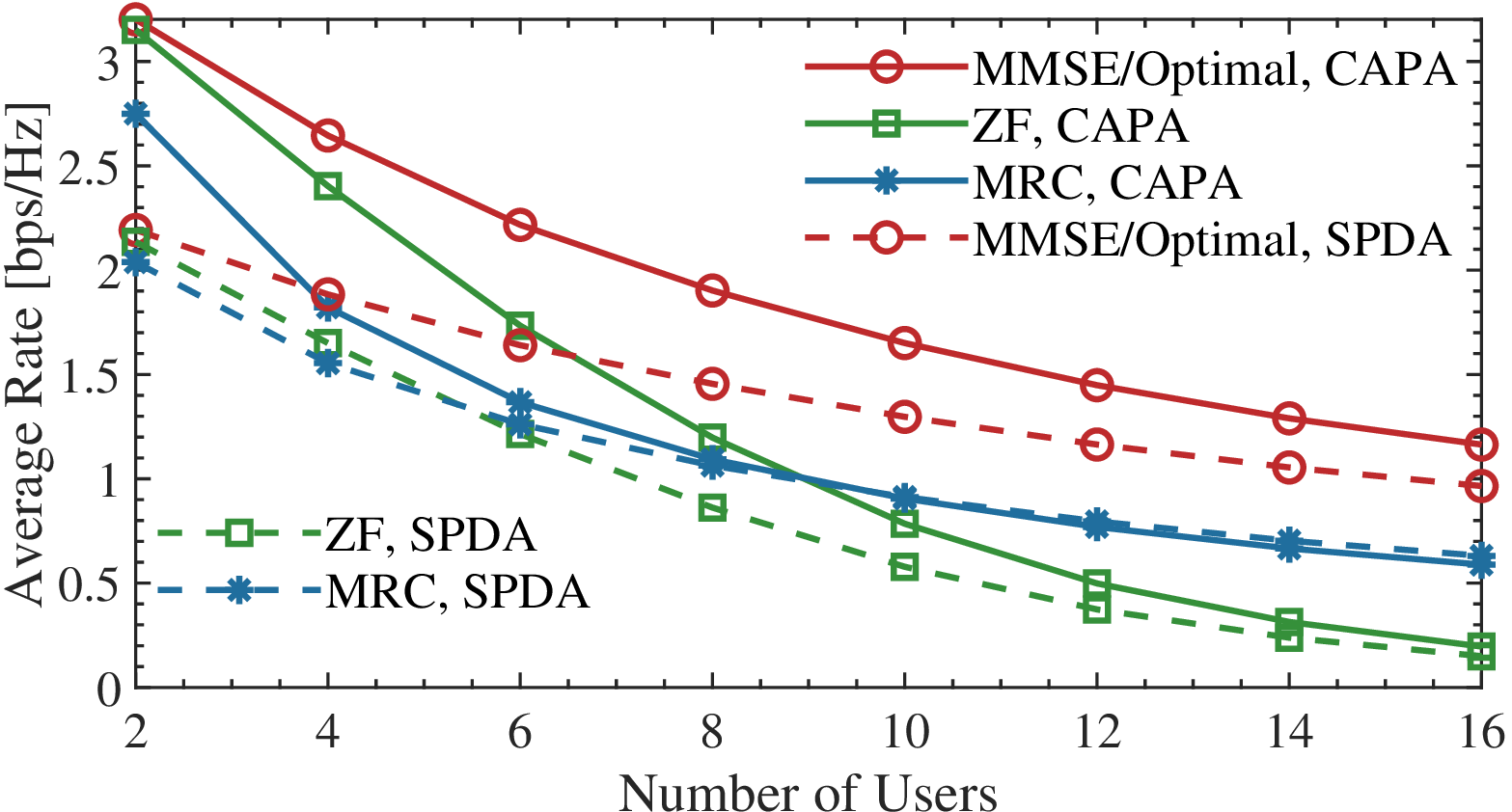}
	   \label{fig3b}	
    }
\caption{Sum-rate versus number of users $K$.}
    \label{Figure3}
\end{figure}

{\figurename} {\ref{fig3a}} illustrates the sum-rates achieved by CAPAs and SPDAs as a function of the number of users. As shown, the sum-rates for MRC and MMSE beamforming increase with the number of users. In contrast, the sum-rate achieved by ZF beamforming initially increases but then decreases as the user count rises. This pattern arises because ZF beamforming aims to set the beamformer orthogonal to the interference subspace while aligning it as closely as possible with the user's channel spatial response, as depicted in {\figurename} {\ref{Relationship_Beamforming}}. With fewer users, it is feasible to align the ZF beamformer more closely with the user's spatial response, which results in a higher sum-rate. However, as the number of users grows, achieving this alignment becomes more challenging due to the need to cancel all IUI, leading to a decline in sum-rate performance. Although the sum-rate for MRC and MMSE beamforming increases with user count, the average rate per user decreases monotonically due to increased IUI, as observed in {\figurename} {\ref{fig3b}}. 

Next, we compare the sum-rate performance of CAPAs and SPDAs. For MMSE and ZF beamforming, the CAPA achieves both higher total and average rates than the SPDA, as seen in {\figurename} {\ref{fig3a}} and {\figurename} {\ref{fig3b}}. However, with MRC beamforming, CAPA only shows a rate advantage when the user count is low. As the number of users increases, the CAPA performs worse than the SPDA. This outcome aligns with the results shown in {\figurename} {\ref{fig1a}}. Specifically, when many users are present, system performance suffers due to IUI. Since MRC beamforming does not effectively mitigate IUI and CAPA receives stronger interference signals than SPDA, CAPA achieves a lower sum-rate in this scenario. Regarding the MSE performance, since it exhibits a trend opposite to that of the sum-rate, we omit the corresponding numerical results here for brevity.

\section{Conclusion}\label{Section: Conclusion}
This paper has designed linear receive beamforming for CAPAs. It was demonstrated that CAPAs, under ZF and MMSE beamforming, outperform SPDAs by achieving higher sum-rates and lower sum-MSEs. However, in some \emph{interference-dominated} scenarios---such as high-SNR regimes---SPDAs may surpass CAPAs in MRC beamforming performance. These findings underscore the importance of effective interference management in uplink CAPA communications. 

The methodologies and mathematical transformations introduced in this work have enabled the derivation of closed-form continuous beamformers and the proof of their optimality under specific constraints, using the framework of integral operators. More importantly, they offer valuable insights into handling complex operator-based operations, such as inversion, which are significantly more challenging than their matrix-based counterparts. These findings are expected to contribute to the broader development of EM signal and information theory \cite{di2024electromagnetic}. Indeed, some recent advancements have already built upon the results presented in this paper. For instance, the established equivalence between MSE-optimal and rate-optimal beamformers, along with the derived estimators, has facilitated the development of a weighted MMSE framework for low-complexity beamforming \cite{wang2025beamforming}.

\begin{appendices}
\section{Proof of \textbf{Lemma \ref{Lemma_MSE_Fundamental}}}\label{Proof_Lemma_MSE_Fundamental}
Given that $s_k\sim{\mathcal{CN}}(0,1)$ ($\forall k\in{\mathcal{K}}$) and ${\mathbbmss{E}}\{s_k^{*}s_{k'}\}=0$ ($\forall k\ne k'$), we use \eqref{Linear_Receive_Beamforming} to rewrite ${\mathbbmss{E}}\{\lvert \beta_k \hat{y}_k - s_k\rvert^2\}$ as follows:
\begin{equation}\label{Proof_Lemma_MSE_Fundamental_Equ1}
\begin{split}
{\mathbbmss{E}}\{\lvert \beta_k \hat{y}_k - s_k\rvert^2\}&=1+\left\lvert\beta_{k}\right\rvert^2G_k\\
&-2\Re\left\{\beta_k\sqrt{P_k}\int_{{\mathcal{A}}}{w}_k^{*}(\mathbf{r})h_k(\mathbf{r}){\rm{d}}{\mathbf{r}}\right\},
\end{split}
\end{equation}
where $G_k\triangleq\sum_{k'=1}^{K}{P_{k'}}\left\lvert\int_{{\mathcal{A}}}{w}_k^{*}
(\mathbf{r})h_{k'}(\mathbf{r}){\rm{d}}{\mathbf{r}}\right\rvert^2+\sigma^2\int_{{\mathcal{A}}}\lvert w_k(\mathbf{r})\rvert^2{\rm{d}}{\mathbf{r}}>0$. Equation \eqref{Proof_Lemma_MSE_Fundamental_Equ1} is a convex function with respect to $\beta_k$. The optimal $\beta_k$ to minimize ${\mathbbmss{E}}\{\lvert \beta_k \hat{y}_k - s_k\rvert^2\}$ is obtained from the first-order optimality condition $\frac{{\rm{d}}}{{\rm{d}}\beta_k}{\mathbbmss{E}}\{\lvert \beta_k \hat{y}_k - s_k\rvert^2\}=0$, which yields
\begin{equation}\nonumber
\begin{split}
\frac{{\rm{d}}}{{\rm{d}}\beta_k}{\mathbbmss{E}}\{\lvert \beta_k \hat{y}_k - s_k\rvert^2\}=\beta_kG_k-\sqrt{P_k}\int_{{\mathcal{A}}}{w}_k(\mathbf{r})h_k^{*}(\mathbf{r}){\rm{d}}{\mathbf{r}}=0.
\end{split}
\end{equation}
Thus, the optimal solution for $\beta_k$ is given by
\begin{align}\label{Optimal_Beta_k_Filtering}
\beta_k^{\star}=\frac{\sqrt{P_k}}
{G_k}\int_{{\mathcal{A}}}{w}_k(\mathbf{r})h_k^{*}(\mathbf{r}){\rm{d}}{\mathbf{r}}.
\end{align}
Substituting $\beta_k=\beta_k^{\star}$ and \eqref{Optimal_Beta_k_Filtering} into \eqref{Proof_Lemma_MSE_Fundamental_Equ1} gives
\begin{subequations}\label{Proof_Lemma_MSE_Fundamental_Equ2}
\begin{align}
&{\mathbbmss{E}}\{\lvert \beta_k^{\star} \hat{y}_k - s_k\rvert^2\}=\frac{-{P_k}\left\lvert\int_{{\mathcal{A}}}{w}_k(\mathbf{r})h_k^{*}(\mathbf{r}){\rm{d}}{\mathbf{r}}\right\rvert^2}
{G_k}+1\\
&=\frac{1}{1+\frac{P_k\lvert\int_{{\mathcal{A}}}{w}_k^{*}(\mathbf{r})h_k(\mathbf{r}){\rm{d}}{\mathbf{r}}\rvert^2}{G_k-{P_k}\left\lvert\int_{{\mathcal{A}}}{w}_k(\mathbf{r})h_k^{*}(\mathbf{r}){\rm{d}}{\mathbf{r}}\right\rvert^2}}
=\frac{1}{1+\gamma_k}.
\end{align}
\end{subequations}
Combining \eqref{Optimal_Beta_k_Filtering} and \eqref{Proof_Lemma_MSE_Fundamental_Equ2} completes the proof.
\section{A Full Derivation of Equation \eqref{Matrix_Inversion_Result_2}}\label{Proof_Lemma_Matrix_Blockwise_Inversion}
Consider the block matrix $\left[\begin{smallmatrix}{\mathbf{A}}&{\mathbf{B}}\\
{\mathbf{C}}&{\mathbf{D}}\end{smallmatrix}\right]$, where ${\mathbf{A}}$ and ${\mathbf{D}}$ are square blocks of arbitrary size, and ${\mathbf{B}}$ and ${\mathbf{C}}$ are conformable for partitioning. The inverse of this matrix is given by \cite{bernstein2009matrix}
\begin{align}\label{Matrix_Blockwise_Inversion_Formula}
\left[\begin{smallmatrix}
{\mathbf{A}}&{\mathbf{B}}\\
{\mathbf{C}}&{\mathbf{D}}
\end{smallmatrix}
\right]^{-1}=\left[\begin{smallmatrix}
{\mathbf{E}}^{-1}&-{\mathbf{E}}^{-1}{\mathbf{B}}{\mathbf{D}}^{-1}\\
-{\mathbf{D}}^{-1}{\mathbf{C}}{\mathbf{E}}^{-1}&{\mathbf{D}}^{-1}+{\mathbf{D}}^{-1}{\mathbf{C}}{\mathbf{E}}^{-1}{\mathbf{B}}{\mathbf{D}}^{-1}
\end{smallmatrix}\right],
\end{align}
where ${\mathbf{E}}={\mathbf{A}}-{\mathbf{B}}{\mathbf{D}}^{-1}{\mathbf{C}}$, and both ${\mathbf{D}}$ and ${\mathbf{E}}$ must be invertible. To compute the inverse of ${\overline{\mathbf{R}}}_k=\left[\begin{smallmatrix}
r_{k,k}&{{\mathbf{r}}}_{k,k}^{\mathsf{H}}\\
{{\mathbf{r}}}_{k,k}&{\mathbf{R}}_k
\end{smallmatrix}
\right]$, we set ${\mathbf{A}}=r_{k,k}$, ${\mathbf{B}}={\mathbf{r}}_{k,k}^{\mathsf{H}}$, ${\mathbf{C}}={\mathbf{r}}_{k,k}$, and ${\mathbf{D}}={\mathbf{R}}_k$. By \textbf{\emph{Assumption-1}}, ${\mathbf{D}}={\mathbf{R}}_k\succ{\mathbf{0}}$. We then introduce \textbf{Lemma \ref{Lemma_ZF_SNR_Gain_Loss}} to facilitate the subsequent derivations.
\vspace{-5pt}
\begin{lemma}\label{Lemma_ZF_SNR_Gain_Loss}
Given the channel responses $\{h_k({\mathbf{r}})\}_{k=1}^{K}$, it holds that $r_{k,k}>{{{\mathbf{r}}}_{k,k}^{\mathsf{H}}{\mathbf{R}}_k^{-1}{{\mathbf{r}}}_{k,k}}\geq0 $. 
\end{lemma}
\vspace{-5pt}
\begin{IEEEproof}
Please refer to Appendix \ref{Proof_Lemma_ZF_SNR_Gain_Loss} for more details.
\end{IEEEproof}
\textbf{Lemma \ref{Lemma_ZF_SNR_Gain_Loss}} implies ${\mathbf{A}}-{\mathbf{B}}{\mathbf{D}}^{-1}{\mathbf{C}}=r_{k,k}-{\mathbf{r}}_{k,k}^{\mathsf{H}}{\mathbf{R}}_k^{-1}{\mathbf{r}}_{k,k}>0$. The above arguments confirm that both ${\mathbf{D}}={\mathbf{R}}_k$ and ${\mathbf{A}}-{\mathbf{B}}{\mathbf{D}}^{-1}{\mathbf{C}}=r_{k,k}-{\mathbf{r}}_{k,k}^{\mathsf{H}}{\mathbf{R}}_k^{-1}{\mathbf{r}}_{k,k}$ are invertible. With these conditions satisfied, the inverse ${\overline{\mathbf{R}}}_k^{-1}$ can be computed directly using \eqref{Matrix_Blockwise_Inversion_Formula}, from which the final results follow directly.
\section{The Gram-Schmidt Process}\label{Proof_Gram-Schmidt}
Given $K$ spatial channel response functions $\{h_k({\mathbf{r}})\}_{k=1}^{K}$ defined over ${\mathbf{r}}\in{\mathcal{A}}$. Let ${\mathcal{W}}$ denote the signal subspace spanned by $\{h_k({\mathbf{r}})\}_{k=1}^{K}$. The Gram-Schmidt orthonormalization constructs an orthonormal basis $\{\chi_{k}({\mathbf{r}})\}_{k=1}^{K}$ for ${\mathcal{W}}$ through the following iterative steps:
\begin{align}\nonumber
\chi_{k}({\mathbf{r}})=\frac{\hat{\chi_{k}}({\mathbf{r}})}{\sqrt{\epsilon_k}},~ \hat{\chi_{k}}({\mathbf{r}}) = h_k({\mathbf{r}})-\sum_{k'=1}^{k-1}
\frac{p_{k,k'}}{\epsilon_{k'}}\chi_{k'}({\mathbf{r}}),
\end{align}
where $\epsilon_k=\int_{\mathcal{A}}\lvert\hat{\chi_{k}}({\mathbf{r}})\rvert^2{\rm{d}}{\mathbf{r}}$ and $p_{k,k'}=\int_{\mathcal{A}}h_k({\mathbf{r}})\chi_{k'}^{*}({\mathbf{r}}){\rm{d}}{\mathbf{r}}$.
\section{A Full Derivation of Equation \eqref{ZF_Projection_Kernel}}\label{Proof_Lemma_ZF_Projection_Kernel}
Substituting $h_{k'}({\mathbf{r}})={\bm\psi}_k({\mathbf{r}}){\bm\phi}_{k,k'}$ (from \eqref{Function_Space_Expansion}) into $P_{k}({\mathbf{r}},{\mathbf{r}}')={\mathbf{h}}_k({\mathbf{r}}){\mathbf{R}}_k^{-1}{\mathbf{h}}_k^{\mathsf{H}}({\mathbf{r}}')$ (from \eqref{ZF_Projection_Kernel_Most_Original}) yields
\begin{align}\label{Proof_Lemma_ZF_Projection_Kernel_Equation1}
P_{k}({\mathbf{r}},{\mathbf{r}}')={\bm\psi}_k({\mathbf{r}}){\bm\Phi}_k{\mathbf{R}}_k^{-1}{\bm\Phi}_k^{\mathsf{H}}{\bm\psi}_k^{\mathsf{H}}({\mathbf{r}}'),
\end{align}
with ${\bm\Phi}_k\triangleq[{\bm\phi}_{k,1},\ldots,{\bm\phi}_{k,k-1},{\bm\phi}_{k,k+1},\ldots,{\bm\phi}_{k,K}]\in{\mathbbmss{C}}^{K_1\times K_1}$. Moreover, the matrix ${{\mathbf{R}}}_k= \int_{\mathcal{A}}{\mathbf{h}}_k^{\mathsf{H}}({\mathbf{r}}){\mathbf{h}}_k({\mathbf{r}}){\rm{d}}{\mathbf{r}}$ satisfies
\begin{align}\label{Proof_Lemma_ZF_Projection_Kernel_Equ2}
{{\mathbf{R}}}_k={\bm\Phi}_k^{\mathsf{H}}\int_{{\mathcal{A}}}{\bm\psi}_k^{\mathsf{H}}({\mathbf{r}}){\bm\psi}_k({\mathbf{r}}){\rm{d}}{\mathbf{r}}{\bm\Phi}_k
\overset{\natural}{=}{\bm\Phi}_k^{\mathsf{H}}{\bm\Phi}_k,
\end{align} 
where step $\natural$ holds with \eqref{Orthonormal_Basis}. By \emph{\textbf{Assumption-1}}, ${\bm\Phi}_k$ is a full-rank square matrix. Its singular value decomposition is ${\bm\Phi}_k={\mathbf{U}}_{{\bm\Phi}_k}{\bm\Lambda}_{{\bm\Phi}_k}{\mathbf{V}}_{{\bm\Phi}_k}$, where ${\mathbf{U}}_{{\bm\Phi}_k}\in{\mathbbmss{C}}^{{K_1}\times{K_1}}$ and ${\mathbf{V}}_{{\bm\Phi}_k}\in{\mathbbmss{C}}^{{K_1}\times{K_1}}$ are unitary matrices, and ${\bm\Lambda}_{{\bm\Phi}_k}\in{\mathbbmss{C}}^{{K_1}\times{K_1}}$ is a diagonal matrix with ${\bm\Lambda}_{{\bm\Phi}_k}\succ {\mathbf{0}}$. Consequently, we have ${\mathbf{R}}_k^{-1}=({\bm\Phi}_k^{\mathsf{H}}{\bm\Phi}_k)^{-1}={\mathbf{V}}_{{\bm\Phi}_k}^{-1}{\bm\Lambda}_{{\bm\Phi}_k}^{-2}
({\mathbf{V}}_{{\bm\Phi}_k}^{\mathsf{H}})^{-1}$, which, together with ${\bm\Phi}_k={\mathbf{U}}_{{\bm\Phi}_k}{\bm\Lambda}_{{\bm\Phi}_k}{\mathbf{V}}_{{\bm\Phi}_k}$, yields
\begin{subequations}\label{Proof_Lemma_ZF_Projection_Kernel_Equ4}
\begin{align}
&{\bm\Phi}_k{\mathbf{R}}_k^{-1}{\bm\Phi}_k^{\mathsf{H}}={\bm\Phi}_k({\bm\Phi}_k^{\mathsf{H}}{\bm\Phi}_k)^{-1}{\bm\Phi}_k^{\mathsf{H}}\\
&={\mathbf{U}}_{{\bm\Phi}_k}{\bm\Lambda}_{{\bm\Phi}_k}^{1+1-2}{\mathbf{U}}_{{\bm\Phi}_k}^{\mathsf{H}}
={\mathbf{U}}_{{\bm\Phi}_k}{\mathbf{U}}_{{\bm\Phi}_k}^{\mathsf{H}}={\mathbf{I}}_{K_1}.
\end{align}
\end{subequations}
Inserting \eqref{Proof_Lemma_ZF_Projection_Kernel_Equ4} into \eqref{Proof_Lemma_ZF_Projection_Kernel_Equation1} directly produces the result in \eqref{ZF_Projection_Kernel}.
\section{Proof of \textbf{Lemma} \ref{Lemma_ZF_SNR_Gain_Loss}}\label{Proof_Lemma_ZF_SNR_Gain_Loss}
Based on \emph{\textbf{Assumption-1}}, $h_k({\mathbf{r}})$ is not parallel to any $\{h_{k'}({\mathbf{r}})\}_{k'\in{\mathcal{K}}_k}$. From a function space perspective, $h_k({\mathbf{r}})$ can be decomposed into a linear combination of the orthonormal basis $\{\psi_{k'}({\mathbf{r}})\}_{k'\in{\mathcal{K}}_k}$, which spans the interference subspace ${\mathcal{W}}_k$ (generated by $\{h_{k'}({\mathbf{r}})\}_{k'\in{\mathcal{K}}_k}$), and a normalized signal $\psi_k({\mathbf{r}})$ orthogonal to ${\mathcal{W}}_k$. It follows that
\begin{align}\nonumber
h_k({\mathbf{r}})=\sum\nolimits_{k'\in{\mathcal{K}}_k}\phi_{k,k'}\psi_{k'}({\mathbf{r}})+\phi_{k,k}\psi_{k}({\mathbf{r}}),
\end{align}
where $\phi_{k,k'}=\int_{\mathcal{A}}h_k({\mathbf{r}})\psi_{k'}^{*}({\mathbf{r}}){\rm{d}}{\mathbf{r}}$ ($\forall k'\in{\mathcal{K}}$), $\int_{\mathcal{A}}\psi_{k}({\mathbf{r}})\psi_{k'}^{*}({\mathbf{r}}){\rm{d}}{\mathbf{r}}=\delta_{k,k'}$ ($\forall k'\in{\mathcal{K}}$), and $\phi_{k,k}\ne0$. Using the orthogonality of $\{\psi_{k'}({\mathbf{r}})\}_{k'=1}^{K}$, we have
\begin{align}\label{ZF_Power_Loss_Factor_Proof_Final1}
r_{k,k}=\int_{\mathcal{A}}\lvert h_k({\mathbf{r}})\rvert^2{\rm{d}}{\mathbf{r}}=\lVert {\bm\phi}_k\rVert^2+\lvert {\phi}_{k,k}\rvert^2,
\end{align}
where ${\bm\phi}_k=[\phi_{k,1},\ldots,\phi_{k,k-1},\phi_{k,k+1},\ldots,\phi_{k,K}]^{\mathsf{T}}\in{\mathbbmss{C}}^{{K_1}\times1}$. Following the steps used to derive \eqref{Proof_Lemma_ZF_Projection_Kernel_Equ2}, we obtain
\begin{align}\nonumber
{\mathbf{r}}_{k,k}=[r_{1,k},\ldots,r_{k-1,k},r_{k+1,k},\ldots,r_{K,k}]^{\mathsf{T}}={\bm\Phi}_k^{\mathsf{H}}{\bm\phi}_k.
\end{align}
Using ${\mathbf{R}}_k={\bm\Phi}_k^{\mathsf{H}}{\bm\Phi}_k$ and the result in \eqref{Proof_Lemma_ZF_Projection_Kernel_Equ4}, we obtain
\begin{align}\label{ZF_Power_Loss_Factor_Proof_Final2}
{\mathbf{r}}_{k,k}^{\mathsf{H}}{\mathbf{R}}_k^{-1}{\mathbf{r}}_{k,k}={\bm\phi}_k^{\mathsf{H}}{\bm\Phi}_k({\bm\Phi}_k^{\mathsf{H}}{\bm\Phi}_k)^{-1}{\bm\Phi}_k^{\mathsf{H}}{\bm\phi}_k=\lVert {\bm\phi}_k\rVert^2.
\end{align}
Since $\phi_{k,k}\ne0$, it follows from \eqref{ZF_Power_Loss_Factor_Proof_Final1} and \eqref{ZF_Power_Loss_Factor_Proof_Final2} that
\begin{align}\label{ZF_Power_Loss_Factor_Proof_Final}
r_{k,k}-{\mathbf{r}}_{k,k}^{\mathsf{H}}{\mathbf{R}}_k^{-1}{\mathbf{r}}_{k,k}=\lvert {\phi}_{k,k}\rvert^2>0.
\end{align}
Combined with ${\mathbf{r}}_{k,k}^{\mathsf{H}}{\mathbf{R}}_k^{-1}{\mathbf{r}}_{k,k}\geq0$, this completes the proof.
\section{A Full Derivation of Equations \eqref{Channel_No_User_k_Correlation_Matrix_With_Power_Trans1} and \eqref{Channel_No_User_k_Correlation_Matrix_With_Power_Trans2}}\label{Section: Analysis of the SINR: Preliminaries1}
\subsubsection*{Proof of Equation \eqref{Channel_No_User_k_Correlation_Matrix_With_Power_Trans1}}
By definition, $C_k({\mathbf{r}},{\mathbf{r}}')= \delta({\mathbf{r}}-{\mathbf{r}}')+{\mathbf{h}}_k({\mathbf{r}}){\mathbf{P}}_k{\mathbf{h}}_k^{\mathsf{H}}({\mathbf{r}}')$. Combining this with \eqref{MMSE_Proof_Pilot2}, we derive
\begin{equation}\label{Lemma_Squared_Root_Inversion_Result_Step1}
\begin{split}
\int_{{\mathcal{A}}}{\overline{B}}_{k}({\mathbf{r}}_1,{\mathbf{r}})C_{k}({\mathbf{r}},{\mathbf{r}}'){\rm{d}}{\mathbf{r}}
&=\delta({\mathbf{r}}_1-{\mathbf{r}}')\\
&-{\mathbf{h}}_k({\mathbf{r}}_1){\mathbf{P}}_k^{\frac{1}{2}}\hat{\mathbf{B}}_k{\mathbf{P}}_k^{\frac{1}{2}}{\mathbf{h}}_k^{\mathsf{H}}({\mathbf{r}}'),
\end{split}
\end{equation}
where $\hat{\mathbf{B}}_k={\overline{{\mathbf{B}}}_{k}}-{\mathbf{I}}_{K_1}+{\overline{{\mathbf{B}}}_{k}}{\mathbf{C}}_k\in{\mathbbmss{C}}^{{K_1}\times{K_1}}$ and ${\mathbf{C}}_k={\mathbf{P}}_k^{\frac{1}{2}}{\mathbf{R}}_k{\mathbf{P}}_k^{\frac{1}{2}}$ (as defined in \eqref{Channel_No_User_k_Correlation_Matrix_With_Power}). Inserting \eqref{Lemma_Squared_Root_Inversion_Result_Step1} into \eqref{Channel_No_User_k_Correlation_Matrix_With_Power_Trans1} gives
\begin{equation}\label{Lemma_Squared_Root_Inversion_Result_Step2}
\begin{split}
\int_{{\mathcal{A}}}\eqref{Lemma_Squared_Root_Inversion_Result_Step1}\times{\overline{B}}_{k}({\mathbf{r}}',{\mathbf{r}}_2){\rm{d}}{\mathbf{r}}'
&=\delta({\mathbf{r}}_1-{\mathbf{r}}_2)\\
&-{\mathbf{h}}_k({\mathbf{r}}_1){\mathbf{P}}_k^{\frac{1}{2}}\check{\mathbf{B}}_k{\mathbf{P}}_k^{\frac{1}{2}}{\mathbf{h}}_k^{\mathsf{H}}({\mathbf{r}}_2),
\end{split}
\end{equation}
where $\check{\mathbf{B}}_k\in{\mathbbmss{C}}^{{K_1}\times{K_1}}$ is given as follows:
\begin{equation}\label{Lemma_Squared_Root_Inversion_Result_Matrix_Form}
\begin{split}
\check{\mathbf{B}}_k&=\hat{\mathbf{B}}_k+{\overline{{\mathbf{B}}}_{k}}-
{\hat{\mathbf{B}}_k}{\mathbf{C}}_k{\overline{{\mathbf{B}}}_{k}}=2{\overline{{\mathbf{B}}}_{k}}-{\mathbf{I}}_{K_1}\\
&+{\overline{{\mathbf{B}}}_{k}}{\mathbf{C}}_k-({\overline{{\mathbf{B}}}_{k}}-{\mathbf{I}}_{K_1}+{\overline{{\mathbf{B}}}_{k}}{\mathbf{C}}_k){\mathbf{C}}_k{\overline{{\mathbf{B}}}_{k}}.
\end{split}
\end{equation}
Leveraging the EVDs ${\mathbf{C}}_k={\mathbf{U}}_{{\mathbf{C}}_k}{\bm\Lambda}_{{\mathbf{C}}_k}{\mathbf{U}}_{{\mathbf{C}}_k}^{\mathsf{H}}$ and ${\overline{{\mathbf{B}}}_{k}}={\mathbf{U}}_{{{{\mathbf{C}}}_{k}}}{{\bm\Lambda}}_{\overline{{\mathbf{B}}}_{k}}{\mathbf{U}}_{{{\mathbf{C}}}_{k}}^{\mathsf{H}}$ along with the unitary property ${\mathbf{U}}_{{\mathbf{C}}_k}{\mathbf{U}}_{{\mathbf{C}}_k}^{\mathsf{H}}={\mathbf{U}}_{{\mathbf{C}}_k}^{\mathsf{H}}{\mathbf{U}}_{{\mathbf{C}}_k}={\mathbf{I}}_{K_1}$, we simplify \eqref{Lemma_Squared_Root_Inversion_Result_Matrix_Form} to $\check{\mathbf{B}}_k={\mathbf{U}}_{{\mathbf{C}}_k}{{\bm\Lambda}}_{\check{\mathbf{B}}_k}{\mathbf{U}}_{{\mathbf{C}}_k}^{\mathsf{H}}$, where ${{\bm\Lambda}}_{\check{\mathbf{B}}_k}\in{\mathbbmss{C}}^{{K_1}\times{K_1}}$ is a diagonal matrix with entries:
\begin{equation}\label{Lemma_Squared_Root_Inversion_Result_Matrix_Form_Final_Before}
\begin{split}
[{{\bm\Lambda}}_{\check{\mathbf{B}}_k}]_{n,n}&=-[{\bm\Lambda}_{{\mathbf{C}}_k}]_{n,n}([{\bm\Lambda}_{{\mathbf{C}}_k}]_{n,n}+1)[{{\bm\Lambda}}_{\overline{{\mathbf{B}}}_{k}}]_{n,n}^2\\
&+2([{\bm\Lambda}_{{\mathbf{C}}_k}]_{n,n}+1)[{{\bm\Lambda}}_{\overline{{\mathbf{B}}}_{k}}]_{n,n}-1.
\end{split}
\end{equation}
Substituting $[{{\bm\Lambda}}_{\overline{{\mathbf{B}}}_{k}}]_{n,n}=\frac{1+\sqrt{1+[{\bm\Lambda}_{{\mathbf{C}}_k}]_{n,n}}}{[{\bm\Lambda}_{{\mathbf{C}}_k}]_{n,n}\sqrt{1+[{\bm\Lambda}_{{\mathbf{C}}_k}]_{n,n}}}$ into \eqref{Lemma_Squared_Root_Inversion_Result_Matrix_Form_Final_Before} and simplifying reveals $[{{\bm\Lambda}}_{\check{\mathbf{B}}_k}]_{n,n}=0$ for $n=1,\ldots,K_1$. Consequently, ${{\bm\Lambda}}_{\check{\mathbf{B}}_k}={\mathbf{0}}$ and $\check{\mathbf{B}}_k={\mathbf{0}}$. Combining this result with \eqref{Lemma_Squared_Root_Inversion_Result_Step2} directly leads to the conclusion in \eqref{Channel_No_User_k_Correlation_Matrix_With_Power_Trans1}.
\subsubsection*{Proof of Equation \eqref{Channel_No_User_k_Correlation_Matrix_With_Power_Trans2}}
Combining \eqref{MMSE_Proof_Pilot1} and \eqref{MMSE_Proof_Pilot2}, we derive
\begin{equation}\label{Lemma_Squared_Root_Matrix_Form}
\begin{split}
\int_{{\mathcal{A}}}{\overline{B}}_{k}({\mathbf{r}}_1,{\mathbf{r}})B_{k}({\mathbf{r}},{\mathbf{r}}_2){\rm{d}}{\mathbf{r}}
&=\delta({\mathbf{r}}_1-{\mathbf{r}}_2)\\
&-{\mathbf{h}}_k({\mathbf{r}}_1){\mathbf{P}}_k^{\frac{1}{2}}\breve{\mathbf{B}}_k{\mathbf{P}}_k^{\frac{1}{2}}{\mathbf{h}}_k^{\mathsf{H}}({\mathbf{r}}'),
\end{split}
\end{equation}
where $\breve{\mathbf{B}}_k={\overline{\mathbf{B}}}_k+{\mathbf{B}}_k-{\overline{\mathbf{B}}}_k{\mathbf{C}}_k{\mathbf{B}}_k\in{\mathbbmss{C}}^{{K_1}\times{K_1}}$. Following the simplification method applied to \eqref{Lemma_Squared_Root_Inversion_Result_Matrix_Form}, we rewrite \eqref{Lemma_Squared_Root_Matrix_Form} as $\breve{\mathbf{B}}_k={\mathbf{U}}_{{\mathbf{C}}_k}{{\bm\Lambda}}_{\breve{\mathbf{B}}_k}{\mathbf{U}}_{{\mathbf{C}}_k}^{\mathsf{H}}$, where ${{\bm\Lambda}}_{\breve{\mathbf{B}}_k}={{\bm\Lambda}}_{\overline{{\mathbf{B}}}_{k}}+{{\bm\Lambda}}_{{{\mathbf{B}}}_{k}}
-{{\bm\Lambda}}_{\overline{{\mathbf{B}}}_{k}}{\bm\Lambda}_{{\mathbf{C}}_k}{{\bm\Lambda}}_{{{\mathbf{B}}}_{k}}\in{\mathbbmss{C}}^{{K_1}\times{K_1}}$ is a diagonal matrix with entries:
\begin{equation}\nonumber
\begin{split}
[{{\bm\Lambda}}_{\breve{\mathbf{B}}_k}]_{n,n}&=[{{\bm\Lambda}}_{\overline{{\mathbf{B}}}_{k}}]_{n,n}+[{{\bm\Lambda}}_{{{\mathbf{B}}}_{k}}]_{n,n}\\
&-[{{\bm\Lambda}}_{\overline{{\mathbf{B}}}_{k}}]_{n,n}[{\bm\Lambda}_{{\mathbf{C}}_k}]_{n,n}[{{\bm\Lambda}}_{{{\mathbf{B}}}_{k}}]_{n,n}.
\end{split}
\end{equation}
Inserting $[{{\bm\Lambda}}_{\overline{{\mathbf{B}}}_{k}}]_{n,n}=\frac{1+\sqrt{1+[{\bm\Lambda}_{{\mathbf{C}}_k}]_{n,n}}}{[{\bm\Lambda}_{{\mathbf{C}}_k}]_{n,n}\sqrt{1+[{\bm\Lambda}_{{\mathbf{C}}_k}]_{n,n}}}$ and $[{{\bm\Lambda}}_{{{\mathbf{B}}}_{k}}]_{n,n}=\frac{1+\sqrt{1+[{\bm\Lambda}_{{\mathbf{C}}_k}]_{n,n}}}{[{\bm\Lambda}_{{\mathbf{C}}_k}]_{n,n}}$ into \eqref{Lemma_Squared_Root_Inversion_Result_Matrix_Form_Final_Before} results in $[{{\bm\Lambda}}_{\breve{\mathbf{B}}_k}]_{n,n}=0$ for $n=1,\ldots,K_1$. Consequently, ${{\bm\Lambda}}_{\breve{\mathbf{B}}_k}={\mathbf{0}}$, which leads to 
\begin{align}\label{Channel_No_User_k_Correlation_Matrix_With_Power_Trans2_Half}
\int_{{\mathcal{A}}}{\overline{B}}_{k}({\mathbf{r}}_1,{\mathbf{r}})B_{k}({\mathbf{r}},{\mathbf{r}}_2){\rm{d}}{\mathbf{r}}=\delta({\mathbf{r}}_1-{\mathbf{r}}_2)
\end{align}
and $(\int_{{\mathcal{A}}}{\overline{B}}_{k}({\mathbf{r}}_1,{\mathbf{r}})B_{k}({\mathbf{r}},{\mathbf{r}}_2){\rm{d}}{\mathbf{r}})^{*}
=(\delta({\mathbf{r}}_1-{\mathbf{r}}_2))^{*}=\delta({\mathbf{r}}_2-{\mathbf{r}}_1)$.

Since ${\overline{\mathbf{B}}}_{k}$ and ${{\mathbf{B}}}_{k}$ are Hermitian matrices, it holds that ${\overline{B}}_{k}({\mathbf{r}},{\mathbf{r}}')={\overline{B}}_{k}^{*}({\mathbf{r}}',{\mathbf{r}})$ and $B_{k}({\mathbf{r}},{\mathbf{r}}')=B_{k}^{*}({\mathbf{r}}',{\mathbf{r}})$. This implies $(\int_{{\mathcal{A}}}{\overline{B}}_{k}({\mathbf{r}}_1,{\mathbf{r}})B_{k}({\mathbf{r}},{\mathbf{r}}_2){\rm{d}}{\mathbf{r}})^{*}=\int_{{\mathcal{A}}}{\overline{B}}_{k}^{*}({\mathbf{r}}_1,{\mathbf{r}})B_{k}^{*}({\mathbf{r}},{\mathbf{r}}_2){\rm{d}}{\mathbf{r}}
=\int_{{\mathcal{A}}}B_{k}({\mathbf{r}}_2,{\mathbf{r}}){\overline{B}}_{k}({\mathbf{r}},{\mathbf{r}}_1){\rm{d}}{\mathbf{r}}$. Taken together, we have
\begin{align}\nonumber
\int_{{\mathcal{A}}}B_{k}({\mathbf{r}}_2,{\mathbf{r}}){\overline{B}}_{k}({\mathbf{r}},{\mathbf{r}}_1){\rm{d}}{\mathbf{r}}=\delta({\mathbf{r}}_2-{\mathbf{r}}_1).
\end{align}
Combined with \eqref{Channel_No_User_k_Correlation_Matrix_With_Power_Trans2_Half}, this result directly proves \eqref{Channel_No_User_k_Correlation_Matrix_With_Power_Trans2}.
\section{A Full Derivation of Equations \eqref{Channel_No_User_k_Correlation_Matrix_With_Power_Trans_Result2} and \eqref{Channel_No_User_k_Correlation_Matrix_With_Power_Trans_Result3}}\label{Proof_Lemma_Channel_No_User_k_Correlation_Matrix_With_Power_Trans_Result23}
Substituting \eqref{Equivalent_Trans_Problem_2} into \eqref{Channel_No_User_k_Correlation_Matrix_With_Power_Trans_Result2_Before} and leveraging the Hermitian property ${\overline{B}}_{k}({\mathbf{r}},{\mathbf{r}}')={\overline{B}}_{k}^{*}({\mathbf{r}}',{\mathbf{r}})$, we obtain
\begin{align}\label{Channel_No_User_k_Correlation_Matrix_With_Power_Trans_Result2_After}
f_{\mathsf{dm}}({w}_k(\mathbf{r}))=\iint_{{\mathcal{A}}^2}u_k^{*}(\mathbf{r})\hat{C}_k({\mathbf{r}},{\mathbf{r}}')u_k({\mathbf{r}}'){\rm{d}}{\mathbf{r}}'{\rm{d}}{\mathbf{r}},
\end{align}
where $\hat{C}_k({\mathbf{r}},{\mathbf{r}}')=\iint_{{\mathcal{A}}^2}{\overline{B}}_{k}({\mathbf{r}},{\mathbf{r}}_1){C}_k({\mathbf{r}}_1,{\mathbf{r}}_2)
{\overline{B}}_{k}({\mathbf{r}}_2,{\mathbf{r}}'){\rm{d}}{\mathbf{r}}_1{\rm{d}}{\mathbf{r}}_2$. From \eqref{Channel_No_User_k_Correlation_Matrix_With_Power_Trans1}, $\hat{C}_k({\mathbf{r}},{\mathbf{r}}')=\delta({\mathbf{r}}-{\mathbf{r}}')$. Combining this with \eqref{Channel_No_User_k_Correlation_Matrix_With_Power_Trans_Result2_After} directly produces the final result in \eqref{Channel_No_User_k_Correlation_Matrix_With_Power_Trans_Result2}.

Similarly, substituting \eqref{Equivalent_Trans_Problem_2} into \eqref{Channel_No_User_k_Correlation_Matrix_With_Power_Trans_Result3_Before} yields
\begin{align}\label{Channel_No_User_k_Correlation_Matrix_With_Power_Trans_Result3_After}
f_{\mathsf{nm}}({w}_k(\mathbf{r}))
=\left\lvert\int_{{\mathcal{A}}}\int_{{\mathcal{A}}} h_k(\mathbf{r}) {\overline{B}}_{k}^{*}({\mathbf{r}},{\mathbf{r}}_1){\rm{d}}{\mathbf{r}}{u}_k^{*}({\mathbf{r}}_1){\rm{d}}{\mathbf{r}}_1\right\rvert^2.
\end{align}
Since ${\overline{B}}_{k}({\mathbf{r}},{\mathbf{r}}')={\overline{B}}_{k}^{*}({\mathbf{r}}',{\mathbf{r}})$, we derive
\begin{align}
\int_{{\mathcal{A}}}h_{k}(\mathbf{r}){\overline{B}}_{k}^{*}({\mathbf{r}},{\mathbf{r}}_1){\rm{d}}{\mathbf{r}}=
\int_{{\mathcal{A}}}{\overline{B}}_{k}({\mathbf{r}}_1,{\mathbf{r}})h_{k}(\mathbf{r}){\rm{d}}{\mathbf{r}}={v}_k({\mathbf{r}}_1)\nonumber.
\end{align}
Inserting this identity into \eqref{Channel_No_User_k_Correlation_Matrix_With_Power_Trans_Result3_After} completes the proof of \eqref{Channel_No_User_k_Correlation_Matrix_With_Power_Trans_Result3}.
\section{Proofs of \textbf{Theorem \ref{Theorem_MMSE_Rate_Optimal}} and \textbf{Corollary \ref{Corollary_SINR_Per_User_Optimal_Beamforming}}}\label{Proof_Theorem_MMSE_Rate_Optimal}
Substituting \eqref{MMSE_Proof_Pilot2} into $\int_{{\mathcal{A}}} {\overline{B}}_{k}({\mathbf{r}},{\mathbf{r}}')
{\overline{B}}_{k}({\mathbf{r}}',{\mathbf{r}}''){\rm{d}}{\mathbf{r}}'$, we derive
\begin{equation}\label{Proof_Theorem_MMSE_Rate_Optimal_Step1}
\begin{split}
\int_{{\mathcal{A}}} {\overline{B}}_{k}({\mathbf{r}},{\mathbf{r}}')
{\overline{B}}_{k}({\mathbf{r}}',{\mathbf{r}}''){\rm{d}}{\mathbf{r}}'&=\delta({\mathbf{r}}-{\mathbf{r}}'')\\
&-{\mathbf{h}}_k({\mathbf{r}}){\mathbf{P}}_k^{\frac{1}{2}}\tilde{\mathbf{B}}_k{\mathbf{P}}_k^{\frac{1}{2}}{\mathbf{h}}_k^{\mathsf{H}}({\mathbf{r}}''),
\end{split}
\end{equation}
where $\tilde{\mathbf{B}}_k=2\overline{\mathbf{B}}_k-\overline{\mathbf{B}}_k{\mathbf{C}}_k\overline{\mathbf{B}}_k\in{\mathbbmss{C}}^{{K_1}\times{K_1}}$. Following the simplification method applied to \eqref{Lemma_Squared_Root_Inversion_Result_Matrix_Form}, we rewrite $\tilde{\mathbf{B}}_k$ as $\tilde{\mathbf{B}}_k={\mathbf{U}}_{{\mathbf{C}}_k}{{\bm\Lambda}}_{\tilde{\mathbf{B}}_k}{\mathbf{U}}_{{\mathbf{C}}_k}^{\mathsf{H}}$, where ${{\bm\Lambda}}_{\tilde{\mathbf{B}}_k}=2{{\bm\Lambda}}_{\overline{{\mathbf{B}}}_{k}}-{{\bm\Lambda}}_{\overline{{\mathbf{B}}}_{k}}{\bm\Lambda}_{{\mathbf{C}}_k}{{\bm\Lambda}}_{{{\mathbf{B}}}_{k}}\in{\mathbbmss{C}}^{{K_1}\times{K_1}}$ is a diagonal matrix with entries:
\begin{align}\nonumber
[{{\bm\Lambda}}_{\tilde{\mathbf{B}}_k}]_{n,n}=2[{{\bm\Lambda}}_{\overline{{\mathbf{B}}}_{k}}]_{n,n}-[{{\bm\Lambda}}_{\overline{{\mathbf{B}}}_{k}}]_{n,n}^2[{\bm\Lambda}_{{\mathbf{C}}_k}]_{n,n},
\end{align}
Using $[{{\bm\Lambda}}_{\overline{{\mathbf{B}}}_{k}}]_{n,n}=\frac{1+\sqrt{1+[{\bm\Lambda}_{{\mathbf{C}}_k}]_{n,n}}}{[{\bm\Lambda}_{{\mathbf{C}}_k}]_{n,n}\sqrt{1+[{\bm\Lambda}_{{\mathbf{C}}_k}]_{n,n}}}$, this simplifies to
\begin{align}\nonumber
[{{\bm\Lambda}}_{\tilde{\mathbf{B}}_k}]_{n,n}=(1+[{\bm\Lambda}_{{\mathbf{C}}_k}]_{n,n})^{-1}.
\end{align}
Therefore, ${{\bm\Lambda}}_{\tilde{\mathbf{B}}_k}=({\mathbf{I}}_{K_1}+{\bm\Lambda}_{{\mathbf{C}}_k})^{-1}$, and $\tilde{\mathbf{B}}_k=({\mathbf{I}}_{K_1}+{\mathbf{C}}_k)^{-1}$. Substituting this into \eqref{Proof_Theorem_MMSE_Rate_Optimal_Step1} yields
\begin{equation}\label{Corollary_Inversion_Operator_Eq1_Further_Results}
\begin{split}
&\int_{{\mathcal{A}}} {\overline{B}}_{k}({\mathbf{r}},{\mathbf{r}}')
{\overline{B}}_{k}({\mathbf{r}}',{\mathbf{r}}''){\rm{d}}{\mathbf{r}}'=\delta({\mathbf{r}}-{\mathbf{r}}'')\\
&-{\mathbf{h}}_k({\mathbf{r}}){\mathbf{P}}_k^{\frac{1}{2}}({\mathbf{I}}_{K_1}+{\mathbf{C}}_k)^{-1}{\mathbf{P}}_k^{\frac{1}{2}}{\mathbf{h}}_k^{\mathsf{H}}({\mathbf{r}}'').
\end{split}
\end{equation}
Inserting \eqref{Corollary_Inversion_Operator_Eq1_Further_Results} into \eqref{Rayleigh_Quotient_Optimal_Solution_Step2_Eq1} and evaluating the integral produces the results in \textbf{Theorem \ref{Theorem_MMSE_Rate_Optimal}}. 

Furthermore, by substituting ${v}_k({\mathbf{r}})= \int_{{\mathcal{A}}}{\overline{B}}_{k}({\mathbf{r}},{\mathbf{r}}')h_{k}({\mathbf{r}}'){\rm{d}}{\mathbf{r}}'$ and ${\overline{B}}_{k}^{*}({\mathbf{r}},{\mathbf{r}}')={\overline{B}}_{k}({\mathbf{r}}',{\mathbf{r}})$ into \eqref{MMSE_SINR_Maximization_Final}, the SINR achieved by the optimal beamformer becomes:
\begin{align}
\gamma_{{\mathsf{MMSE}},k}=\frac{\iiint_{{\mathcal{A}}^3}h_{k}^{*}({\mathbf{r}})
{\overline{B}}_{k}({\mathbf{r}},{\mathbf{r}}')
{\overline{B}}_{k}({\mathbf{r}}',{\mathbf{r}}'')h_{k}({\mathbf{r}}''){\rm{d}}{\mathbf{r}}{\rm{d}}{\mathbf{r}}''{\rm{d}}{\mathbf{r}}'}{\sigma^2/P_{k}}.\nonumber
\end{align}
Combining this with \eqref{Corollary_Inversion_Operator_Eq1_Further_Results} and the definition ${{\mathbf{r}}}_{k,k'}=\int_{\mathcal{A}}{\mathbf{h}}_k^{\mathsf{H}}({\mathbf{r}})h_{k'}({\mathbf{r}}){\rm{d}}{\mathbf{r}}$, we directly obtain \textbf{Corollary \ref{Corollary_SINR_Per_User_Optimal_Beamforming}}.
\section{Proof of $\alpha_{{\mathsf{MMSE}},k}\in[0,1)$}\label{Proof_MMSE_Power_Loss_Factor}
Applying the Woodbury matrix identity \cite{bernstein2009matrix}, we derive
\begin{align}
({\mathbf{P}}_k^{-1}+{\mathbf{R}}_k)^{-1}&=({\mathbf{R}}_k+{\mathbf{I}}_{K_1}{\mathbf{P}}_k^{-1}{\mathbf{I}}_{K_1})^{-1}\nonumber\\
&={\mathbf{R}}_k^{-1}-{\mathbf{R}}_k^{-1}({\mathbf{P}}_k+{\mathbf{R}}_k^{-1}){\mathbf{R}}_k^{-1}.\nonumber
\end{align}
From this, it follows that
\begin{align}
{\mathbf{R}}_k^{-1}-({\mathbf{P}}_k^{-1}+{\mathbf{R}}_k)^{-1}={\mathbf{R}}_k^{-1}
({\mathbf{P}}_k+{\mathbf{R}}_k^{-1}){\mathbf{R}}_k^{-1}.\nonumber
\end{align}
Since ${\mathbf{R}}_k\succ{\mathbf{0}}$ and ${\mathbf{P}}_k\succ{\mathbf{0}}$, we conclude
\begin{align}\nonumber
({\mathbf{P}}_k^{-1}+{\mathbf{R}}_k)^{-1}\succ{\mathbf{0}},~{\mathbf{R}}_k^{-1}-({\mathbf{P}}_k^{-1}+{\mathbf{R}}_k)^{-1}\succ{\mathbf{0}},
\end{align}
which suggests that ${\mathbf{r}}_{k,k}^{\mathsf{H}}({\mathbf{R}}_k^{-1}-({\mathbf{P}}_k^{-1}+{\mathbf{R}}_k)^{-1}){\mathbf{r}}_{k,k}\geq0$. From \eqref{ZF_Power_Loss_Factor_Proof_Final}, we know $a_k=r_{k,k}>{\mathbf{r}}_{k,k}^{\mathsf{H}}{\mathbf{R}}_k^{-1}{\mathbf{r}}_{k,k}$, which yields
\begin{align}\nonumber
a_{k}>{\mathbf{r}}_{k,k}^{\mathsf{H}}{\mathbf{R}}_k^{-1}{\mathbf{r}}_{k,k}\geq{\mathbf{r}}_{k,k}^{\mathsf{H}}({\mathbf{P}}_k^{-1}+{\mathbf{R}}_k)^{-1}{\mathbf{r}}_{k,k}\geq0.
\end{align}
Therefore, $\alpha_{{\mathsf{MMSE}},k}=\frac{1}{a_{k}}{\mathbf{r}}_{k,k}^{\mathsf{H}}({\mathbf{P}}_k^{-1}+{\mathbf{R}}_k)^{-1}{\mathbf{r}}_{k,k}\in[0,1)$. 
\section{Proof of \textbf{Lemma \ref{Lemma_Optimal_MMSE_Indirectky}}}\label{Proof_Lemma_Optimal_MMSE_Indirectky}
Inserting \eqref{Rate_Optimal_Beamformer_General_Matrix} into the numerator of \eqref{MSE_Fundamental} gives
\begin{align}\label{Proof_Lemma_Optimal_MMSE_Indirectky_Step1}
\int_{{\mathcal{A}}}{w}_k(\mathbf{r})h_k^{*}(\mathbf{r}){\rm{d}}{\mathbf{r}}
=a_k-{\mathbf{r}}_{k,k}^{\mathsf{H}}({\mathbf{P}}_k^{-1}+{\mathbf{R}}_k)^{-1}{\mathbf{r}}_{k,k}.
\end{align}
Furthermore, the denominator of \eqref{MSE_Fundamental} satisfies
\begin{equation}\nonumber
{\text{Denominator~of}}~\eqref{MSE_Fundamental}=\sigma^2f_{\mathsf{dm}}({w}_k(\mathbf{r}))\left(1+\gamma_k\right).
\end{equation}
Using the results from \eqref{Channel_No_User_k_Correlation_Matrix_With_Power_Trans_Result3} and \eqref{MMSE_SINR_Maximization_Final}, we have
\begin{align}
f_{\mathsf{dm}}(w_{{\mathsf{MMSE}},k}(\mathbf{r}))&=\int_{{\mathcal{A}}}\lvert{v}_k({\mathbf{r}})\rvert^2{\rm{d}}{\mathbf{r}}
=\frac{\gamma_{{\mathsf{MMSE}},k}}{{P_{k}}/{\sigma^2}}\nonumber\\
&=a_k-{\mathbf{r}}_{k,k}^{\mathsf{H}}({\mathbf{P}}_k^{-1}+{\mathbf{R}}_k)^{-1}{\mathbf{r}}_{k,k},\nonumber
\end{align}
where the last equality follows from \eqref{SINR_Per_User_Optimal_Beamforming_Final_Result}. This allows rewriting the denominator of \eqref{MSE_Fundamental} as follows:
\begin{align}\label{Proof_Lemma_Optimal_MMSE_Indirectky_Step1_2}
{\text{Denominator~of}}~\eqref{MSE_Fundamental}=\sigma^2\frac{\gamma_{{\mathsf{MMSE}},k}}{{P_{k}}/{\sigma^2}}\left(1+\gamma_{{\mathsf{MMSE}},k}\right).
\end{align} 
Substituting \eqref{Proof_Lemma_Optimal_MMSE_Indirectky_Step1}, \eqref{Proof_Lemma_Optimal_MMSE_Indirectky_Step1_2}, and \eqref{SINR_Per_User_Optimal_Beamforming_Final_Result} into \eqref{MSE_Fundamental} gives $\beta_k^{\star}=\beta_{{\mathsf{MMSE}},k}$. Inserting this result and \eqref{Rate_Optimal_Beamformer_General_Matrix} into $w_{{\mathsf{MMSE}},k}^{\star}({\mathbf{r}})=\beta_{{\mathsf{MMSE}},k}w_{{\mathsf{MMSE}},k}(\mathbf{r})$ completes the proof. 
\section{Proof of \textbf{Theorem \ref{Theorem_MMSE_Uplink_Direct_Solution}}}\label{Proof_Theorem_MMSE_Uplink_Direct_Solution}
The proof of \textbf{Theorem \ref{Theorem_MMSE_Uplink_Direct_Solution}} requires the following lemma.
\vspace{-5pt}
\begin{lemma}\label{Lemma_Quadrature_Inversion}
Let $C(\mathbf{r}-\mathbf{r}')\triangleq\delta(\mathbf{r}-\mathbf{r}')+{\mathbf{h}}({\mathbf{r}}){\mathbf{P}}{\mathbf{h}}^{\mathsf{H}}({\mathbf{r}}')$. The following identity holds
\begin{align}
\int_{{\mathcal{A}}}{\overline{C}}({\mathbf{r}}_1,{\mathbf{r}})C({\mathbf{r}},{\mathbf{r}}_2){\rm{d}}{\mathbf{r}}
=\int_{{\mathcal{A}}}C({\mathbf{r}}_2,{\mathbf{r}}){\overline{C}}({\mathbf{r}},{\mathbf{r}}_1){\rm{d}}{\mathbf{r}}
=\delta({\mathbf{r}}_1-{\mathbf{r}}_2),\nonumber
\end{align}
where ${\overline{C}}({\mathbf{r}},{\mathbf{r}}')\triangleq\delta({\mathbf{r}}-{\mathbf{r}}')-
{\mathbf{h}}({\mathbf{r}}){\mathbf{P}}^{\frac{1}{2}}{\tilde{\mathbf{C}}}{\mathbf{P}}^{\frac{1}{2}}{\mathbf{h}}^{\mathsf{H}}({\mathbf{r}}')$. 
\end{lemma}
\vspace{-5pt}
\begin{IEEEproof}
This lemma follows from methods analogous to those in Appendices \ref{Section: Analysis of the SINR: Preliminaries1} and \ref{Proof_Theorem_MMSE_Rate_Optimal}.
\end{IEEEproof}
Inserting \eqref{Linear_Receive_Beamforming} into \eqref{MMSE_Formulation} expands the MSE as follows:
\begin{align}
{\overline{\mathsf{MSE}}}_k&=\sum_{k'=1}^{K}P_{k'}\left\lvert\int_{{\mathcal{A}}}{w}_k^{*}(\mathbf{r})h_{k'}(\mathbf{r}){\rm{d}}{\mathbf{r}}\right\rvert^2+\sigma^2\int_{{\mathcal{A}}}\lvert w_k(\mathbf{r})\rvert^2{\rm{d}}{\mathbf{r}}\nonumber\\
&-2\Re\left\{\sqrt{P_k}\int_{{\mathcal{A}}}{w}_k^{*}(\mathbf{r})h_k(\mathbf{r}){\rm{d}}{\mathbf{r}}\right\}+1.\label{MMSE_Formulation_Objective}
\end{align}
The problem in \eqref{MMSE_Formulation} belongs to functional programming. Equation \eqref{MMSE_Formulation_Objective} reveals that ${\overline{\mathsf{MSE}}}_k$ is convex in ${w}_k(\mathbf{r})$. Consequently, the optimal solution $w_{{\mathsf{MMSE}},k}^{\diamond}({\mathbf{r}})$ satisfies the first-order optimality condition as follows:
\begin{equation}\nonumber
\begin{split}
\frac{{\rm{d}}}{{\rm{d}}w_k({\mathbf{r}})}{\overline{\mathsf{MSE}}}_k&\overset{\flat}{=}
\sum_{k'=1}^{K}P_{k'}\int_{{\mathcal{A}}}{w}_k(\mathbf{r}')h_{k'}^{*}(\mathbf{r}'){\rm{d}}{\mathbf{r}}'h_{k'}(\mathbf{r})\\
&-\sqrt{P_k}h_k(\mathbf{r})+\sigma^2w_k(\mathbf{r})=0,
\end{split}
\end{equation}
where step $\flat$ employs the CoV method from \cite{wang2024beamforming}. The MSE-optimal beamformer thus satisfies
\begin{align}
\int_{{\mathcal{A}}}\left(\delta(\mathbf{r}-\mathbf{r}')+{\mathbf{h}}({\mathbf{r}}){\mathbf{P}}{\mathbf{h}}^{\mathsf{H}}({\mathbf{r}}')\right)w_k({\mathbf{r}}'){\rm{d}}{\mathbf{r}}'
=\frac{\sqrt{P_k}}{\sigma^2}h_k(\mathbf{r}).\nonumber
\end{align}
Let $W^{\diamond}(\mathbf{r},\mathbf{r}')$ denote the inverse operator of $\delta(\mathbf{r}-\mathbf{r}')+{\mathbf{h}}({\mathbf{r}}){\mathbf{P}}{\mathbf{h}}^{\mathsf{H}}({\mathbf{r}}')$, which is defined such that
\begin{align}
\int_{{\mathcal{A}}}W^{\diamond}(\mathbf{r}'',\mathbf{r})(\delta(\mathbf{r}-\mathbf{r}')+{\mathbf{h}}({\mathbf{r}}){\mathbf{P}}{\mathbf{h}}^{\mathsf{H}}({\mathbf{r}}')){\rm{d}}{\mathbf{r}}
=\delta(\mathbf{r}''-\mathbf{r}').\nonumber
\end{align}
This implies $w_{{\mathsf{MMSE}},k}^{\diamond}({\mathbf{r}})=\int_{{\mathcal{A}}}W^{\diamond}(\mathbf{r},\mathbf{r}')\frac{\sqrt{P_k}}{\sigma^2}h_k(\mathbf{r}'){\rm{d}}{\mathbf{r}}'$. By \textbf{Lemma \ref{Lemma_Quadrature_Inversion}}, $W^{\diamond}(\mathbf{r}'',\mathbf{r})={\overline{C}}(\mathbf{r}'',\mathbf{r})$, which directly yields the final result.
\section{Proof of \textbf{Corollary \ref{Corollary_MMSE_Uplink_Direct_Indirect_Comparison}}}\label{Proof_Corollary_MMSE_Uplink_Direct_Indirect_Comparison}
The proof of \textbf{Corollary \ref{Corollary_MMSE_Uplink_Direct_Indirect_Comparison}} requires the following lemma.
\vspace{-5pt}
\begin{lemma}\label{Proof_Corollary_MMSE_Uplink_Direct_Indirect_Comparison_Step0}
The operator $W^{\diamond}(\mathbf{r}'',\mathbf{r})={\overline{C}}(\mathbf{r}'',\mathbf{r})$ also satisfies
\begin{equation}\label{Proof_Corollary_MMSE_Uplink_Direct_Indirect_Comparison_Step1}
\begin{split}
&W^{\diamond}(\mathbf{r}'',\mathbf{r})=W_k^{\diamond}(\mathbf{r}'',\mathbf{r})\\&-
\frac{\iint_{{\mathcal{A}}^2}W_k^{\diamond}(\mathbf{r}'',\mathbf{r}_1)h_k({\mathbf{r}}_1)h_k^{*}(\mathbf{r}_2)W_k^{\diamond}
(\mathbf{r}_2,\mathbf{r}){\rm{d}}{\mathbf{r}}_1{\rm{d}}{\mathbf{r}}_2}{\frac{\sigma^2}{P_k}+
\iint_{{\mathcal{A}}^2}h_k^{*}({\mathbf{r}}_2)W_k^{\diamond}(\mathbf{r}_2,\mathbf{r}_1)h_k(\mathbf{r}_1){\rm{d}}{\mathbf{r}}_1{\rm{d}}{\mathbf{r}}_2},
\end{split}
\end{equation}
where $W_k^{\diamond}(\mathbf{r}'',\mathbf{r})$ represents the inverse operator of $C_k({\mathbf{r}},{\mathbf{r}}')= \delta({\mathbf{r}}-{\mathbf{r}}')+{\mathbf{h}}_k({\mathbf{r}}){\mathbf{P}}_k{\mathbf{h}}_k^{\mathsf{H}}({\mathbf{r}}')$.
\end{lemma}
\vspace{-5pt}
\begin{IEEEproof}
To prove \textbf{Lemma \ref{Proof_Corollary_MMSE_Uplink_Direct_Indirect_Comparison_Step0}}, we need to verify $\int_{{\mathcal{A}}}\eqref{Proof_Corollary_MMSE_Uplink_Direct_Indirect_Comparison_Step1}\times C(\mathbf{r}-\mathbf{r}'){\rm{d}}\mathbf{r}=\delta(\mathbf{r}''-\mathbf{r}')$, namely 
\begin{align}
\int_{{\mathcal{A}}}\eqref{Proof_Corollary_MMSE_Uplink_Direct_Indirect_Comparison_Step1}\times \left(\frac{P_{k}}{\sigma^2}h_{k}(\mathbf{r})h_{k}^{*}(\mathbf{r}')+C_k({\mathbf{r}},{\mathbf{r}}')\right){\rm{d}}\mathbf{r}=\delta(\mathbf{r}''-\mathbf{r}').\nonumber
\end{align}
Expanding the left-hand side of the above expression and simplifying with the following identities:
\begin{align}\nonumber
&\int_{{\mathcal{A}}}W_k^{\diamond}(\mathbf{r}'',\mathbf{r})C_k({\mathbf{r}},{\mathbf{r}}'){\rm{d}}{\mathbf{r}}
=\delta(\mathbf{r}''-\mathbf{r}'),\\
&\iint_{{\mathcal{A}}^2}h_k^{*}(\mathbf{r}_2)W_k^{\diamond}
(\mathbf{r}_2,\mathbf{r})C_k({\mathbf{r}},{\mathbf{r}}'){\rm{d}}{\mathbf{r}}{\rm{d}}{\mathbf{r}}_2=h_k^{*}({\mathbf{r}}'),\nonumber
\end{align}
directly yields the result.
\end{IEEEproof}
Upon substituting \eqref{Proof_Corollary_MMSE_Uplink_Direct_Indirect_Comparison_Step1} into $w_{{\mathsf{MMSE}},k}^{\diamond}({\mathbf{r}})=\int_{{\mathcal{A}}}W^{\diamond}(\mathbf{r},\mathbf{r}')\frac{\sqrt{P_k}}{\sigma^2}h_k(\mathbf{r}'){\rm{d}}{\mathbf{r}}'$ and simplifying, we have
\begin{align}
w_{{\mathsf{MMSE}},k}^{\diamond}({\mathbf{r}})=\frac{\frac{{\sqrt{P_k}}}{\sigma^2}\int_{{\mathcal{A}}}W_k^{\diamond}(\mathbf{r},\mathbf{r}_1)h_k(\mathbf{r}_1){\rm{d}}{\mathbf{r}}_1}
{1+\frac{P_k}{\sigma^2}
\iint_{{\mathcal{A}}^2}h_k^{*}({\mathbf{r}}_2)W_k^{\diamond}(\mathbf{r}_2,\mathbf{r}_1)h_k(\mathbf{r}_1){\rm{d}}{\mathbf{r}}_1{\rm{d}}{\mathbf{r}}_2}.\nonumber
\end{align}
By applying the method used in \textbf{Lemma \ref{Lemma_Quadrature_Inversion}}, we obtain
\begin{align}
W_k^{\diamond}(\mathbf{r},\mathbf{r}')&=\delta({\mathbf{r}}-{\mathbf{r}}')-{\mathbf{h}}_k({\mathbf{r}}){\mathbf{P}}_k^{\frac{1}{2}}({\mathbf{I}}_{K_1}
+{\mathbf{C}}_k)^{-1}{\mathbf{P}}_k^{\frac{1}{2}}{\mathbf{h}}_k^{\mathsf{H}}({\mathbf{r}}')\nonumber\\
&=\delta({\mathbf{r}}-{\mathbf{r}}')-{\mathbf{h}}_k({\mathbf{r}})({\mathbf{P}}_k^{-1}
+{\mathbf{R}}_k)^{-1}{\mathbf{h}}_k^{\mathsf{H}}({\mathbf{r}}'),\nonumber
\end{align}
where ${\mathbf{C}}_k={\mathbf{P}}_k^{\frac{1}{2}}{\mathbf{R}}_k{\mathbf{P}}_k^{\frac{1}{2}}$ (from \eqref{Channel_No_User_k_Correlation_Matrix_With_Power}). Consequently,
\begin{align}
&\int_{{\mathcal{A}}}W_k^{\diamond}(\mathbf{r},\mathbf{r}_1)h_k(\mathbf{r}_1){\rm{d}}{\mathbf{r}}_1={{h}}_k({\mathbf{r}})-{\mathbf{h}}_k({\mathbf{r}})\breve{\mathbf{r}}_{k},\nonumber\\
&\iint_{{\mathcal{A}}^2}h_k^{*}({\mathbf{r}}_2)W_k^{\diamond}(\mathbf{r}_2,\mathbf{r}_1)h_k(\mathbf{r}_1){\rm{d}}{\mathbf{r}}_1{\rm{d}}{\mathbf{r}}_2=a_k-{\mathbf{r}}_{k,k}^{\mathsf{H}}\breve{\mathbf{r}}_{k},\nonumber
\end{align}
with $\breve{\mathbf{r}}_{k}=({\mathbf{P}}_k^{-1}
+{\mathbf{R}}_k)^{-1}{\mathbf{r}}_{k,k}\in{\mathbbmss{C}}^{{K_1}\times1}$. Therefore, the optimal beamformer satisfies
\begin{align}
w_{{\mathsf{MMSE}},k}^{\diamond}({\mathbf{r}})=\frac{\sqrt{P_k}(h_k({\mathbf{r}})-{\mathbf{h}}_k({\mathbf{r}})({\mathbf{P}}_k^{-1}+{\mathbf{R}}_k)^{-1}{\mathbf{r}}_{k,k})}
{P_{k}(a_k-{\mathbf{r}}_{k,k}^{\mathsf{H}}({\mathbf{P}}_k^{-1}+{\mathbf{R}}_k)^{-1}{\mathbf{r}}_{k,k})+\sigma^2},\nonumber
\end{align} 
which completes the proof.
\section{Proof of \textbf{Corollary \ref{Corollary_MMSE_Uplink_Direct_Indirect_Comparison_Beamforming_Rewritten}}}\label{Proof_Corollary_MMSE_Uplink_Direct_Indirect_Comparison_Beamforming_Rewritten}
By omitting $\frac{\sqrt{P_k}}{\sigma^2}$ in \eqref{MMSE_Uplink_Direct_Solution_Eq2} and using the fact that ${\tilde{\mathbf{C}}}=({\mathbf{I}}_{K}+{\mathbf{P}}^{\frac{1}{2}}{\mathbf{R}}{\mathbf{P}}^{\frac{1}{2}})^{-1}={\mathbf{P}}^{-\frac{1}{2}}({\mathbf{P}}^{-1}+{\mathbf{R}}){\mathbf{P}}^{-\frac{1}{2}}$, we obtain 
\begin{align}\nonumber
w_{{\mathsf{MMSE}},k}(\mathbf{r})=h_k({\mathbf{r}})-{\mathbf{h}}({\mathbf{r}})({\mathbf{P}}^{-1}+{\mathbf{R}})^{-1}{\mathbf{r}}_k.
\end{align}
By writing it in vector form, we have
\begin{align}\nonumber
{{\mathbf{w}}}_{{\mathsf{MMSE}}}(\mathbf{r})&=[w_{{\mathsf{MMSE}},1}(\mathbf{r}),\ldots,w_{{\mathsf{MMSE}},K}(\mathbf{r})]\\
&={\mathbf{h}}({\mathbf{r}})({\mathbf{I}}_K-({\mathbf{P}}^{-1}+{\mathbf{R}})^{-1}{\mathbf{R}}).\nonumber
\end{align}
Since ${\mathbf{I}}_K=({\mathbf{P}}^{-1}+{\mathbf{R}})^{-1}({\mathbf{P}}^{-1}+{\mathbf{R}})$, we obtain
\begin{align}\nonumber
{\mathbf{I}}_K-({\mathbf{P}}^{-1}+{\mathbf{R}})^{-1}{\mathbf{R}}&=({\mathbf{P}}^{-1}+{\mathbf{R}})^{-1}{\mathbf{P}}^{-1}\\
&=({\mathbf{I}}_K+{\mathbf{P}}{\mathbf{R}})^{-1},\nonumber
\end{align}
which yields ${{\mathbf{w}}}_{{\mathsf{MMSE}}}(\mathbf{r})
={\mathbf{h}}({\mathbf{r}})({\mathbf{I}}_K+{\mathbf{P}}{\mathbf{R}})^{-1}$ and
\begin{equation}\nonumber
\begin{split}
w_{{\mathsf{MMSE}},k}(\mathbf{r})&=[{{\mathbf{w}}}_{{\mathsf{MMSE}}}(\mathbf{r})]_{k}\\
&=\sum\nolimits_{k'=1}^{K}[({\mathbf{I}}_K+{\mathbf{P}}{\mathbf{R}})^{-1}]_{k',k}h_{k'}({\mathbf{r}}).
\end{split}
\end{equation}
Combining the above results completes the proof.
\end{appendices}
\bibliographystyle{IEEEtran}
\bibliography{mybib}
\end{document}